\documentclass[reprint,onecolumn,superscriptaddress,floatfix,preprintnumbers,letterpaper]{revtex4-1}

\usepackage{array}
\usepackage{dcolumn}
\usepackage{bm}
\usepackage[colorlinks, pdfborder={0 0 0}, plainpages=false]{hyperref}
\usepackage{graphicx}
\usepackage{xspace}
\usepackage[usenames,dvipsnames]{color}
\usepackage{amssymb}
\usepackage[normalem]{ulem} 
\usepackage{lineno}
\usepackage{letltxmacro}
\usepackage{amsfonts}
\usepackage{amsmath}
\usepackage{amssymb}
\usepackage{bm}
\usepackage{dcolumn}
\usepackage{epsfig}
\usepackage{graphicx}
\usepackage{graphics}
\usepackage[latin1]{inputenc}
\usepackage{latexsym}
\usepackage{rotating}
\usepackage{hyperref}
\usepackage[usenames]{color}
\usepackage{float}
\usepackage{xspace} 
\usepackage{mathrsfs}
\usepackage{subfigure}
\usepackage{multirow}
\usepackage{booktabs}
\usepackage{soul}

\usepackage{colortbl}
\usepackage{etoolbox}
\usepackage[table]{xcolor}

\makeatletter
\def\frontmatter@maketitle{%
  \@author@finish
  \title@column\titleblock@produce
  \suppressfloats[t]%
  \let\and\relax
  \let\affiliation\@gobble
  \let\author\@gobble
  \let\@AAC@list\@empty
  \let\@AFF@list\@empty
  \let\@AFG@list\@empty
  \let\@AF@join\@AF@join@error
  \let\email\@gobble
  \let\@address\@empty
  \let\maketitle\relax
  \let\thanks\@gobble
  \let\abstract\@undefined\let\endabstract\@undefined
  \titlepage@sw{%
  }{}%
}%

\makeatother

\def\prd{{\it Phys. Rev.} D~}

\def\apjl{{\it Astrophys. J.} Lett~}
\def\apj{{\it Astrophys. J.}}

\def\mnras{{\it MNRAS~}} 

\def\aap{{\it A\&A~}}

\def\aj{{\it The Astrophysical Journal~}}
\def\pasp{{\it Publications of the Astronomical Society of the Pacific~}}

\def\lesssim{\mathrel{\hbox{\rlap{\hbox{\lower4pt\hbox{$\sim$}}}\hbox{$<$}}}}
\def\gtrsim{\mathrel{\hbox{\rlap{\hbox{\lower4pt\hbox{$\sim$}}}\hbox{$>$}}}}
\def\alt{\mathrel{\hbox{\rlap{\hbox{\lower4pt\hbox{$\sim$}}}\hbox{$<$}}}}
\def\agt{\mathrel{\hbox{\rlap{\hbox{\lower4pt\hbox{$\sim$}}}\hbox{$>$}}}}

\usepackage{mathtools}

\def\gta{\ifmmode {\mathbin{\lower 3pt\hbox   
    {$\,\rlap{\raise 5pt\hbox{$\char'076$}}\mathchar"7218\,$}}}
    \else {${\mathbin{\lower 3pt\hbox
    {$\rlap{\raise 5pt\hbox{$\char'076$}}\mathchar"7218\,$}}}
    $}\fi}
\def\lta{\ifmmode {\,\mathbin{\lower 3pt\hbox   
    {$\,\rlap{\raise 5pt\hbox{$\char'074$}}\mathchar"7218\,$}}}
    \else {${\mathbin{\lower 3pt\hbox
    {$\rlap{\raise 5pt\hbox{$\char'074$}}\mathchar"7218\,$}}}
    $}\fi}

\newcommand{\msun}{{\rm M}_{\odot}}
\newcommand{\beq}{\begin{equation}}
\newcommand{\eeq}{\end{equation}}
\newcommand{\bea}{\begin{eqnarray}}
\newcommand{\eea}{\end{eqnarray}}

\definecolor{darkperiwinkle}{RGB}{102, 102, 128}

\newcommand{\NCSA}{\affiliation{NCSA, University of Illinois at Urbana-Champaign, Urbana, Illinois 61801, USA}}
\newcommand{\ANCSA}{\affiliation{Department of Astronomy, University of Illinois at Urbana-Champaign, Urbana, Illinois 61801, USA}}
\newcommand{\PNCSA}{\affiliation{Department of Physics, University of Illinois at Urbana-Champaign, Urbana, Illinois 61801, USA}}
\newcommand{\CNCSA}{\affiliation{Department of Computer Science, University of Illinois at Urbana-Champaign, Urbana, Illinois 61801, USA}}
\newcommand{\ENCSA}{\affiliation{Department of Electrical \& Computer Engineering, University of Illinois at Urbana-Champaign, Urbana, Illinois 61801, USA}}
\newcommand{\iNCSA}{\affiliation{School of Information Sciences, University of Illinois at Urbana-Champaign, Urbana, Illinois 61801, USA}}
\newcommand{\TSNCSA}{\affiliation{Office of the CIO, University of Illinois at Urbana-Champaign, Urbana, Illinois 61801, USA}}
\newcommand{\CCS}{\affiliation{Computational Physics and Methods, Los Alamos National Laboratory, Los Alamos, New Mexico 87545, USA}}
\newcommand{\CTA}{\affiliation{Center for Theoretical Astrophysics, Los Alamos National Laboratory, Los Alamos, New Mexico 87545, USA}}

\newcommand{\CNLS}{\affiliation{Center for Nonlinear Studies, Los Alamos National Laboratory, Los Alamos, New Mexico 87545, USA}}
\newcommand{\ALCF}{\affiliation{Argonne National Laboratory, Leadership Computing Facility, Lemont, Illinois 60439, USA}}
\newcommand{\UMDCP}{\affiliation{University of Maryland, College Park, Maryland 80742, USA}}
\newcommand{\LCO}{\affiliation{Las Cumbres Observatory, 6740 Cortona Drive, Suite 102, Goleta, California 93117 USA}}
\newcommand{\Caltech}{\affiliation{California Institute of Technology, 1200 E California Blvd, Pasadena, California 91125, USA}}
\newcommand{\CaltechIPAC}{\affiliation{Caltech/IPAC-NExScI, 1200 E California Blvd, Pasadena, California 91125, USA}}
\newcommand{\CD}{\affiliation{Center for Data Driven Discovery, Caltech, Pasadena, California 91125, USA}}
\newcommand{\OKC}{\affiliation{The Oskar Klein Centre for Cosmoparticle Physics, Stockholm University, AlbaNova, Stockholm SE-106 91, Sweden}}
\newcommand{\UChicago}{\affiliation{The University of Chicago, Chicago, Illinois 60605, USA}}
\newcommand{\UDEL}{\affiliation{Department of Physics and Astronomy, University of Delaware, Newark, Delaware, 19716, USA}}
\newcommand{\UDELA}{\affiliation{Joseph R. Biden Jr,. School of Public Policy and Administration, University of Delaware, Newark, Delaware, 19716, USA}}
\newcommand{\UDELAW}{\affiliation{Data Science Institute, University of Delaware, Newark, Delaware, 19716, USA}}
\newcommand{\NYUCUSP}{\affiliation{Center for Urban Science and Progress, New York University, 370 Jay St, Brooklyn, NY 11201, USA}}
\newcommand{\NVI}{\affiliation{NVIDIA, 2788 San Tomas Expressway Santa Clara, California, 95050}}
\newcommand{\ORNL}{\affiliation{National Center for Computational Sciences, Oak Ridge National Laboratory, Tennessee, USA}}
\newcommand{\CAR}{\affiliation{Cardiff University, Cardiff CF24 3AA, United Kingdom}}
\newcommand{\IBMRES}{\affiliation{IBM T.J. Watson Research Center, Yorktown Heights, New York 10598, USA}}
\newcommand{\IBMSYS}{\affiliation{IBM Systems, Austin, Texas 78717, USA}}
\newcommand{\DIRAC}{\affiliation{DIRAC Institute, Department of Astronomy, University of Washington, 3910 15th Avenue NE, Seattle, Washington 98195, USA}}
\newcommand{\ANL}{\affiliation{Argonne National Laboratory, Lemont, Illinois 60439, USA}}
\newcommand{\CSL}{\affiliation{Coordinated Science Laboratory, University of Illinois at Urbana-Champaign, Urbana, Illinois 61801, USA}}
\newcommand{\WVU}{\affiliation{Department of Mathematics, West Virginia University, Morgantown, West Virginia 26506, USA}}
\newcommand{\GWAC}{\affiliation{Center for Gravitational Waves and Cosmology, West Virginia University, Chestnut Ridge Research Building, Morgantown, West Virginia 26505, USA}}
\newcommand{\CarnegieObs}{\affiliation{The Observatories of the Carnegie Institution for Science, 813 Santa Barbara St., Pasadena, California 91101, USA}}

\definecolor{light-gray}{gray}{0.9}

\begin{document}

\title{Deep Learning for Multi-Messenger Astrophysics\\ A Gateway for Discovery in the Big Data Era}

\author{Gabrielle Allen}\NCSA\ANCSA
\author{Igor Andreoni}\Caltech
\author{Etienne Bachelet}\LCO
\author{G. Bruce Berriman}\CaltechIPAC
\author{Federica B. Bianco}\UDEL\UDELA\UDELAW\NYUCUSP
\author{Rahul Biswas}\OKC
\author{Matias Carrasco Kind}\NCSA\ANCSA
\author{Kyle Chard}\UChicago
\author{Minsik Cho}\IBMSYS
\author{Philip S. Cowperthwaite}\CarnegieObs
\author{Zachariah B. Etienne}\WVU\GWAC
\author{Daniel George}\NCSA\ANCSA
\author{Tom Gibbs}\NVI
\author{Matthew Graham}\Caltech\CD
\author{William Gropp}\NCSA\CNCSA
\author{Anushri Gupta}\NCSA\ANCSA
\author{Roland Haas}\NCSA
\author{E. A. Huerta} \thanks{Corresponding author}\email{elihu@illinois.edu}\NCSA\ANCSA
\author{Elise Jennings}\ALCF
\author{Daniel S. Katz}\NCSA\CNCSA\ENCSA\iNCSA
\author{Asad Khan}\NCSA\PNCSA
\author{Volodymyr Kindratenko}\NCSA\ENCSA
\author{William T. C. Kramer}\NCSA\CNCSA
\author{Xin Liu}\ANCSA\NCSA
\author{Ashish Mahabal}\Caltech\CD
\author{Kenton McHenry}\NCSA\CNCSA
\author{J. M. Miller}\CCS\CTA\CNLS
\author{M. S. Neubauer}\NCSA\PNCSA
\author{Steve Oberlin}\NVI
\author{Alexander R. Olivas Jr}\UMDCP
\author{Shawn Rosofsky}\NCSA\PNCSA
\author{Milton Ruiz}\PNCSA
\author{Aaron Saxton}\NCSA
\author{Bernard Schutz}\CAR 
\author{Alex Schwing}\CNCSA\ENCSA\CSL
\author{Ed Seidel}\NCSA\ANCSA\PNCSA
\author{Stuart L. Shapiro}\NCSA\ANCSA\PNCSA
\author{Hongyu Shen}\NCSA\ENCSA
\author{Yue Shen}\ANCSA
\author{Brigitta M. Sip\H{o}cz}\DIRAC
\author{Lunan Sun}\PNCSA
\author{John Towns}\NCSA\TSNCSA
\author{Antonios Tsokaros}\PNCSA
\author{Wei Wei}\NCSA\PNCSA
\author{Jack Wells}\ORNL
\author{Timothy J. Williams}\ANL
\author{Jinjun Xiong}\IBMRES
\author{Zhizhen Zhao}\NCSA\ENCSA\CSL

\date{\today}

\begin{abstract}
\noindent This report provides an overview of recent work that harnesses the Big Data Revolution and Large Scale Computing to address grand computational challenges in Multi-Messenger Astrophysics, with a particular emphasis on real-time discovery campaigns. Acknowledging the transdisciplinary nature of Multi-Messenger Astrophysics, this document has been prepared by members of the physics, astronomy, computer science, data science, software and cyberinfrastructure communities who attended the NSF-, DOE- and NVIDIA-funded ``Deep Learning for Multi-Messenger Astrophysics: Real-time Discovery at Scale" workshop, hosted at the National Center for Supercomputing Applications, October 17-19, 2018. Highlights of this report include unanimous agreement that it is critical to accelerate the development and deployment of novel, signal-processing algorithms that use the synergy between artificial intelligence (AI)  and high performance computing to maximize the potential for scientific discovery with Multi-Messenger Astrophysics. We discuss key aspects to realize this endeavor, namely (i) the design and exploitation of scalable and computationally efficient AI algorithms for Multi-Messenger Astrophysics; (ii) cyberinfrastructure requirements to numerically simulate astrophysical sources, and to process and interpret Multi-Messenger Astrophysics data; (iii) management of gravitational wave detections and triggers to enable electromagnetic and astro-particle  follow-ups; (iv) a vision to harness future developments of machine and deep learning and cyberinfrastructure resources to cope with the scale of discovery in the Big Data Era; (v) and the need to build a community that brings domain experts together with data scientists on equal footing to maximize and accelerate discovery in the nascent field of Multi-Messenger Astrophysics.   
\end{abstract}

\pacs{Valid PACS appear here}
\maketitle


\section{Introduction}
\label{intro}

\noindent The detection of multiple gravitational wave (GW) signals, consistent with the collision of binary black holes (BBHs), has verified Einstein's theory of general relativity with exquisite detail in the most extreme astrophysical environments~\cite{DI:2016,secondBBH:2016,thirddetection,fourth:2017,GW170608,o1o2catalog}.  Furthermore, the observation of two colliding neutron stars (NSs) in GWs and light has initiated the era of Multi-Messenger Astrophysics (MMA), defined as contemporaneous observations of astrophysical phenomena using GWs, electromagnetic radiation, neutrinos and cosmic rays~\cite{bnsdet:2017,2017arXiv171005836T}, which provide complementary insights about the astrophysical processes and environments of MMA sources. Another recent MMA breakthrough was enabled by observations of high-energy gamma rays and cosmic neutrinos from the blazar TXS 0506+056, which provided the first observational evidence that BHs are proton accelerators, and that blazars are central engines of cosmic neutrinos and cosmic rays~\cite{NSFneu,NuetrinoMMA,IceCube:2018Sci}.

A thorough analysis of the data gathered by the LIGO and Virgo detectors during the two previous observing runs indicates that one GW source was detected for every 15 days of analyzed data~\cite{o1o2catalog}. Furthermore, new detections have refined previous calculations for merger rates, which now stand at \(9.7-101 \textrm{Gpc}^{-3} \textrm{yr}^{-1}\) and \(110-3840 \textrm{Gpc}^{-3} \textrm{yr}^{-1}\) for BBH mergers, and NS collisions, at 90\% confidence level. For the yet undetected neutron star-black hole (NSBH) merger scenario, a merger rate upper limit at the 90\% confidence level is estimated to be \(610 \textrm{Gpc}^{-3} \textrm{yr}^{-1}\)~\cite{o1o2catalog,pops:2018}. As the worldwide network of kilometer-scale, GW detectors continues to expand, and each detector gradually reaches design sensitivity, their volumetric sensitivity will continue to grow, thereby increasing these merger rates in upcoming observing runs~\cite{o1o2catalog}.

The combination of GW observations with large scale astronomical surveys has also enabled GW standard-siren measurement of the Hubble constant with and without electromagnetic counterparts~\cite{GWH:NaturA,Maya:2018F,darkgwss:2019}, as envisioned by Schutz~\cite{Schutz:1986Nature}. Once LIGO and its international counterparts observe \({\cal{O}}(100)\) GW170814-type BBH mergers, and these observations are combined with Dark Energy Survey-type photometric redshift galaxy catalogs, it will be possible to conduct statistical measurements of the Hubble constant with \(\sim 5\%\) statistical precision~\cite{darkgwss:2019}. The availability of deeper, more accurate and more numerous photometric-redshift information with next-generation electromagnetic surveys, such as the Large Synoptic Survey Telescope (LSST)~\cite{lsstbook}, combined with the ability of GW detectors to localize signals within ten square degrees by the next decade, implies that Gravitational Wave Cosmology will soon become a standard cosmological probe, as accurate as quasar time-delay strong lensing~\cite{darkgwss:2019,Birrer:2018}.

As the number of MMA observations continues to grow in years to come, it will be possible to conduct statistical analyses to infer the currently unknown matter composition of NSs, the origin of long gamma ray bursts, and the mass distribution of BHs and NSs. We will also learn about mechanisms that may trigger electromagnetic radiation when BHs and NSs collide, and will pinpoint the location of these events with unprecedented accuracy, shedding light on the astrophysical environments where compact objects form and coalesce as a function of cosmic time~\cite{Singer:2014ApJ,marichaMMA:2016,imre:2017}.

In order to fully realize these science goals, it is necessary to address outstanding computational challenges that currently limit the scope of existing MMA searches. In the context of GW discovery campaigns, template matching algorithms used for low-latency analysis can only probe a 4-D signal manifold out of the 9-D parameter space that is available to GW detectors. It is reported in the literature that expanding the existing parameter space into higher-dimensional signal manifolds that are astrophysically motivated is computationally infeasible~\cite{Huerta:2017a,huerta:2018PhRvD,2016PhRvD..94b4012H}. On the other hand, template agnostic algorithms are tailored for burst-like GW sources~\cite{Sergey:2016}, and they may miss \(\gtrsim {\cal{O}}(\textrm{second-long})\) GW signals with moderate signal-to-noise ratios~\cite{secondBBH:2016}.

Similarly, signal-processing algorithms used to identify electromagnetic transients in telescope images are still based on rudimentary machine learning (ML) techniques that do not adequately handle TB-size datasets in real-time analyses. Addressing this limitation is of paramount importance for at least two reasons. First, the construction of LSST will enable the most detailed survey of the southern hemisphere, covering 10 square degrees with every single snapshot taken by the telescope. Every night, LSST will generate over 15 TB of data~\cite{lsstbook}, producing thousands of triggers per second that will encode information about electromagnetic transients and other astrophysical sources in the nearby Universe. Second, by the time the LSST starts producing data, the existing network of ground-based GW detectors will be able to localize MMA sources within ten square degrees. This means that a single LSST observation may contain optical counterparts of GW signals, and it is likely that both LSST and GW detectors will perform simultaneous MMA observations. Therefore, processing both LSST and GW data in real-time will require a different approach to process simultaneously both time-series data and telescope images, leading to the identification of GW sources and their optical counterparts within seconds. For instance, LIGO and Virgo observed the collision of two neutron stars (GW170817) in GWs, and less than two seconds later, the Fermi satellite observed gamma ray emission from this same system, which underscores the speediness with which several windows of observation need to be coordinated to capture key astrophysical information from these objects. It is worth noting that MMA not only encompasses real-time observations, since radio emission from this event peaked more than 160 days after its observation in GWs, and X-ray emission may be observed until 1000 days after the neutron star collision~\cite{alex:2018ApJLA}. In this document, however, we will concentrate on low latency MMA observations.

In summary, to fully realize the MMA science program we need to address two specific problems: (i) the depth and speed of GW searches are insufficient, so we need to develop algorithms that significantly increase their speed and efficacy; and (ii) to effectively search for optical counterparts of GWs, we need to develop new algorithms to process a large number of triggers generated by LSST-type surveys, and then search for an electromagnetic counterpart in telescope images that span \(\sim 10\, \textrm{sq. degrees}\).  

Similarly, understanding the physics of MMA sources is of paramount importance to inform the development of algorithms to extract as much information as possible from real-time observations, and to define the scope and length of broadband electromagnetic and astro-particle follow-up observations. GW170817 made evident the need to further strengthen community efforts to accelerate the development of numerical relativity software stacks that are essential both for LIGO data analyses searches, and to obtain insights about the nature of the post-merger remnant from electromagnetic follow-up studies. 

In order to start addressing these pressing issues, synergistic activities between domain experts and members of the data science, high performance computing (HPC), and cyberinfrastructure communities must be organized and pursued in earnest. This document provides a vision to achieve this goal, building upon ongoing activities that are successfully harnessing the Big Data Revolution to address computational challenges in GW astrophysics, large scale astronomical surveys and neutrino observations.  This document is organized as follows. Section~\ref{sources} provides a brief status update of numerical methods used to simulate MMA sources. Section~\ref{nr_future} describes key improvements that are needed in existing numerical relativity software stacks to gain a better understanding of the physical processes that lead to GW, electromagnetic and astro-particle production in the merger of compact binary systems. Section~\ref{optical} outlines existing challenges in optical follow-ups of GW sources, and recent developments on the use of machine and deep learning to optimize these searches. Section~\ref{gws} presents a summary of machine and deep learning applications for GW data analysis. Section~\ref{gw_future} provides a vision to build upon this work to increase the depth and speed of GW searches in the context of MMA discovery campaigns. The cyberinfrastruture requirements to achieve convergence between deep learning, large scale computing and MMA is described in Section~\ref{cyberi}. Section~\ref{needs} identifies key milestones, both technical and sociological, that need to be accomplished to maximize the scientific return of MMA. Acknowledging the transdisciplinary nature of the MMA science program, Section~\ref{comm} provides a vision to build an MMA community that enables domain experts from disparate fields of research to contribute their expertise on an equal footing, and the need to create synergies with industry partners that in addition to co-funding data science activities at the heart of MMA, may also provide avenues to diversify the career paths of MMA researchers. We summarize the findings of this report in Section~\ref{end}.


\section{Numerical Relativity Simulations of Multi-Messenger Sources}
\label{sources}

Modeling an MMA source requires a pipeline rather than a single simulation. The expected messengers from a compact binary merger are, in order: GW emission from the inspiral and merger of a compact binary system, gamma ray emission from the central post-merger engine, and an optical and infrared afterglow from the radioactive decay of r-process elements in ejected material. Each of these processes requires one, if not many simulations. Accurately modeling the GW emission requires capturing the general relativistic effects. Modeling the formation of the jet and ejected material requires magnetic fields. Modeling the r-process elements requires neutrino transport and nuclear reaction networks.

The inspiral and merger of two compact objects can be modeled via numerical relativity (for a review of numerical relativity see~\cite{Shibata2011LLR,Faber2012LRR,BrugmenScienceReview18,DuezReview19} and references therein). This calculation provides the GW signal and the morphology and composition of the dynamical ejecta. Calculations with general relativistic magnetohydrodynamics (GRMHD) are now routine. For instance, the GRMHD simulation of merging NSBH and NSNS systems reportedly in~\cite{prs15,Ruiz:2016rai} showed for the first time that these systems can launch jets and be the progenitors of short gamma ray bursts (sGRBs). In the BHNS scenario, it was found that a jet can be launched if a) the NS is endowed with a magnetic field that extends from the stellar interior into the exterior, as in a radio pulsar, b) the tilt angle between the magnetic moment and the total angular momentum of the system is small, and c) the initial BH spin is $a/M_{\rm BH}\gtrsim 0.4$~\cite{Ruiz:2018wah}. In the NSNS scenario, the GRMHD studies reported in~\cite{Ruiz:2016rai,Ruiz:2017inq}, in which the NS is modeled as an irrotational $\Gamma=2$ polytrope, showed that NSNS systems may launch an incipient jet whether or not the seeded poloidal magnetic field is confined or not to the NS interior, as long as the binary forms a hypermassive NS that undergoes delayed collapse to a BH. The lifetime of the jets [$\Delta t\sim 100(M_{\rm NS}/1.625M_\odot){\rm ms}$] and the outgoing electromagnetic luminosities [$L_{jet}\sim 10^{51}\rm erg/s$] in the above cases turn out to be consistent with sGRBs~\cite{Bhat:2016odd,Lien:2016zny,Svinkin:2016fho}. Based on the GRMHD simulations in~\cite{Ruiz:2017due}, it has been argued that in order to have a sGRB as in GW170817 the merger remnant is likely a hypermassive NS that undergoes delayed collapse. Furthermore, the GW170817 data can be combined with causality arguments to establish a NS upper mass limit in the range \(2.16\msun-2.28\msun\).

The inclusion of more sophisticated neutrino and microphysics, however, is in its infancy. The most common approach to ignore neutrinos or use neutrino leakage and the state of the art is a moment scheme~\cite{Deaton:2013,Neilson2014leakage,Wanajo_2014,RadiceDynamicalMass,FoucartPostMerger}---although neutrinos have been modeled more accurately with Monte Carlo methods in post-processing~\cite{RichersSedonu,FoucarMCMoment}. After the merger, the remnant-disk-wind system must be modeled to accurately capture the formation of the jet and the composition and morphology of the gravitationally unbound material driven off of the disk. This requires general relativity in a fixed, rather than dynamic, background, magnetic fields, and neutrino transport. Recently there has been a significant effort to model this system in three dimensions~\cite{FoucartPostMerger,NouriPostMerger, SiegelMetzger3DBNS, FernandezLongTermGRMHD}. The launching and propagation of the jet itself requires additional specialized modeling as well (see~\cite{BaiottiRezzollaReview,KUMAR20151} and references therein for a review).

Rapid, or r-process, nucleosynthesis drives the formation of heavy elements in the ejected material. The r-process is generally modeled in post-processing via codes such as~\cite{LippunerSkynet}. These nucleosynthetic yields are then fed into radiative transport models to extract the light curve for the kilonova, yielding results such as those presented in~\cite{KasenOpacitiesBNS} and~\cite{TanvirHST}.

While NSNS mergers have been confirmed as progenitors of GW and EM emission, and NSBH mergers are expected to also fall into this category, there have been just a handful of studies that explore the feasibility of observing EM emission from BBH mergers~\cite{ConBurGold16,VerTavUrs17}. For instance, recent GRMDH simulations suggest that a circumbinary disk accretion onto non-spinning, stellar-mass black hole binaries may launch magnetically driven jets (or a collimated, mildly relativistic outflow which is at least partially magnetically dominated) whose Poynting luminosity is of order $0.1\%$ of the accretion power, i.e., BBH systems could serve both as radio, X- and gamma-ray engines. Future MMA observations will confirm or rule out these predictions.


\section{Future developments of numerical relativity simulations of MMA sources}
\label{nr_future}

Moving towards a more realistic treatment of MMA sources requires a number of improvements in existing software stacks. For instance, a realistic nuclear equation of the state with density, temperature and composition dependence to better capture not only the inspiral and merger effects, but more importantly to study the remnant with its subsequent BH-disk evolution. In the late merger phase, neutrino transport and realistic magnetic fields are important elements to simulate the sGRB and kilonova scenarios reliably~\cite{rad:2018ApJR}, as well as the implementation of electromagnetic radiative transport to obtain realistic estimates of the disk geometry, the mass accretion rate and electromagnetic luminosity and spectra.

Astrophysics modeling occurs on many disparate length and energy scales, ranging from the very small to the very large. In many cases, the relevant scales can be modeled via, e.g., mesh refinement~\cite{BERGER198964}. Unfortunately, if the small scales are relevant everywhere within a simulation domain, global simulations can be prohibitively expensive. One such example system is magnetically driven turbulence in accretion disks. It is well known that the magneto-rotational instability can drive turbulence and momentum transport within a disk~\cite{BH91,BH98}. This transport can be modeled as an effective ``alpha" viscosity~\cite{ShakuraSunyaevAlpha} that contains only local shear stresses. This viscosity can be matched to high-resolution models of turbulence within the disk~\cite{HawleyGammieBalbusShearingBox}.

This approach has proven to be extraordinarily powerful. However, these methods fail to capture many relevant effects of the disk-wind system, such as jets, magnetically driven wind, and magnetic arrestment. One potential avenue for improvement is to adapt the more sophisticated models developed by the engineering community, such as large eddy simulations and Reynolds-averaged Navier-Stokes models. These models incorporate many more locally-defined viscous stress and transport terms, but require more sophisticated tuning~\cite{ZHIYIN201511} .

The engineering community has recently begun exploring the possibility of using tuned subgrid models for turbulence to train neural networks for turbulence and the field has exploded~ \cite{DZS15AIAA,PD16JCOMP,WWX17PRF,KHC17APS,MS17JFM,MJ17,BHP17,Hennigh17,XFCT18,MGD18,KHMC18,WBT18}. Although these techniques are in their infancy, they have the potential to enable accurate subgrid modeling in regimes that were previously intractable.

One promising application for ML in simulating MMA sources is the use of ML to model subgrid physics that cannot currently be modeled from first principles at acceptable computational costs~\cite{weinan2017deep,berg2018unified,chen2018neural}. For example fully resolved simulations that capture turbulent motion during supernovae or neutron star collision simulations~\cite{Mosta:2015ucs,Kiuchi:2017zzg,Radice:2015qva} are too expensive to employ for parameter study simulations. Instead the use of subgrid models tuned via a small number of high resolution simulations has become increasingly popular in astrophysical simulations~\cite{Giacomazzo:2014qba}. These type of models have been used for a long time in computational fluid dynamics where recently the use of ML to model closure parameters of the system has seen increasing use.

These methods thus require access to advanced cyberinfrastructure platforms to handle both the few high resolution, parallel simulations used to tune the subgroup model as well as infrastructure to construct the model and eventually infrastructure for a large set of simulations to study MMA sources. Full, high resolution, parallel simulations to tune subgrid models will require tens of millions of node hours or more, which can only be delivered by high-performance resources on par or better than current NSF leadership compute resources like Blue Waters or the future Frontera system. 


\section{Needs for low-latency identification of the optical counterparts of MMA sources}
\label{optical}

The optical counterparts of GWs emitted during the merger of NSNS or NSBH systems are known as kilonovae or macronovae~\cite{Metzger:2012,Barnes2013,Kasen2013,2015IJMPD..2430012R,villar:2017ApJ,2017arXiv171005931M}. This emission spans the ultra violet (UV), optical and the near infrared (NIR) bands, and encodes key insights regarding ejected material that is powered by radioactive decays of r-process nuclei~\cite{SiegelMetzger3DBNS,Tanaka_2018}. As had been expected, the landmark discovery GW170817 demonstrated that such sources can power sGRBs structured jets~\cite{alex:2018ApJLA}. Future MMA observations of these sources will provide new and detailed insights about the astrophysical scenarios that lead to the emission or suppression of GRB jets.

The optical counterpart is one of the crucial messengers of MMA observations, since detection in the UV/Optical/NIR is the best, and perhaps, the only method of localization of the source sufficient for unambiguously identifying the host galaxy~\cite{2041-8205-848-2-L17}. The detection is aided by a suitable pipeline of transient detection, e.g.,~\cite{2015AJ....150..172K}, including automated classification of astrophysical vs algorithmic artifacts, e.g.,~\citep{2015AJ....150...82G}. Such an unambiguous identification allows an easy determination of the redshift, and therefore their use in a distance-redshift relationship to infer cosmological parameters in conjunction with the luminosity distance determined from the standard siren property of GWs~\cite{Schutz:1986Nature,2003CQGra..20S..65H}. This is currently anticipated with much excitement as an independent method of inferring the Hubble constant in the local universe, since a tension of the order of three sigma exists in the inferred value from high-redshift cosmic microwave background data, and distance ladder measurements~\cite{2018arXiv180203404F,2018Natur.562..545C,2018arXiv181111723M}. This science requires the construction of complete galaxy catalogs, an activity in which machine and deep learning (DL) is in earnest exploration~\cite{asad:2018K,Dom:2018MNRAS,Dom:2018D}. Such a precise localization is also crucial for triggering deep multi-wavelength follow-ups and spectroscopy studies, since relevant telescopes have small fields of view. Such monitoring is essential to extract physical properties of the merger, and detailed modeling of the sources such as the physics of the jet. The location of the source within its host galaxy can provide useful information about the evolution of the progenitor system,  and even its dynamics in the host galaxy~\cite{2041-8205-848-2-L28}. Multi-wavelength follow-up enables our understanding of physical processes that take place during MMA events. For example, those include the production of heavy elements and diagnostics regarding the mass and composition of the ejected material, properties of the circumstellar medium, and whether jets are generated during the event~\cite{Molley:2018Natur}.

High confidence identification of the optical counterparts of GWs is a challenging task. These transients have peculiar optical properties, namely, they are fast, dim, and rare. Furthermore, the area within which LIGO-type detectors can localize MMA triggers includes many other unrelated optical transients (including artifacts and supernovae) whose properties will be consistent with optical counterparts of GW events. Due to their fast decay rate (the initial optical emission of the counterpart of GW170817 faded more than 1 mag/day,  and  was  followed  by  a  longer-lived  red transient~\cite{drout:2017SCIENCE}) and to the particular usefulness of early observations in constraining the physical properties of the transient, they need to be located within a few hours of the compact binary merger, and then followed up using relatively competitive spectroscopic resources, and observations on the entire electromagnetic (EM) spectrum (for an actual detection scenario, see Figure 2 in~\cite{mma:2017arXiv}, which itemizes the timeline of GW and broadband EM observations of GW170817). In practice, this requires prompt response to the trigger to initiate a multi-filters discovery imaging campaign that will cover large sky areas using large field-of-view, deep-imaging telescopes, or campaigns imaging preselected galaxies (selected for their morphological properties and distance) with deep-imaging telescopes. In both cases, the campaigns are conducted using fully automated discovery pipelines that perform image subtraction. ML is already a standard component of such pipelines~\cite{igor:2017PASA,gold:2015AJG}, and plays a critical role to tell apart real astrophysical sources from noise artifacts. Rapid integration with archive data will be valuable in pinpointing the optical counterparts. In addition to post-processing applications, neural networks are now being used as key discovery methods in new pipelines, in the optical~\cite{seda:2018MNRAS} as well as at other frequencies. In fact, image-subtraction based campaigns require the existence of adequate templates over the entire are of sky to be searched, as inadequate or missing templates would lead to an overwhelming amount of false positive artifacts, and DL techniques can be faster and more accurate than image-subtraction methods for transient discovery, especially when the amount of data to be processed grows at TB scales. Such techniques are being tested to be fully implemented in future large surveys and MMA follow-up programs. Finally, aside from using these techniques for identifying a real astrophysical source, and then using spectroscopy to make the positive identification as a kilonova, future analyses may also utilize photometric classification methods for identification. Such possibilities have been studied in quite some detail for other transients like supernovae~\citep{2010PASP..122.1415K,2016ApJS..225...31L,2017ApJ...837L..28C} or methods applied to real data~\citep{2013ApJ...763...88C,2018ApJ...857...51J}. The possibility of identifying kilonovae using this approach from wide field surveys like LSST using complete light curves in a serendipitous survey have been discussed in~\cite{2017CQGra..34j4001R,2018ApJ...852L...3S, 2018arXiv181103098C,2018arXiv181210492S}. Graph neural networks have also been used to improve neutrino detection with the IceCube observatory. This detection scenario combines two challenging problems, namely, the sparse nature of the neutrino signals, the irregular geometry of the detectors, and the large asymmetry between positive and negative events, in which case the pattern classification problem demands false positive rates of order \(10^{-6}\)~\cite{choma2018graph}.


\section{State-of-the-art deep learning algorithms for gravitational wave detection}
\label{gws}

While DL algorithms have been used for classification studies of time-series signals~\cite{ismail:2018I}, the first demonstration that these algorithms could be used both for classification and regression of time-series data was presented in~\cite{geodf:2017a}, with an application for the detection and parameter estimation of BBH mergers whose GWs are embedded in Gaussian noise. Follow-up studies demonstrated that this methodology is also applicable for the detection, parameter estimation and denoising of BBH mergers, detected by the LIGO and Virgo detectors, whose GWs are embedded in non-Gaussian and non-stationary LIGO data~\cite{geodf:2017b,hshen:2017,wei:2019W,Rebei:2018R,shengeorge:PhysRevD97}. These studies have sparked the interest of the community, leading to several developments on ML and DL for GW data analysis and source modeling~\cite{AlvinC:2018,2018GN,Fan:2018,Gonza:2018,2018GN,Fuji:2018,LiYu:2017,Nakano:2018}.

Central developments in this field require the use of Bayesian neural networks to enable parameter estimation analyses endowed with statistical errors~\cite{NIPS2016_6117,gal2016dropout,gal2015bayesian}. In order to probe a higher-dimensional signal manifold, beyond the 4-D parameter space covered with traditional detection algorithms, it will be essential to combine DL and HPC to use distributed training at scale. Such an approach is needed both to conduct large scale parameter sweeps to determine the optimal architecture of neural network models. Once a neural network model is chosen, the use of HPC platforms is vital to have the flexibility to train neural network models using TB-size waveform template banks to achieve state-of-the-art accuracy for detection and characterization of GW signals in actual detection scenarios. Reducing the training stage from weeks to minutes using distributed training in HPC platforms not only enables the design and training of robust neural network models, and detailed uncertainty quantification studies to assess the robustness of the model, but also guarantees that once a model is fully trained, one can use information about the sensitivity of the GW detectors, during an ongoing detection search, to enhance the accuracy of neural network models for low-latency GW detection and parameter estimation studies.


\section{Future developments of deep learning algorithms for gravitational wave detection}
\label{gw_future}

Existing DL-based signal processing tools have only been explored in the context of BBH mergers, assuming \({\cal{O}}(\textrm{second-long})\) GWs. There is a pressing need to expand the scope of these methods to cover the 9-D signal manifold that is available to existing GW detectors. 

Furthermore, it is critical to design neural network models adequate for the detection and characterization of NSNS and NSBH mergers. Obvious challenges related to this work involve the fact that these neural network models will have to be trained with significantly longer waveform signals. Since existing DL algorithms~\cite{geodf:2017b} are capable of processing over 4,000 seconds of data within a second, one expects that even if DL algorithms for NSNS observations are \(\sim 1,000\textrm{x}\) slower than those used for BBH detection, one may still be able to extract NSNS and NSBH signals from GW data in low latency. 

Other developments that involve MMA observations concern GW observations of supernovae, and other burst-like GW sources. A study of this nature is a straightforward extension of the algorithms already developed for GW detection, denoising and unsupervised clustering that have been developed for the GW observation of BBH mergers~\cite{dgNIPS,shengeorge:PhysRevD97}.


\section{Cyberinfrastructure requirements for the development of deep learning algorithms for MMA searches}
\label{cyberi}

As pointed out in the previous sections, DL is steadily being implemented in all aspects of the MMA analysis pipeline.  Key features of DL algorithms that promote them as tools to address existing grand computational challenges in MMA include their scalability, and the fact that once these algorithms are fully trained, they require minimal computational resources for inference, or in different words, in an actual MMA discovery campaign. 

With packages like \texttt{TensorFlow}~\cite{abadi2016tensorflow} and \texttt{PyTorch}~\cite{paszke2017automatic}, much of the GPU/CPU optimization has been done for us. However MMA will continue to push the computation and data limits of any cyberinfrastructure. Scaling these limits can broadly be described in two ways: horizontal and vertical scaling. Vertical scaling is generally perceived as ``make the CPU/GPU processing faster", and horizontal scaling as ``give me more CPU/GPUs." Different stages of a DL pipeline have varying success taking advantage of these two scaling strategies. The major components of a DL pipeline are data preparation, model training, model inference, and results analysis. For this part of the discussion we ignore data preparation and results analysis as they are typically either done offline or are integrated into the DL model and thus fall into the category of training or inference. 

Training is by far the most computationally intensive step. It requires a significant amount of interprocess/thread communication and sequential calculations. Most mature and well studied models can take advantage of horizontal scaling by using large training batch sizes. However, new models that are being developed and studied can suffer from poor generalization with horizontal scaling and they require careful data-specific scaling strategies. Training requires a rounded balance of horizontal and vertical scaling. Inference is commonly an embarrassingly parallel process. Each input can be processed completely independently of any other input. Depending on the vertical scale of computation, the bottleneck can often be data retrieval and I/O latency and bandwidth. 

In a production grade cyberinfrastructure, source and revision control is an integrated feature. Studying DL models for MMA revealed many best practices that should be carried forward into an eventual production cyberinfrastructure.  With the most recent DL programing models, e.g., \texttt{TensorFlow} and \texttt{PyTorch}, a DL model is made up of code plus numeric weights. \texttt{TensorFlow} popularized the concept of a model graph and created file formats to store such objects, but a DL model is still created and maintained as traditional developer code. As the project matures, errors and improvements on models will be discovered. The best practices are to treat the code as part of a model as if it was any other software project with source/revision control such as github and a gitflow style development cycle. The tricky part is traceability of model weights. Storing the literal model weight is always a valid approach. However, the weights by themselves give very little diagnostic insight and so is a poor candidate to source control. From an information theoretic perspective, the training data, training method, and hyper parameters, and the information on DL software framework are equivalent to the raw model weights. This report has emphasized the need for a common public datastore. We propose source controlling the metadata needed to fetch training data from the eventual common public datastore.

Cyberinfrastructure for MMA analytics needs to support large collections of disparate datasets, and the ability to apply complex and compute-intensive analytics algorithms on large portions of the data. This type of cyberinfrastructure can be satisfied with an low interconnect latency HPC-like system, with adequately provisioned data and compute resources. On the other side, the infrastructure must support interactive access to the data, and the ability to apply analytics tools to subsets of the data in real-time for a large number of users. 

At the time of this writing (January 2019) the resources available from the NSF via XSEDE offer a decreasing number of resources appropriate to support MMA efforts.  At present, XSEDE compute resources provide 2.8 PFlop/s of single precision GPU compute through the Bridges and Comet clusters~\cite{comet,bridges}. With the impending retirement of several resources there will be a continued decline in the availability of GPU resources.  A current solicitation will bring new resources online in the 2020 time frame, but it is unknown as to what the configuration of these resources will be (as they will be selected based on proposals that have not yet been written) and whether they will effectively support the computing modalities needed by MMA. 

The NSF-funded Frontera supercomputer focuses on CPU performance providing 35-40 PFlop/s through its Intel Xeon CPUs, and only 8 PF/s of 32-bit precision compute through a small GPU section of the system~\cite{frontera}. While explicitly designed with ML in mind, MMA ML applications will easily consume more compute resources than available through Frontera's GPU section. Other NSF-funded investments include the DL project at NCSA, which consists of a computer system dedicated to supporting DL research at the University of Illinois at Urbana-Champaign.  The system consists of 16 IBM Power9 servers with four NVIDIA V100 GPUs, or 64 GPUs in total. The nodes are interconnected with dual-channel EDR IB. The system supports \texttt{TensorFlow} and other DL frameworks in a container-based environment. 

In summary, the landscape of existing and planned NSF-funded HPC cyberinfrastructure indicates that there is a pressing need to expand these resources to meet MMA cyberinfrastructure needs~\cite{HuertaES,8109152,2017Weitzel,shifter}. In addition, it is unclear as to whether there are sufficient technical staff available with whom to partner and gain the necessary expertise to further develop these techniques.  Given the much broader applicability of ML/DL/AI approaches, the agencies are advised to invest in the resources and services necessary to enable the success of MMA and other community efforts reliant on these approaches.  This goes beyond simply provisioning GPU systems and the technical staff required to support them.  Appropriate storage and data sharing environments that can support multidisciplinary collaborations along with adequate networking infrastructure to support both bulk data transport and rapid notifications is required.  There must also be exploration of newer technologies, e.g., tensor processing units, quantum information systems, and of their applicability to these challenges. 

On the other hand, Department of Energy (DoE) HPC and leadership computing centers provide access to state-of-the-art computing platforms such as Summit and Titan at Oak Ridge National Laboratory which are equipped with over 27k NVIDIA Volta V100 GPUs, and over 18k K20X GPUs, respectively. Theta at Argonne National Laboratory, though not providing GPUs, has the infrastructure in place to conduct DL research at scale. In 2018, the DoE's Innovative and Novel Computational Impact on Theory and Experiment (INCITE) and Argonne Leadership Computing Facility Data Science Program (ADSP) provided computing resources for data science projects with an emphasis on machine and DL research at scale, filling in a critical void in the US HPC landscape for this emergent area of research, which are essential for the realization of the MMA science program.


\section{Future needs of Multi-Messenger Astrophysics} 
\label{needs}

To maximize the scientific return of MMA in the near future, progress on several fronts has to be made. It is of primary importance to obtain alerts from GW detectors as soon as  possible, but also from high energy surveys such as Fermi, to the optical follow up community to select the most promising targets. A list of potential candidates is essential to enable every team to rank the targets by priority, given their observing capabilities and science goals. A common tool for galaxy cross matching is highly desirable, as currently this task is duplicated by each team.  In addition, research into a ML classifier for MMA is highly desirable for an early classification.

For observational programs, it grows more and more important that all steps from target selection, to the planning and execution of observations be done programmatically. This avoids, in principle, any human biases and mistakes and could resolve some potential conflicts. The two first steps can be achieved with new generation of observing software, the so-called Target and Observation Manager (TOM) systems~\cite{street:2018SPIE}. These ensure an efficient way to manage and submit observations at large scale. These tools are specially designed to have the flexibility to interact with any instrument/telescopes in order to maximize the efficiency of the observing process.  The systems provide an efficient platform for communication between observers.

There needs to be discussion of the needs of the future generation of telescopes that will be used for MMA follow-up. With the upcoming new generation of large telescopes (LSST, Thirty Meter Telescope, Extremely Large Telescope, etc.), it is crucial to design future instruments to accurately characterize GW events. Moreover, the amount of telescope time that will be dedicated to MMA has to be decided. The observing strategy, mechanism and software infrastructure should be flexible enough to allow a rapid reaction to any potential targets of primary importance.

Ideally, a common, public database would be build to centralize all information available for any targets. This will minimize the duplication of common tasks (such as data collection, galaxy cross-matching, ranking, etc.) among follow-up teams, therefore allow their focus to be dedicated to producing the best science. This can be coupled with classification algorithms and catalogs cross-matching. The infrastructure of the LSST data management system and the alert brokers under development  are good examples. 

The development and the generalization of TOM tools in parallel to the robotization of telescopes is a important step that has to be explored to maximize the science return of future MMA events. The architecture and operational practices  of primary detection instrument should be examined for improvements for MMA use cases.  For example, while primarily a general-purpose astronomy instrument,  LSST is able to respond promptly to GW detections. Access to these observations  for the general community  is currently seen as implemented thought community event brokers. The match of this architecture to MMA use case should be validated. 

Finally, it is crucial that funding agencies explore the best way to manage competitive proposals that try to achieve the same scientific goals. Given the scientific value of these future events, several teams are likely to simultaneously request the same observations at the same facilities, creating inefficient duplication and potential for conflict. With increasing rates of target discovery this inefficiency could become a significant drain on limited observing resources. While DL/ML software can manage the technical issues relevant to these problems, the political aspects remain to be addressed.

\section{Community Building}
\label{comm}

MMA, like many other fields, is inherently both collaborative and competitive. Competition can be both constructive and destructive.  Destructive competition inhibits collaboration and teamwork and in the burgeoning MMA field, is counterproductive.  However, because MMA is transdisciplinary in nature, and its science goals will not be realized by any one individual or small group, similar to how it necessarily requires large groups several years to put multiple satellites in orbit, build vast surface arrays, or instrument a volume of Antarctic ice, both collaboration and constructive competition must be encouraged and developed.  In this section, we discuss several aspects of the future MMA community that promote such collaboration and constructive competition, including software and services, workforce development, and social and technical mechanisms for community building.

\subsection{Community software and services}

While web-based services will remain useful, the predominant mechanism for accessing and analyzing data in the immediate future will likely be through the command-line in a Python environment. Technologies such Jupyter Notebooks are enabling the sharing and understanding of complex workflows. We encourage the MMA community to develop Python interfaces to all software packages. In addition the MMA community should be asked to share their workflows via Notebooks. In so far as is possible, we encourage software teams to license their software with a permissive license to optimize sharing and re-use of code. Housing the code in a common repository (unless forbidden for legal reasons) will support discovery and use of the data. Policies for use of the repository will require care and attention, but it will be understood that the software providers own the code they contribute. Successful projects such as Astropy~\cite{astropy} and yt~\cite{ytproject} provide examples of how community projects can be managed, how credit and reward can be assigned, and how conflict can be managed. And in some cases, MMA work can be merged into these existing packages.

Astronomy archives will be developing data access services aimed at maximizing science return from MMA. The NASA Extragalactic Database (NED) is actively developing an EM-GW follow-up service for the advanced LIGO-Virgo third observing run in Spring 2019, to optimize fast EM follow-ups to GW events. Its design has been informed by NEDs support for follow-up to the GW170817 event. The MMA community should be active in identifying follow-up services, their requirements and their performance, and in evaluating the services.

The reproducibility of results is one of the fundamental tenets of the scientific method.  When it comes to complex software stacks, it is important to keep in mind that reproducibility within systematic and statistical errors is what is important, as opposed to bitwise reproducibility, which is the standard in software engineering.  IceCube has designed a system, soon to be released as open source~\cite{icerepo,Porter:2008ar}, that will detect systematic deviations due to improvements, bug fixes, or optimizations in the underlying algorithms but will be robust to purely statistical deviations.  This system has the potential to detect changes in continuous distributions beyond those due purely to statistics and below the level of human observations.  This was originally designed to detect small changes in large, complex simulation chains, but can be adapted to analysis chains as well.  It will be critical that this software be easy to use correctly and difficult to use incorrectly, to facilitate early adoption.  Near the end of an analysis, there is often little enthusiasm for validation and verification and less care is given to this component of the process.  This can often make results difficult for others to reproduce, even within the same experiment.  Investing in validation and verification reduces the amount of time it takes for new people to get started.  Reproducing previous results can be a great way for a new person to gain the necessary confidence to adapt and extend the analysis, resulting in better science quicker for the entire MMA community.

\subsection{Workforce development}
 
Current large scientific collaborations are composed of two largely separate communities: scientists and engineers.  Though scientists have the technical background and aptitude to communicate needs and requirements to engineers, they often lack the mindset, skills, and training to develop and design as engineers do.  In large projects, a significant fraction of the codebase is designed, developed, and maintained by students and postdocs who have very little formal training in software engineering best practices.  Their software is often integrated into complex processing stacks, intended to work for years and even decades.  Critical to the long-term performance of complex processing chains is the training of the workforce largely responsible for design, development, and maintenance.

Modern ML/DL courses offered by Udacity, for example, tend to focus on industrial applications like facial recognition and natural language processing.  The overlap with scientific applications can range from significant, e.g., facial recognition and astronomical image categorization, to orthogonal, e.g., natural language processing. Issues such as error propagation through DL architectures, critical to scientific applications, are largely ignored in industry. An initial proposal for a DL curriculum, which could serve as an extension to Software Carpentry~\cite{scar}, should include software engineering best practices, and several programming languages. It is well known that there are far fewer academic positions than degrees earned and many colleagues therefore end up in industry.  There is an opportunity to engage and compel our industrial partners to provide support in training their future workforce.  It is both cost effective and mutually beneficial.

A variety of activities may be pursued to transfer skills and to on-board new members, e.g., in-person bootcamps, either separate or in conjunction with major conferences.  NCSA would be an ideal venue, but the community should be open to other hosts. The Software Carpentry project is one successful example. Engaging such projects to provide training needed for MMA is likely to be the optimum approach. Engaging such projects to provide training needed for MMA is likely to be the optimum approach. It is important to engage industry partners in these activities. More astronomers and physicists leave academia for industry, so it's in industries' best interest to invest in their workforce training early. 

\subsection{Towards a cohesive community} 

Developing a cohesive community depends on many complex sociological factors, implemented in policies and practices as well as culture. Individuals first need to feel a sense of community, a group that is working together at some level, and to feel that they are or want to be part of that community.  Next, in order for a community to thrive, individuals need to find it rewarding (i.e., to have incentives, whether intrinsic or extrinsic) to become or to remain part of the community, to contribute back, and to urge fellow colleagues to join.  Though modern technology has made it relatively inexpensive to provide the tools to build and maintain global online communities, these tools are just tools; they are not the community. The community must form and persists, and it is important to recognize the value that face-to-face meetings, including conferences, workshops, bootcamps, can bring to community identity. Funding for these events must be sought from NSF, DOE, and industry partners. 

Defining teams as early as possible and providing the tools, e.g., on-boarding, to easily get up and running, will provide individuals with the opportunity for optimal career advancement, while also optimally advancing the goals of the community.  As different groups work together, driving scientific and technological innovation for MMA, it will be essential to implement reward mechanisms to attract and retain talent through career advancement. Furthermore, as individuals and teams engage with this community, it is essential to establish transparent mechanisms to enable them to achieve their goals, while contributing back to the team, thereby sustaining and encouraging team structures.

Many of the challenges described in this article, such as a building and sustaining cohesive community, workforce training, development and advancement; access and delivery of large volume data, low-latency data analysis, and employing ML/DL/AI techniques with the associated cyberinfrastructure required to minimize time-to-insight, are shared with other data-intensive scientific domains such as high-energy physics (HEP). Through a process involving 18 workshops over two years, key national and international partners from HEP, computer science, industry, and data science were brought together to generate over eight community position papers, including a software institute Strategic Plan~\cite{S2I2HEP-SPj} and a Community Whitepaper~\cite{CWP} as a roadmap for HEP software and computing R\&D over the next decade. The MMA community should ascertain what can be learned from community processes such as this, and seek out shared software and cyberinfrastructure solutions to challenges in common to maximize the scientific return-on-investment from MMA data. This process is already in motion with the NSF-funded project ``Community planning for Scalable Cyberinfrastructure to support Multi-Messenger Astrophysics"~\cite{whitepaper:SCIMMA}. 

\section{Conclusions}
\label{end}

Representatives from the astronomy, physics, computer science, HPC, software, and data communities gathered together at NCSA to discuss recent accomplishments of the MMA science program, and to identify grand computational and theoretical challenges that need to be addressed to maximize the potential for discovery of future MMA discovery campaigns. 

This community expressed unanimous agreement that there is a pressing need to accelerate the development of ML and DL algorithms for the GW, EM, and neutrino detection of MMA sources in real-time and at scale. MMA observations of GW170817 have been transformative, and have identified three pillars that require immediate attention for the full realization of the MMA science program: 1)  increase the speed, accuracy and robustness of numerical relativity simulations of MMA sources; 2) increase the speed and depth of signal processing algorithms for real-time detection of GWs and their counterparts in light and astro-particles; and 3) identify cyberinfrastructure resources to simulate and search for MMA sources in ever increasing and disparate datasets. 

There is concern regarding the availability of future NSF-funded cyberinfrastructure facilities for MMA data analytics. The convergence of DL with HPC to train deeper and more accurate models with TB-size training data sets is critical to ensuring that neural network models cover as deep a parameter space as possible. Once these models are fully trained, they can be used for inference searches using minimal computational resources. We suggest reaching out to DOE HPC platforms to obtain computational resources for DL at scale, given that DOE labs such as Oak Ridge and Argonne provide state-of-the-art resources for ML/DL research.

Beyond scientific development, this team recognizes the need for policy making regarding data acquisition and  data sharing at astronomical observatories and methods for fast interaction between the GW and the astronomy communities to maximize scientific discovery. We also emphasize the need to create and nurture an MMA community that designs mechanisms to reward its members, and to grow in an organic manner, providing career paths for its members beyond academia. There are successful examples in astronomy (e.g., Astropy~\cite{astropy}, yt~\cite{ytproject} and the LSST Dark Energy Science Collaboration~\cite{desc}) that can act as models for building a community. We suggest creating synergies with industry partners that, in addition to co-funding data science workshops and bootcamps for MMA researchers, may also recruit members of this community.

\section{Acknowledgements}
We gratefully acknowledge support from NVIDIA, Argonne Computing Leadership Facility and the National Science Foundation through grant NSF-1848815.
\bibliography{references,ref_two}

\begin{thebibliography}{152}%
\makeatletter
\providecommand \@ifxundefined [1]{%
 \@ifx{#1\undefined}
}%
\providecommand \@ifnum [1]{%
 \ifnum #1\expandafter \@firstoftwo
 \else \expandafter \@secondoftwo
 \fi
}%
\providecommand \@ifx [1]{%
 \ifx #1\expandafter \@firstoftwo
 \else \expandafter \@secondoftwo
 \fi
}%
\providecommand \natexlab [1]{#1}%
\providecommand \enquote  [1]{``#1''}%
\providecommand \bibnamefont  [1]{#1}%
\providecommand \bibfnamefont [1]{#1}%
\providecommand \citenamefont [1]{#1}%
\providecommand \href@noop [0]{\@secondoftwo}%
\providecommand \href [0]{\begingroup \@sanitize@url \@href}%
\providecommand \@href[1]{\@@startlink{#1}\@@href}%
\providecommand \@@href[1]{\endgroup#1\@@endlink}%
\providecommand \@sanitize@url [0]{\catcode `\\12\catcode `\$12\catcode
  `\&12\catcode `\#12\catcode `\^12\catcode `\_12\catcode `\%12\relax}%
\providecommand \@@startlink[1]{}%
\providecommand \@@endlink[0]{}%
\providecommand \url  [0]{\begingroup\@sanitize@url \@url }%
\providecommand \@url [1]{\endgroup\@href {#1}{\urlprefix }}%
\providecommand \urlprefix  [0]{URL }%
\providecommand \Eprint [0]{\href }%
\providecommand \doibase [0]{http://dx.doi.org/}%
\providecommand \selectlanguage [0]{\@gobble}%
\providecommand \bibinfo  [0]{\@secondoftwo}%
\providecommand \bibfield  [0]{\@secondoftwo}%
\providecommand \translation [1]{[#1]}%
\providecommand \BibitemOpen [0]{}%
\providecommand \bibitemStop [0]{}%
\providecommand \bibitemNoStop [0]{.\EOS\space}%
\providecommand \EOS [0]{\spacefactor3000\relax}%
\providecommand \BibitemShut  [1]{\csname bibitem#1\endcsname}%
\let\auto@bib@innerbib\@empty
\bibitem [{\citenamefont {{Abbott}}\ \emph
  {et~al.}(2016{\natexlab{a}})\citenamefont {{Abbott}}, \citenamefont
  {{Abbott}}, \citenamefont {{Abbott}}, \citenamefont {{Abernathy}},
  \citenamefont {{Acernese}}, \citenamefont {{Ackley}}, \citenamefont
  {{Adams}}, \citenamefont {{Adams}}, \citenamefont {{Addesso}}, \citenamefont
  {{Adhikari}},\ and\ \citenamefont {et~al.}}]{DI:2016}%
  \BibitemOpen
  \bibfield  {author} {\bibinfo {author} {\bibfnamefont {B.~P.}\ \bibnamefont
  {{Abbott}}}, \bibinfo {author} {\bibfnamefont {R.}~\bibnamefont {{Abbott}}},
  \bibinfo {author} {\bibfnamefont {T.~D.}\ \bibnamefont {{Abbott}}}, \bibinfo
  {author} {\bibfnamefont {M.~R.}\ \bibnamefont {{Abernathy}}}, \bibinfo
  {author} {\bibfnamefont {F.}~\bibnamefont {{Acernese}}}, \bibinfo {author}
  {\bibfnamefont {K.}~\bibnamefont {{Ackley}}}, \bibinfo {author}
  {\bibfnamefont {C.}~\bibnamefont {{Adams}}}, \bibinfo {author} {\bibfnamefont
  {T.}~\bibnamefont {{Adams}}}, \bibinfo {author} {\bibfnamefont
  {P.}~\bibnamefont {{Addesso}}}, \bibinfo {author} {\bibfnamefont {R.~X.}\
  \bibnamefont {{Adhikari}}}, \ and\ \bibinfo {author} {\bibnamefont
  {et~al.}},\ }\href {\doibase 10.1103/PhysRevLett.116.061102} {\bibfield
  {journal} {\bibinfo  {journal} {Physical Review Letters}\ }\textbf {\bibinfo
  {volume} {116}},\ \bibinfo {eid} {061102} (\bibinfo {year}
  {2016}{\natexlab{a}})},\ \Eprint {http://arxiv.org/abs/1602.03837}
  {arXiv:1602.03837 [gr-qc]} \BibitemShut {NoStop}%
\bibitem [{\citenamefont {{Abbott}}\ \emph
  {et~al.}(2016{\natexlab{b}})\citenamefont {{Abbott}}, \citenamefont
  {{Abbott}}, \citenamefont {{Abbott}}, \citenamefont {{Abernathy}},
  \citenamefont {{Acernese}}, \citenamefont {{Ackley}}, \citenamefont
  {{Adams}}, \citenamefont {{Adams}}, \citenamefont {{Addesso}}, \citenamefont
  {{Adhikari}},\ and\ \citenamefont {et~al.}}]{secondBBH:2016}%
  \BibitemOpen
  \bibfield  {author} {\bibinfo {author} {\bibfnamefont {B.~P.}\ \bibnamefont
  {{Abbott}}}, \bibinfo {author} {\bibfnamefont {R.}~\bibnamefont {{Abbott}}},
  \bibinfo {author} {\bibfnamefont {T.~D.}\ \bibnamefont {{Abbott}}}, \bibinfo
  {author} {\bibfnamefont {M.~R.}\ \bibnamefont {{Abernathy}}}, \bibinfo
  {author} {\bibfnamefont {F.}~\bibnamefont {{Acernese}}}, \bibinfo {author}
  {\bibfnamefont {K.}~\bibnamefont {{Ackley}}}, \bibinfo {author}
  {\bibfnamefont {C.}~\bibnamefont {{Adams}}}, \bibinfo {author} {\bibfnamefont
  {T.}~\bibnamefont {{Adams}}}, \bibinfo {author} {\bibfnamefont
  {P.}~\bibnamefont {{Addesso}}}, \bibinfo {author} {\bibfnamefont {R.~X.}\
  \bibnamefont {{Adhikari}}}, \ and\ \bibinfo {author} {\bibnamefont
  {et~al.}},\ }\href {\doibase 10.1103/PhysRevLett.116.241103} {\bibfield
  {journal} {\bibinfo  {journal} {Physical Review Letters}\ }\textbf {\bibinfo
  {volume} {116}},\ \bibinfo {eid} {241103} (\bibinfo {year}
  {2016}{\natexlab{b}})},\ \Eprint {http://arxiv.org/abs/1606.04855}
  {arXiv:1606.04855 [gr-qc]} \BibitemShut {NoStop}%
\bibitem [{\citenamefont {{Abbott}}\ \emph
  {et~al.}(2017{\natexlab{a}})\citenamefont {{Abbott}}, \citenamefont
  {{Abbott}}, \citenamefont {{Abbott}}, \citenamefont {{Abernathy}},
  \citenamefont {{Acernese}}, \citenamefont {{Ackley}}, \citenamefont
  {{Adams}}, \citenamefont {{Adams}}, \citenamefont {{Addesso}}, \citenamefont
  {{Adhikari}} \emph {et~al.}}]{thirddetection}%
  \BibitemOpen
  \bibfield  {author} {\bibinfo {author} {\bibfnamefont {B.~P.}\ \bibnamefont
  {{Abbott}}}, \bibinfo {author} {\bibfnamefont {R.}~\bibnamefont {{Abbott}}},
  \bibinfo {author} {\bibfnamefont {T.~D.}\ \bibnamefont {{Abbott}}}, \bibinfo
  {author} {\bibfnamefont {M.~R.}\ \bibnamefont {{Abernathy}}}, \bibinfo
  {author} {\bibfnamefont {F.}~\bibnamefont {{Acernese}}}, \bibinfo {author}
  {\bibfnamefont {K.}~\bibnamefont {{Ackley}}}, \bibinfo {author}
  {\bibfnamefont {C.}~\bibnamefont {{Adams}}}, \bibinfo {author} {\bibfnamefont
  {T.}~\bibnamefont {{Adams}}}, \bibinfo {author} {\bibfnamefont
  {P.}~\bibnamefont {{Addesso}}}, \bibinfo {author} {\bibfnamefont {R.~X.}\
  \bibnamefont {{Adhikari}}},  \emph {et~al.},\ }\href {\doibase
  10.1103/PhysRevLett.118.221101} {\bibfield  {journal} {\bibinfo  {journal}
  {Physical Review Letters}\ }\textbf {\bibinfo {volume} {118}},\ \bibinfo
  {pages} {221101} (\bibinfo {year} {2017}{\natexlab{a}})}\BibitemShut
  {NoStop}%
\bibitem [{\citenamefont {{Abbott}}\ \emph
  {et~al.}(2017{\natexlab{b}})\citenamefont {{Abbott}}, \citenamefont
  {{Abbott}}, \citenamefont {{Abbott}}, \citenamefont {{Acernese}},
  \citenamefont {{Ackley}}, \citenamefont {{Adams}}, \citenamefont {{Adams}},
  \citenamefont {{Addesso}}, \citenamefont {{Adhikari}}, \citenamefont
  {{Adya}},\ and\ \citenamefont {et~al.}}]{fourth:2017}%
  \BibitemOpen
  \bibfield  {author} {\bibinfo {author} {\bibfnamefont {B.~P.}\ \bibnamefont
  {{Abbott}}}, \bibinfo {author} {\bibfnamefont {R.}~\bibnamefont {{Abbott}}},
  \bibinfo {author} {\bibfnamefont {T.~D.}\ \bibnamefont {{Abbott}}}, \bibinfo
  {author} {\bibfnamefont {F.}~\bibnamefont {{Acernese}}}, \bibinfo {author}
  {\bibfnamefont {K.}~\bibnamefont {{Ackley}}}, \bibinfo {author}
  {\bibfnamefont {C.}~\bibnamefont {{Adams}}}, \bibinfo {author} {\bibfnamefont
  {T.}~\bibnamefont {{Adams}}}, \bibinfo {author} {\bibfnamefont
  {P.}~\bibnamefont {{Addesso}}}, \bibinfo {author} {\bibfnamefont {R.~X.}\
  \bibnamefont {{Adhikari}}}, \bibinfo {author} {\bibfnamefont {V.~B.}\
  \bibnamefont {{Adya}}}, \ and\ \bibinfo {author} {\bibnamefont {et~al.}},\
  }\href {\doibase 10.1103/PhysRevLett.119.141101} {\bibfield  {journal}
  {\bibinfo  {journal} {Physical Review Letters}\ }\textbf {\bibinfo {volume}
  {119}},\ \bibinfo {eid} {141101} (\bibinfo {year} {2017}{\natexlab{b}})},\
  \Eprint {http://arxiv.org/abs/1709.09660} {arXiv:1709.09660 [gr-qc]}
  \BibitemShut {NoStop}%
\bibitem [{\citenamefont {{Abbott}}\ \emph
  {et~al.}(2017{\natexlab{c}})\citenamefont {{Abbott}}, \citenamefont
  {{Abbott}}, \citenamefont {{Abbott}}, \citenamefont {{Acernese}},
  \citenamefont {{Ackley}}, \citenamefont {{Adams}}, \citenamefont {{Adams}},
  \citenamefont {{Addesso}}, \citenamefont {{Adhikari}}, \citenamefont
  {{Adya}},\ and\ \citenamefont {et~al.}}]{GW170608}%
  \BibitemOpen
  \bibfield  {author} {\bibinfo {author} {\bibfnamefont {B.~P.}\ \bibnamefont
  {{Abbott}}}, \bibinfo {author} {\bibfnamefont {R.}~\bibnamefont {{Abbott}}},
  \bibinfo {author} {\bibfnamefont {T.~D.}\ \bibnamefont {{Abbott}}}, \bibinfo
  {author} {\bibfnamefont {F.}~\bibnamefont {{Acernese}}}, \bibinfo {author}
  {\bibfnamefont {K.}~\bibnamefont {{Ackley}}}, \bibinfo {author}
  {\bibfnamefont {C.}~\bibnamefont {{Adams}}}, \bibinfo {author} {\bibfnamefont
  {T.}~\bibnamefont {{Adams}}}, \bibinfo {author} {\bibfnamefont
  {P.}~\bibnamefont {{Addesso}}}, \bibinfo {author} {\bibfnamefont {R.~X.}\
  \bibnamefont {{Adhikari}}}, \bibinfo {author} {\bibfnamefont {V.~B.}\
  \bibnamefont {{Adya}}}, \ and\ \bibinfo {author} {\bibnamefont {et~al.}},\
  }\href {\doibase 10.3847/2041-8213/aa9f0c} {\bibfield  {journal} {\bibinfo
  {journal} {\apjl}\ }\textbf {\bibinfo {volume} {851}},\ \bibinfo {eid} {L35}
  (\bibinfo {year} {2017}{\natexlab{c}})},\ \Eprint
  {http://arxiv.org/abs/1711.05578} {arXiv:1711.05578 [astro-ph.HE]}
  \BibitemShut {NoStop}%
\bibitem [{\citenamefont {{The LIGO Scientific Collaboration}}\ \emph
  {et~al.}(2018)\citenamefont {{The LIGO Scientific Collaboration}},
  \citenamefont {{the Virgo Collaboration}} \emph {et~al.}}]{o1o2catalog}%
  \BibitemOpen
  \bibfield  {author} {\bibinfo {author} {\bibnamefont {{The LIGO Scientific
  Collaboration}}}, \bibinfo {author} {\bibnamefont {{the Virgo
  Collaboration}}},  \emph {et~al.},\ }\href@noop {} {\bibfield  {journal}
  {\bibinfo  {journal} {arXiv e-prints}\ ,\ \bibinfo {eid} {arXiv:1811.12907}}
  (\bibinfo {year} {2018})},\ \Eprint {http://arxiv.org/abs/1811.12907}
  {arXiv:1811.12907 [astro-ph.HE]} \BibitemShut {NoStop}%
\bibitem [{\citenamefont {{Abbott}}\ \emph
  {et~al.}(2017{\natexlab{d}})\citenamefont {{Abbott}}, \citenamefont
  {{Abbott}}, \citenamefont {{Abbott}}, \citenamefont {{Acernese}},
  \citenamefont {{Ackley}}, \citenamefont {{Adams}}, \citenamefont {{Adams}},
  \citenamefont {{Addesso}}, \citenamefont {{Adhikari}}, \citenamefont
  {{Adya}},\ and\ \citenamefont {et~al.}}]{bnsdet:2017}%
  \BibitemOpen
  \bibfield  {author} {\bibinfo {author} {\bibfnamefont {B.~P.}\ \bibnamefont
  {{Abbott}}}, \bibinfo {author} {\bibfnamefont {R.}~\bibnamefont {{Abbott}}},
  \bibinfo {author} {\bibfnamefont {T.~D.}\ \bibnamefont {{Abbott}}}, \bibinfo
  {author} {\bibfnamefont {F.}~\bibnamefont {{Acernese}}}, \bibinfo {author}
  {\bibfnamefont {K.}~\bibnamefont {{Ackley}}}, \bibinfo {author}
  {\bibfnamefont {C.}~\bibnamefont {{Adams}}}, \bibinfo {author} {\bibfnamefont
  {T.}~\bibnamefont {{Adams}}}, \bibinfo {author} {\bibfnamefont
  {P.}~\bibnamefont {{Addesso}}}, \bibinfo {author} {\bibfnamefont {R.~X.}\
  \bibnamefont {{Adhikari}}}, \bibinfo {author} {\bibfnamefont {V.~B.}\
  \bibnamefont {{Adya}}}, \ and\ \bibinfo {author} {\bibnamefont {et~al.}},\
  }\href {\doibase 10.1103/PhysRevLett.119.161101} {\bibfield  {journal}
  {\bibinfo  {journal} {Physical Review Letters}\ }\textbf {\bibinfo {volume}
  {119}},\ \bibinfo {eid} {161101} (\bibinfo {year} {2017}{\natexlab{d}})},\
  \Eprint {http://arxiv.org/abs/1710.05832} {arXiv:1710.05832 [gr-qc]}
  \BibitemShut {NoStop}%
\bibitem [{\citenamefont {{The LIGO Scientific Collaboration}}\ \emph
  {et~al.}(2017)\citenamefont {{The LIGO Scientific Collaboration}},
  \citenamefont {{the Virgo Collaboration}}, \citenamefont {{Abbott}},
  \citenamefont {{Abbott}}, \citenamefont {{Abbott}}, \citenamefont
  {{Acernese}}, \citenamefont {{Ackley}}, \citenamefont {{Adams}},
  \citenamefont {{Adams}}, \citenamefont {{Addesso}},\ and\ \citenamefont
  {et~al.}}]{2017arXiv171005836T}%
  \BibitemOpen
  \bibfield  {author} {\bibinfo {author} {\bibnamefont {{The LIGO Scientific
  Collaboration}}}, \bibinfo {author} {\bibnamefont {{the Virgo
  Collaboration}}}, \bibinfo {author} {\bibfnamefont {B.~P.}\ \bibnamefont
  {{Abbott}}}, \bibinfo {author} {\bibfnamefont {R.}~\bibnamefont {{Abbott}}},
  \bibinfo {author} {\bibfnamefont {T.~D.}\ \bibnamefont {{Abbott}}}, \bibinfo
  {author} {\bibfnamefont {F.}~\bibnamefont {{Acernese}}}, \bibinfo {author}
  {\bibfnamefont {K.}~\bibnamefont {{Ackley}}}, \bibinfo {author}
  {\bibfnamefont {C.}~\bibnamefont {{Adams}}}, \bibinfo {author} {\bibfnamefont
  {T.}~\bibnamefont {{Adams}}}, \bibinfo {author} {\bibfnamefont
  {P.}~\bibnamefont {{Addesso}}}, \ and\ \bibinfo {author} {\bibnamefont
  {et~al.}},\ }\href@noop {} {\bibfield  {journal} {\bibinfo  {journal} {ArXiv
  e-prints}\ } (\bibinfo {year} {2017})},\ \bibinfo {note} {arXiv:1710.05836
  [astro-ph.HE]},\ \Eprint {http://arxiv.org/abs/1710.05836} {arXiv:1710.05836
  [astro-ph.HE]} \BibitemShut {NoStop}%
\bibitem [{\citenamefont {{National Science Foundation}}(2018)}]{NSFneu}%
  \BibitemOpen
  \bibfield  {author} {\bibinfo {author} {\bibnamefont {{National Science
  Foundation}}},\ }\href@noop {} {\enquote {\bibinfo {title} {Neutrino
  observation points to one source of high-energy cosmic rays},}\ } (\bibinfo
  {year} {2018}),\ \bibinfo {note}
  {\url{https://nsf.gov/news/news_summ.jsp?cntn_id=295955}}\BibitemShut
  {NoStop}%
\bibitem [{\citenamefont {{IceCube Collaboration}}(2018)}]{NuetrinoMMA}%
  \BibitemOpen
  \bibfield  {author} {\bibinfo {author} {\bibnamefont {{IceCube
  Collaboration}}},\ }\href {\doibase 10.1126/science.aat2890} {\bibfield
  {journal} {\bibinfo  {journal} {Science}\ }\textbf {\bibinfo {volume}
  {361}},\ \bibinfo {pages} {147} (\bibinfo {year} {2018})}\BibitemShut
  {NoStop}%
\bibitem [{\citenamefont {{IceCube Collaboration}}\ \emph
  {et~al.}(2018)\citenamefont {{IceCube Collaboration}}, \citenamefont
  {{Aartsen}}, \citenamefont {{Ackermann}}, \citenamefont {{Adams}},
  \citenamefont {{Aguilar}}, \citenamefont {{Ahlers}}, \citenamefont
  {{Ahrens}}, \citenamefont {{Al Samarai}}, \citenamefont {{Altmann}},
  \citenamefont {{Andeen}},\ and\ \citenamefont {et~al.}}]{IceCube:2018Sci}%
  \BibitemOpen
  \bibfield  {author} {\bibinfo {author} {\bibnamefont {{IceCube
  Collaboration}}}, \bibinfo {author} {\bibfnamefont {M.~G.}\ \bibnamefont
  {{Aartsen}}}, \bibinfo {author} {\bibfnamefont {M.}~\bibnamefont
  {{Ackermann}}}, \bibinfo {author} {\bibfnamefont {J.}~\bibnamefont
  {{Adams}}}, \bibinfo {author} {\bibfnamefont {J.~A.}\ \bibnamefont
  {{Aguilar}}}, \bibinfo {author} {\bibfnamefont {M.}~\bibnamefont {{Ahlers}}},
  \bibinfo {author} {\bibfnamefont {M.}~\bibnamefont {{Ahrens}}}, \bibinfo
  {author} {\bibfnamefont {I.}~\bibnamefont {{Al Samarai}}}, \bibinfo {author}
  {\bibfnamefont {D.}~\bibnamefont {{Altmann}}}, \bibinfo {author}
  {\bibfnamefont {K.}~\bibnamefont {{Andeen}}}, \ and\ \bibinfo {author}
  {\bibnamefont {et~al.}},\ }\href {\doibase 10.1126/science.aat1378}
  {\bibfield  {journal} {\bibinfo  {journal} {Science}\ }\textbf {\bibinfo
  {volume} {361}},\ \bibinfo {eid} {eaat1378} (\bibinfo {year} {2018})},\
  \Eprint {http://arxiv.org/abs/1807.08816} {arXiv:1807.08816 [astro-ph.HE]}
  \BibitemShut {NoStop}%
\bibitem [{\citenamefont {{The LIGO Scientific Collaboration}}\ and\
  \citenamefont {{the Virgo Collaboration}}(2018)}]{pops:2018}%
  \BibitemOpen
  \bibfield  {author} {\bibinfo {author} {\bibnamefont {{The LIGO Scientific
  Collaboration}}}\ and\ \bibinfo {author} {\bibnamefont {{the Virgo
  Collaboration}}},\ }\href@noop {} {\bibfield  {journal} {\bibinfo  {journal}
  {arXiv e-prints}\ ,\ \bibinfo {eid} {arXiv:1811.12940}} (\bibinfo {year}
  {2018})},\ \Eprint {http://arxiv.org/abs/1811.12940} {arXiv:1811.12940
  [astro-ph.HE]} \BibitemShut {NoStop}%
\bibitem [{\citenamefont {{Abbott}}\ \emph
  {et~al.}(2017{\natexlab{e}})\citenamefont {{Abbott}} \emph
  {et~al.}}]{GWH:NaturA}%
  \BibitemOpen
  \bibfield  {author} {\bibinfo {author} {\bibfnamefont {B.~P.}\ \bibnamefont
  {{Abbott}}} \emph {et~al.},\ }\href {\doibase 10.1038/nature24471} {\bibfield
   {journal} {\bibinfo  {journal} {\nat}\ }\textbf {\bibinfo {volume} {551}},\
  \bibinfo {pages} {85} (\bibinfo {year} {2017}{\natexlab{e}})}\BibitemShut
  {NoStop}%
\bibitem [{\citenamefont {{Fishbach}}\ \emph {et~al.}(2018)\citenamefont
  {{Fishbach}}, \citenamefont {{Gray}}, \citenamefont {{Maga{\~n}a Hernandez}},
  \citenamefont {{Qi}}, \citenamefont {{Sur}}, \citenamefont {{members of the
  LIGO Scientific Collaboration}},\ and\ \citenamefont {{the Virgo
  Collaboration}}}]{Maya:2018F}%
  \BibitemOpen
  \bibfield  {author} {\bibinfo {author} {\bibfnamefont {M.}~\bibnamefont
  {{Fishbach}}}, \bibinfo {author} {\bibfnamefont {R.}~\bibnamefont {{Gray}}},
  \bibinfo {author} {\bibfnamefont {I.}~\bibnamefont {{Maga{\~n}a Hernandez}}},
  \bibinfo {author} {\bibfnamefont {H.}~\bibnamefont {{Qi}}}, \bibinfo {author}
  {\bibfnamefont {A.}~\bibnamefont {{Sur}}}, \bibinfo {author} {\bibnamefont
  {{members of the LIGO Scientific Collaboration}}}, \ and\ \bibinfo {author}
  {\bibnamefont {{the Virgo Collaboration}}},\ }\href@noop {} {\bibfield
  {journal} {\bibinfo  {journal} {ArXiv e-prints}\ ,\ \bibinfo {eid}
  {arXiv:1807.05667}} (\bibinfo {year} {2018})},\ \Eprint
  {http://arxiv.org/abs/1807.05667} {arXiv:1807.05667 [astro-ph.CO]}
  \BibitemShut {NoStop}%
\bibitem [{\citenamefont {{The DES Collaboration}}\ \emph
  {et~al.}(2019)\citenamefont {{The DES Collaboration}}, \citenamefont {{the
  LIGO Scientific Collaboration}},\ and\ \citenamefont {{the Virgo
  Collaboration}}}]{darkgwss:2019}%
  \BibitemOpen
  \bibfield  {author} {\bibinfo {author} {\bibnamefont {{The DES
  Collaboration}}}, \bibinfo {author} {\bibnamefont {{the LIGO Scientific
  Collaboration}}}, \ and\ \bibinfo {author} {\bibnamefont {{the Virgo
  Collaboration}}},\ }\href@noop {} {\bibfield  {journal} {\bibinfo  {journal}
  {arXiv e-prints}\ ,\ \bibinfo {eid} {arXiv:1901.01540}} (\bibinfo {year}
  {2019})},\ \Eprint {http://arxiv.org/abs/1901.01540} {arXiv:1901.01540
  [astro-ph.CO]} \BibitemShut {NoStop}%
\bibitem [{\citenamefont {{Schutz}}(1986)}]{Schutz:1986Nature}%
  \BibitemOpen
  \bibfield  {author} {\bibinfo {author} {\bibfnamefont {B.~F.}\ \bibnamefont
  {{Schutz}}},\ }\href {\doibase 10.1038/323310a0} {\bibfield  {journal}
  {\bibinfo  {journal} {\nat}\ }\textbf {\bibinfo {volume} {323}},\ \bibinfo
  {pages} {310} (\bibinfo {year} {1986})}\BibitemShut {NoStop}%
\bibitem [{\citenamefont {{LSST Science Collaboration}}\ \emph
  {et~al.}(2009)\citenamefont {{LSST Science Collaboration}}, \citenamefont
  {{Abell}}, \citenamefont {{Allison}}, \citenamefont {{Anderson}},
  \citenamefont {{Andrew}}, \citenamefont {{Angel}}, \citenamefont {{Armus}},
  \citenamefont {{Arnett}}, \citenamefont {{Asztalos}}, \citenamefont
  {{Axelrod}},\ and\ \citenamefont {et~al.}}]{lsstbook}%
  \BibitemOpen
  \bibfield  {author} {\bibinfo {author} {\bibnamefont {{LSST Science
  Collaboration}}}, \bibinfo {author} {\bibfnamefont {P.~A.}\ \bibnamefont
  {{Abell}}}, \bibinfo {author} {\bibfnamefont {J.}~\bibnamefont {{Allison}}},
  \bibinfo {author} {\bibfnamefont {S.~F.}\ \bibnamefont {{Anderson}}},
  \bibinfo {author} {\bibfnamefont {J.~R.}\ \bibnamefont {{Andrew}}}, \bibinfo
  {author} {\bibfnamefont {J.~R.~P.}\ \bibnamefont {{Angel}}}, \bibinfo
  {author} {\bibfnamefont {L.}~\bibnamefont {{Armus}}}, \bibinfo {author}
  {\bibfnamefont {D.}~\bibnamefont {{Arnett}}}, \bibinfo {author}
  {\bibfnamefont {S.~J.}\ \bibnamefont {{Asztalos}}}, \bibinfo {author}
  {\bibfnamefont {T.~S.}\ \bibnamefont {{Axelrod}}}, \ and\ \bibinfo {author}
  {\bibnamefont {et~al.}},\ }\href@noop {} {\bibfield  {journal} {\bibinfo
  {journal} {ArXiv e-prints}\ } (\bibinfo {year} {2009})},\ \Eprint
  {http://arxiv.org/abs/0912.0201} {arXiv:0912.0201 [astro-ph.IM]} \BibitemShut
  {NoStop}%
\bibitem [{\citenamefont {{Birrer}}\ \emph {et~al.}(2018)\citenamefont
  {{Birrer}}, \citenamefont {{Treu}}, \citenamefont {{Rusu}}, \citenamefont
  {{Bonvin}}, \citenamefont {{Fassnacht}}, \citenamefont {{Chan}},
  \citenamefont {{Agnello}}, \citenamefont {{Shajib}}, \citenamefont {{Chen}},
  \citenamefont {{Auger}}, \citenamefont {{Courbin}}, \citenamefont
  {{Hilbert}}, \citenamefont {{Sluse}}, \citenamefont {{Suyu}}, \citenamefont
  {{Wong}}, \citenamefont {{Marshall}}, \citenamefont {{Lemaux}},\ and\
  \citenamefont {{Meylan}}}]{Birrer:2018}%
  \BibitemOpen
  \bibfield  {author} {\bibinfo {author} {\bibfnamefont {S.}~\bibnamefont
  {{Birrer}}}, \bibinfo {author} {\bibfnamefont {T.}~\bibnamefont {{Treu}}},
  \bibinfo {author} {\bibfnamefont {C.~E.}\ \bibnamefont {{Rusu}}}, \bibinfo
  {author} {\bibfnamefont {V.}~\bibnamefont {{Bonvin}}}, \bibinfo {author}
  {\bibfnamefont {C.~D.}\ \bibnamefont {{Fassnacht}}}, \bibinfo {author}
  {\bibfnamefont {J.~H.~H.}\ \bibnamefont {{Chan}}}, \bibinfo {author}
  {\bibfnamefont {A.}~\bibnamefont {{Agnello}}}, \bibinfo {author}
  {\bibfnamefont {A.~J.}\ \bibnamefont {{Shajib}}}, \bibinfo {author}
  {\bibfnamefont {G.~C.~F.}\ \bibnamefont {{Chen}}}, \bibinfo {author}
  {\bibfnamefont {M.}~\bibnamefont {{Auger}}}, \bibinfo {author} {\bibfnamefont
  {F.}~\bibnamefont {{Courbin}}}, \bibinfo {author} {\bibfnamefont
  {S.}~\bibnamefont {{Hilbert}}}, \bibinfo {author} {\bibfnamefont
  {D.}~\bibnamefont {{Sluse}}}, \bibinfo {author} {\bibfnamefont {S.~H.}\
  \bibnamefont {{Suyu}}}, \bibinfo {author} {\bibfnamefont {K.~C.}\
  \bibnamefont {{Wong}}}, \bibinfo {author} {\bibfnamefont {P.}~\bibnamefont
  {{Marshall}}}, \bibinfo {author} {\bibfnamefont {B.~C.}\ \bibnamefont
  {{Lemaux}}}, \ and\ \bibinfo {author} {\bibfnamefont {G.}~\bibnamefont
  {{Meylan}}},\ }\href@noop {} {\bibfield  {journal} {\bibinfo  {journal}
  {arXiv e-prints}\ ,\ \bibinfo {eid} {arXiv:1809.01274}} (\bibinfo {year}
  {2018})},\ \Eprint {http://arxiv.org/abs/1809.01274} {arXiv:1809.01274
  [astro-ph.CO]} \BibitemShut {NoStop}%
\bibitem [{\citenamefont {{Singer}}\ \emph {et~al.}(2014)\citenamefont
  {{Singer}}, \citenamefont {{Price}}, \citenamefont {{Farr}}, \citenamefont
  {{Urban}}, \citenamefont {{Pankow}}, \citenamefont {{Vitale}}, \citenamefont
  {{Veitch}}, \citenamefont {{Farr}}, \citenamefont {{Hanna}}, \citenamefont
  {{Cannon}}, \citenamefont {{Downes}}, \citenamefont {{Graff}}, \citenamefont
  {{Haster}}, \citenamefont {{Mandel}}, \citenamefont {{Sidery}},\ and\
  \citenamefont {{Vecchio}}}]{Singer:2014ApJ}%
  \BibitemOpen
  \bibfield  {author} {\bibinfo {author} {\bibfnamefont {L.~P.}\ \bibnamefont
  {{Singer}}}, \bibinfo {author} {\bibfnamefont {L.~R.}\ \bibnamefont
  {{Price}}}, \bibinfo {author} {\bibfnamefont {B.}~\bibnamefont {{Farr}}},
  \bibinfo {author} {\bibfnamefont {A.~L.}\ \bibnamefont {{Urban}}}, \bibinfo
  {author} {\bibfnamefont {C.}~\bibnamefont {{Pankow}}}, \bibinfo {author}
  {\bibfnamefont {S.}~\bibnamefont {{Vitale}}}, \bibinfo {author}
  {\bibfnamefont {J.}~\bibnamefont {{Veitch}}}, \bibinfo {author}
  {\bibfnamefont {W.~M.}\ \bibnamefont {{Farr}}}, \bibinfo {author}
  {\bibfnamefont {C.}~\bibnamefont {{Hanna}}}, \bibinfo {author} {\bibfnamefont
  {K.}~\bibnamefont {{Cannon}}}, \bibinfo {author} {\bibfnamefont
  {T.}~\bibnamefont {{Downes}}}, \bibinfo {author} {\bibfnamefont
  {P.}~\bibnamefont {{Graff}}}, \bibinfo {author} {\bibfnamefont {C.-J.}\
  \bibnamefont {{Haster}}}, \bibinfo {author} {\bibfnamefont {I.}~\bibnamefont
  {{Mandel}}}, \bibinfo {author} {\bibfnamefont {T.}~\bibnamefont {{Sidery}}},
  \ and\ \bibinfo {author} {\bibfnamefont {A.}~\bibnamefont {{Vecchio}}},\
  }\href {\doibase 10.1088/0004-637X/795/2/105} {\bibfield  {journal} {\bibinfo
   {journal} {\apj}\ }\textbf {\bibinfo {volume} {795}},\ \bibinfo {eid} {105}
  (\bibinfo {year} {2014})},\ \Eprint {http://arxiv.org/abs/1404.5623}
  {arXiv:1404.5623 [astro-ph.HE]} \BibitemShut {NoStop}%
\bibitem [{\citenamefont {Branchesi}(2016)}]{marichaMMA:2016}%
  \BibitemOpen
  \bibfield  {author} {\bibinfo {author} {\bibfnamefont {M.}~\bibnamefont
  {Branchesi}},\ }\href {http://stacks.iop.org/1742-6596/718/i=2/a=022004}
  {\bibfield  {journal} {\bibinfo  {journal} {Journal of Physics: Conference
  Series}\ }\textbf {\bibinfo {volume} {718}},\ \bibinfo {pages} {022004}
  (\bibinfo {year} {2016})}\BibitemShut {NoStop}%
\bibitem [{\citenamefont {Bartos}\ \emph {et~al.}(2017)\citenamefont {Bartos},
  \citenamefont {Collaboration},\ and\ \citenamefont
  {Collaboration}}]{imre:2017}%
  \BibitemOpen
  \bibfield  {author} {\bibinfo {author} {\bibfnamefont {I.}~\bibnamefont
  {Bartos}}, \bibinfo {author} {\bibfnamefont {L.~S.}\ \bibnamefont
  {Collaboration}}, \ and\ \bibinfo {author} {\bibfnamefont {V.}~\bibnamefont
  {Collaboration}},\ }\href {http://stacks.iop.org/1742-6596/888/i=1/a=012001}
  {\bibfield  {journal} {\bibinfo  {journal} {Journal of Physics: Conference
  Series}\ }\textbf {\bibinfo {volume} {888}},\ \bibinfo {pages} {012001}
  (\bibinfo {year} {2017})}\BibitemShut {NoStop}%
\bibitem [{\citenamefont {{Huerta}}\ \emph {et~al.}(2017)\citenamefont
  {{Huerta}}, \citenamefont {{Kumar}}, \citenamefont {{Agarwal}}, \citenamefont
  {{George}}, \citenamefont {{Schive}}, \citenamefont {{Pfeiffer}},
  \citenamefont {{Haas}}, \citenamefont {{Ren}}, \citenamefont {{Chu}},
  \citenamefont {{Boyle}}, \citenamefont {{Hemberger}}, \citenamefont
  {{Kidder}}, \citenamefont {{Scheel}},\ and\ \citenamefont
  {{Szilagyi}}}]{Huerta:2017a}%
  \BibitemOpen
  \bibfield  {author} {\bibinfo {author} {\bibfnamefont {E.~A.}\ \bibnamefont
  {{Huerta}}}, \bibinfo {author} {\bibfnamefont {P.}~\bibnamefont {{Kumar}}},
  \bibinfo {author} {\bibfnamefont {B.}~\bibnamefont {{Agarwal}}}, \bibinfo
  {author} {\bibfnamefont {D.}~\bibnamefont {{George}}}, \bibinfo {author}
  {\bibfnamefont {H.-Y.}\ \bibnamefont {{Schive}}}, \bibinfo {author}
  {\bibfnamefont {H.~P.}\ \bibnamefont {{Pfeiffer}}}, \bibinfo {author}
  {\bibfnamefont {R.}~\bibnamefont {{Haas}}}, \bibinfo {author} {\bibfnamefont
  {W.}~\bibnamefont {{Ren}}}, \bibinfo {author} {\bibfnamefont
  {T.}~\bibnamefont {{Chu}}}, \bibinfo {author} {\bibfnamefont
  {M.}~\bibnamefont {{Boyle}}}, \bibinfo {author} {\bibfnamefont {D.~A.}\
  \bibnamefont {{Hemberger}}}, \bibinfo {author} {\bibfnamefont {L.~E.}\
  \bibnamefont {{Kidder}}}, \bibinfo {author} {\bibfnamefont {M.~A.}\
  \bibnamefont {{Scheel}}}, \ and\ \bibinfo {author} {\bibfnamefont
  {B.}~\bibnamefont {{Szilagyi}}},\ }\href {\doibase
  10.1103/PhysRevD.95.024038} {\bibfield  {journal} {\bibinfo  {journal}
  {\prd}\ }\textbf {\bibinfo {volume} {95}},\ \bibinfo {eid} {024038} (\bibinfo
  {year} {2017})},\ \Eprint {http://arxiv.org/abs/1609.05933} {arXiv:1609.05933
  [gr-qc]} \BibitemShut {NoStop}%
\bibitem [{\citenamefont {{Huerta}}\ \emph
  {et~al.}(2018{\natexlab{a}})\citenamefont {{Huerta}}, \citenamefont
  {{Moore}}, \citenamefont {{Kumar}}, \citenamefont {{George}}, \citenamefont
  {{Chua}}, \citenamefont {{Haas}}, \citenamefont {{Wessel}}, \citenamefont
  {{Johnson}}, \citenamefont {{Glennon}}, \citenamefont {{Rebei}},
  \citenamefont {{Holgado}}, \citenamefont {{Gair}},\ and\ \citenamefont
  {{Pfeiffer}}}]{huerta:2018PhRvD}%
  \BibitemOpen
  \bibfield  {author} {\bibinfo {author} {\bibfnamefont {E.~A.}\ \bibnamefont
  {{Huerta}}}, \bibinfo {author} {\bibfnamefont {C.~J.}\ \bibnamefont
  {{Moore}}}, \bibinfo {author} {\bibfnamefont {P.}~\bibnamefont {{Kumar}}},
  \bibinfo {author} {\bibfnamefont {D.}~\bibnamefont {{George}}}, \bibinfo
  {author} {\bibfnamefont {A.~J.~K.}\ \bibnamefont {{Chua}}}, \bibinfo {author}
  {\bibfnamefont {R.}~\bibnamefont {{Haas}}}, \bibinfo {author} {\bibfnamefont
  {E.}~\bibnamefont {{Wessel}}}, \bibinfo {author} {\bibfnamefont
  {D.}~\bibnamefont {{Johnson}}}, \bibinfo {author} {\bibfnamefont
  {D.}~\bibnamefont {{Glennon}}}, \bibinfo {author} {\bibfnamefont
  {A.}~\bibnamefont {{Rebei}}}, \bibinfo {author} {\bibfnamefont {A.~M.}\
  \bibnamefont {{Holgado}}}, \bibinfo {author} {\bibfnamefont {J.~R.}\
  \bibnamefont {{Gair}}}, \ and\ \bibinfo {author} {\bibfnamefont {H.~P.}\
  \bibnamefont {{Pfeiffer}}},\ }\href {\doibase 10.1103/PhysRevD.97.024031}
  {\bibfield  {journal} {\bibinfo  {journal} {\prd}\ }\textbf {\bibinfo
  {volume} {97}},\ \bibinfo {eid} {024031} (\bibinfo {year}
  {2018}{\natexlab{a}})},\ \Eprint {http://arxiv.org/abs/1711.06276}
  {arXiv:1711.06276 [gr-qc]} \BibitemShut {NoStop}%
\bibitem [{\citenamefont {{Harry}}\ \emph {et~al.}(2016)\citenamefont
  {{Harry}}, \citenamefont {{Privitera}}, \citenamefont {{Boh{\'e}}},\ and\
  \citenamefont {{Buonanno}}}]{2016PhRvD..94b4012H}%
  \BibitemOpen
  \bibfield  {author} {\bibinfo {author} {\bibfnamefont {I.}~\bibnamefont
  {{Harry}}}, \bibinfo {author} {\bibfnamefont {S.}~\bibnamefont
  {{Privitera}}}, \bibinfo {author} {\bibfnamefont {A.}~\bibnamefont
  {{Boh{\'e}}}}, \ and\ \bibinfo {author} {\bibfnamefont {A.}~\bibnamefont
  {{Buonanno}}},\ }\href {\doibase 10.1103/PhysRevD.94.024012} {\bibfield
  {journal} {\bibinfo  {journal} {\prd}\ }\textbf {\bibinfo {volume} {94}},\
  \bibinfo {eid} {024012} (\bibinfo {year} {2016})},\ \Eprint
  {http://arxiv.org/abs/1603.02444} {arXiv:1603.02444 [gr-qc]} \BibitemShut
  {NoStop}%
\bibitem [{\citenamefont {{Klimenko}}\ \emph {et~al.}(2016)\citenamefont
  {{Klimenko}}, \citenamefont {{Vedovato}}, \citenamefont {{Drago}},
  \citenamefont {{Salemi}}, \citenamefont {{Tiwari}}, \citenamefont {{Prodi}},
  \citenamefont {{Lazzaro}}, \citenamefont {{Ackley}}, \citenamefont
  {{Tiwari}}, \citenamefont {{Da Silva}},\ and\ \citenamefont
  {{Mitselmakher}}}]{Sergey:2016}%
  \BibitemOpen
  \bibfield  {author} {\bibinfo {author} {\bibfnamefont {S.}~\bibnamefont
  {{Klimenko}}}, \bibinfo {author} {\bibfnamefont {G.}~\bibnamefont
  {{Vedovato}}}, \bibinfo {author} {\bibfnamefont {M.}~\bibnamefont {{Drago}}},
  \bibinfo {author} {\bibfnamefont {F.}~\bibnamefont {{Salemi}}}, \bibinfo
  {author} {\bibfnamefont {V.}~\bibnamefont {{Tiwari}}}, \bibinfo {author}
  {\bibfnamefont {G.~A.}\ \bibnamefont {{Prodi}}}, \bibinfo {author}
  {\bibfnamefont {C.}~\bibnamefont {{Lazzaro}}}, \bibinfo {author}
  {\bibfnamefont {K.}~\bibnamefont {{Ackley}}}, \bibinfo {author}
  {\bibfnamefont {S.}~\bibnamefont {{Tiwari}}}, \bibinfo {author}
  {\bibfnamefont {C.~F.}\ \bibnamefont {{Da Silva}}}, \ and\ \bibinfo {author}
  {\bibfnamefont {G.}~\bibnamefont {{Mitselmakher}}},\ }\href {\doibase
  10.1103/PhysRevD.93.042004} {\bibfield  {journal} {\bibinfo  {journal}
  {\prd}\ }\textbf {\bibinfo {volume} {93}},\ \bibinfo {eid} {042004} (\bibinfo
  {year} {2016})},\ \Eprint {http://arxiv.org/abs/1511.05999} {arXiv:1511.05999
  [gr-qc]} \BibitemShut {NoStop}%
\bibitem [{\citenamefont {{Alexander}}\ \emph {et~al.}(2018)\citenamefont
  {{Alexander}}, \citenamefont {{Margutti}}, \citenamefont {{Blanchard}},
  \citenamefont {{Fong}}, \citenamefont {{Berger}}, \citenamefont {{Hajela}},
  \citenamefont {{Eftekhari}}, \citenamefont {{Chornock}}, \citenamefont
  {{Cowperthwaite}}, \citenamefont {{Giannios}}, \citenamefont {{Guidorzi}},
  \citenamefont {{Kathirgamaraju}}, \citenamefont {{MacFadyen}}, \citenamefont
  {{Metzger}}, \citenamefont {{Nicholl}}, \citenamefont {{Sironi}},
  \citenamefont {{Villar}}, \citenamefont {{Williams}}, \citenamefont {{Xie}},\
  and\ \citenamefont {{Zrake}}}]{alex:2018ApJLA}%
  \BibitemOpen
  \bibfield  {author} {\bibinfo {author} {\bibfnamefont {K.~D.}\ \bibnamefont
  {{Alexander}}}, \bibinfo {author} {\bibfnamefont {R.}~\bibnamefont
  {{Margutti}}}, \bibinfo {author} {\bibfnamefont {P.~K.}\ \bibnamefont
  {{Blanchard}}}, \bibinfo {author} {\bibfnamefont {W.}~\bibnamefont {{Fong}}},
  \bibinfo {author} {\bibfnamefont {E.}~\bibnamefont {{Berger}}}, \bibinfo
  {author} {\bibfnamefont {A.}~\bibnamefont {{Hajela}}}, \bibinfo {author}
  {\bibfnamefont {T.}~\bibnamefont {{Eftekhari}}}, \bibinfo {author}
  {\bibfnamefont {R.}~\bibnamefont {{Chornock}}}, \bibinfo {author}
  {\bibfnamefont {P.~S.}\ \bibnamefont {{Cowperthwaite}}}, \bibinfo {author}
  {\bibfnamefont {D.}~\bibnamefont {{Giannios}}}, \bibinfo {author}
  {\bibfnamefont {C.}~\bibnamefont {{Guidorzi}}}, \bibinfo {author}
  {\bibfnamefont {A.}~\bibnamefont {{Kathirgamaraju}}}, \bibinfo {author}
  {\bibfnamefont {A.}~\bibnamefont {{MacFadyen}}}, \bibinfo {author}
  {\bibfnamefont {B.~D.}\ \bibnamefont {{Metzger}}}, \bibinfo {author}
  {\bibfnamefont {M.}~\bibnamefont {{Nicholl}}}, \bibinfo {author}
  {\bibfnamefont {L.}~\bibnamefont {{Sironi}}}, \bibinfo {author}
  {\bibfnamefont {V.~A.}\ \bibnamefont {{Villar}}}, \bibinfo {author}
  {\bibfnamefont {P.~K.~G.}\ \bibnamefont {{Williams}}}, \bibinfo {author}
  {\bibfnamefont {X.}~\bibnamefont {{Xie}}}, \ and\ \bibinfo {author}
  {\bibfnamefont {J.}~\bibnamefont {{Zrake}}},\ }\href {\doibase
  10.3847/2041-8213/aad637} {\bibfield  {journal} {\bibinfo  {journal} {\apj}\
  }\textbf {\bibinfo {volume} {863}},\ \bibinfo {eid} {L18} (\bibinfo {year}
  {2018})},\ \Eprint {http://arxiv.org/abs/1805.02870} {arXiv:1805.02870
  [astro-ph.HE]} \BibitemShut {NoStop}%
\bibitem [{\citenamefont {Shibata}\ and\ \citenamefont
  {Taniguchi}(2011)}]{Shibata2011LLR}%
  \BibitemOpen
  \bibfield  {author} {\bibinfo {author} {\bibfnamefont {M.}~\bibnamefont
  {Shibata}}\ and\ \bibinfo {author} {\bibfnamefont {K.}~\bibnamefont
  {Taniguchi}},\ }\href {\doibase 10.12942/lrr-2011-6} {\bibfield  {journal}
  {\bibinfo  {journal} {Living Reviews in Relativity}\ }\textbf {\bibinfo
  {volume} {14}},\ \bibinfo {pages} {6} (\bibinfo {year} {2011})}\BibitemShut
  {NoStop}%
\bibitem [{\citenamefont {Faber}\ and\ \citenamefont
  {Rasio}(2012)}]{Faber2012LRR}%
  \BibitemOpen
  \bibfield  {author} {\bibinfo {author} {\bibfnamefont {J.~A.}\ \bibnamefont
  {Faber}}\ and\ \bibinfo {author} {\bibfnamefont {F.~A.}\ \bibnamefont
  {Rasio}},\ }\href {\doibase 10.12942/lrr-2012-8} {\bibfield  {journal}
  {\bibinfo  {journal} {Living Reviews in Relativity}\ }\textbf {\bibinfo
  {volume} {15}},\ \bibinfo {pages} {8} (\bibinfo {year} {2012})}\BibitemShut
  {NoStop}%
\bibitem [{\citenamefont {Br{\"u}gmann}(2018)}]{BrugmenScienceReview18}%
  \BibitemOpen
  \bibfield  {author} {\bibinfo {author} {\bibfnamefont {B.}~\bibnamefont
  {Br{\"u}gmann}},\ }\href {\doibase 10.1126/science.aat3363} {\bibfield
  {journal} {\bibinfo  {journal} {Science}\ }\textbf {\bibinfo {volume}
  {361}},\ \bibinfo {pages} {366} (\bibinfo {year} {2018})},\ \Eprint
  {http://arxiv.org/abs/http://science.sciencemag.org/content/361/6400/366.full.pdf}
  {http://science.sciencemag.org/content/361/6400/366.full.pdf} \BibitemShut
  {NoStop}%
\bibitem [{\citenamefont {{Duez}}\ and\ \citenamefont
  {{Zlochower}}(2019)}]{DuezReview19}%
  \BibitemOpen
  \bibfield  {author} {\bibinfo {author} {\bibfnamefont {M.~D.}\ \bibnamefont
  {{Duez}}}\ and\ \bibinfo {author} {\bibfnamefont {Y.}~\bibnamefont
  {{Zlochower}}},\ }\href {\doibase 10.1088/1361-6633/aadb16} {\bibfield
  {journal} {\bibinfo  {journal} {Reports on Progress in Physics}\ }\textbf
  {\bibinfo {volume} {82}},\ \bibinfo {eid} {016902} (\bibinfo {year}
  {2019})},\ \Eprint {http://arxiv.org/abs/1808.06011} {arXiv:1808.06011
  [gr-qc]} \BibitemShut {NoStop}%
\bibitem [{\citenamefont {{Paschalidis}}\ \emph {et~al.}(2015)\citenamefont
  {{Paschalidis}}, \citenamefont {{Ruiz}},\ and\ \citenamefont
  {{Shapiro}}}]{prs15}%
  \BibitemOpen
  \bibfield  {author} {\bibinfo {author} {\bibfnamefont {V.}~\bibnamefont
  {{Paschalidis}}}, \bibinfo {author} {\bibfnamefont {M.}~\bibnamefont
  {{Ruiz}}}, \ and\ \bibinfo {author} {\bibfnamefont {S.~L.}\ \bibnamefont
  {{Shapiro}}},\ }\href {\doibase 10.1088/2041-8205/806/1/L14} {\bibfield
  {journal} {\bibinfo  {journal} {Astrophys. J.}\ }\textbf {\bibinfo {volume}
  {806}},\ \bibinfo {pages} {L14} (\bibinfo {year} {2015})},\ \Eprint
  {http://arxiv.org/abs/1410.7392} {arXiv:1410.7392 [astro-ph.HE]} \BibitemShut
  {NoStop}%
\bibitem [{\citenamefont {Ruiz}\ \emph {et~al.}(2016)\citenamefont {Ruiz},
  \citenamefont {Lang}, \citenamefont {Paschalidis},\ and\ \citenamefont
  {Shapiro}}]{Ruiz:2016rai}%
  \BibitemOpen
  \bibfield  {author} {\bibinfo {author} {\bibfnamefont {M.}~\bibnamefont
  {Ruiz}}, \bibinfo {author} {\bibfnamefont {R.~N.}\ \bibnamefont {Lang}},
  \bibinfo {author} {\bibfnamefont {V.}~\bibnamefont {Paschalidis}}, \ and\
  \bibinfo {author} {\bibfnamefont {S.~L.}\ \bibnamefont {Shapiro}},\ }\href
  {\doibase 10.3847/2041-8205/824/1/L6} {\bibfield  {journal} {\bibinfo
  {journal} {Astrophys. J.}\ }\textbf {\bibinfo {volume} {824}},\ \bibinfo
  {pages} {L6} (\bibinfo {year} {2016})},\ \Eprint
  {http://arxiv.org/abs/1604.02455} {arXiv:1604.02455 [astro-ph.HE]}
  \BibitemShut {NoStop}%
\bibitem [{\citenamefont {{Ruiz}}\ \emph {et~al.}(2018)\citenamefont {{Ruiz}},
  \citenamefont {{Shapiro}},\ and\ \citenamefont {{Tsokaros}}}]{Ruiz:2018wah}%
  \BibitemOpen
  \bibfield  {author} {\bibinfo {author} {\bibfnamefont {M.}~\bibnamefont
  {{Ruiz}}}, \bibinfo {author} {\bibfnamefont {S.~L.}\ \bibnamefont
  {{Shapiro}}}, \ and\ \bibinfo {author} {\bibfnamefont {A.}~\bibnamefont
  {{Tsokaros}}},\ }\href {\doibase 10.1103/PhysRevD.98.123017} {\bibfield
  {journal} {\bibinfo  {journal} {\prd}\ }\textbf {\bibinfo {volume} {98}},\
  \bibinfo {eid} {123017} (\bibinfo {year} {2018})},\ \Eprint
  {http://arxiv.org/abs/1810.08618} {arXiv:1810.08618 [astro-ph.HE]}
  \BibitemShut {NoStop}%
\bibitem [{\citenamefont {Ruiz}\ and\ \citenamefont
  {Shapiro}(2017)}]{Ruiz:2017inq}%
  \BibitemOpen
  \bibfield  {author} {\bibinfo {author} {\bibfnamefont {M.}~\bibnamefont
  {Ruiz}}\ and\ \bibinfo {author} {\bibfnamefont {S.~L.}\ \bibnamefont
  {Shapiro}},\ }\href {\doibase 10.1103/PhysRevD.96.084063} {\bibfield
  {journal} {\bibinfo  {journal} {Phys. Rev.}\ }\textbf {\bibinfo {volume}
  {D96}},\ \bibinfo {pages} {084063} (\bibinfo {year} {2017})},\ \Eprint
  {http://arxiv.org/abs/1709.00414} {arXiv:1709.00414 [astro-ph.HE]}
  \BibitemShut {NoStop}%
\bibitem [{\citenamefont {Bhat}\ \emph {et~al.}(2016)\citenamefont {Bhat} \emph
  {et~al.}}]{Bhat:2016odd}%
  \BibitemOpen
  \bibfield  {author} {\bibinfo {author} {\bibfnamefont {P.~N.}\ \bibnamefont
  {Bhat}} \emph {et~al.},\ }\href {\doibase 10.3847/0067-0049/223/2/28}
  {\bibfield  {journal} {\bibinfo  {journal} {Astrophys. J. Suppl.}\ }\textbf
  {\bibinfo {volume} {223}},\ \bibinfo {pages} {28} (\bibinfo {year} {2016})},\
  \Eprint {http://arxiv.org/abs/1603.07612} {arXiv:1603.07612 [astro-ph.HE]}
  \BibitemShut {NoStop}%
\bibitem [{\citenamefont {Lien}\ \emph {et~al.}(2016)\citenamefont {Lien} \emph
  {et~al.}}]{Lien:2016zny}%
  \BibitemOpen
  \bibfield  {author} {\bibinfo {author} {\bibfnamefont {A.}~\bibnamefont
  {Lien}} \emph {et~al.},\ }\href {\doibase 10.3847/0004-637X/829/1/7}
  {\bibfield  {journal} {\bibinfo  {journal} {Astrophys. J.}\ }\textbf
  {\bibinfo {volume} {829}},\ \bibinfo {pages} {7} (\bibinfo {year} {2016})},\
  \Eprint {http://arxiv.org/abs/1606.01956} {arXiv:1606.01956 [astro-ph.HE]}
  \BibitemShut {NoStop}%
\bibitem [{\citenamefont {Svinkin}\ \emph {et~al.}(2016)\citenamefont
  {Svinkin}, \citenamefont {Frederiks}, \citenamefont {Aptekar}, \citenamefont
  {Golenetskii}, \citenamefont {Pal'shin}, \citenamefont {Oleynik},
  \citenamefont {Tsvetkova}, \citenamefont {Ulanov}, \citenamefont {Cline},\
  and\ \citenamefont {Hurley}}]{Svinkin:2016fho}%
  \BibitemOpen
  \bibfield  {author} {\bibinfo {author} {\bibfnamefont {D.~S.}\ \bibnamefont
  {Svinkin}}, \bibinfo {author} {\bibfnamefont {D.~D.}\ \bibnamefont
  {Frederiks}}, \bibinfo {author} {\bibfnamefont {R.~L.}\ \bibnamefont
  {Aptekar}}, \bibinfo {author} {\bibfnamefont {S.~V.}\ \bibnamefont
  {Golenetskii}}, \bibinfo {author} {\bibfnamefont {V.~D.}\ \bibnamefont
  {Pal'shin}}, \bibinfo {author} {\bibfnamefont {P.~P.}\ \bibnamefont
  {Oleynik}}, \bibinfo {author} {\bibfnamefont {A.~E.}\ \bibnamefont
  {Tsvetkova}}, \bibinfo {author} {\bibfnamefont {M.~V.}\ \bibnamefont
  {Ulanov}}, \bibinfo {author} {\bibfnamefont {T.~L.}\ \bibnamefont {Cline}}, \
  and\ \bibinfo {author} {\bibfnamefont {K.}~\bibnamefont {Hurley}},\ }\href
  {\doibase 10.3847/0067-0049/224/1/10} {\bibfield  {journal} {\bibinfo
  {journal} {Astrophys. J. Suppl.}\ }\textbf {\bibinfo {volume} {224}},\
  \bibinfo {pages} {10} (\bibinfo {year} {2016})},\ \Eprint
  {http://arxiv.org/abs/1603.06832} {arXiv:1603.06832 [astro-ph.HE]}
  \BibitemShut {NoStop}%
\bibitem [{\citenamefont {Ruiz}\ \emph {et~al.}(2018)\citenamefont {Ruiz},
  \citenamefont {Shapiro},\ and\ \citenamefont {Tsokaros}}]{Ruiz:2017due}%
  \BibitemOpen
  \bibfield  {author} {\bibinfo {author} {\bibfnamefont {M.}~\bibnamefont
  {Ruiz}}, \bibinfo {author} {\bibfnamefont {S.~L.}\ \bibnamefont {Shapiro}}, \
  and\ \bibinfo {author} {\bibfnamefont {A.}~\bibnamefont {Tsokaros}},\ }\href
  {\doibase 10.1103/PhysRevD.97.021501} {\bibfield  {journal} {\bibinfo
  {journal} {Phys. Rev.}\ }\textbf {\bibinfo {volume} {D97}},\ \bibinfo {pages}
  {021501} (\bibinfo {year} {2018})},\ \Eprint
  {http://arxiv.org/abs/1711.00473} {arXiv:1711.00473 [astro-ph.HE]}
  \BibitemShut {NoStop}%
\bibitem [{\citenamefont {Deaton}\ \emph {et~al.}(2013)\citenamefont {Deaton},
  \citenamefont {Duez}, \citenamefont {Foucart}, \citenamefont {O'Connor},
  \citenamefont {Ott}, \citenamefont {Kidder}, \citenamefont {Muhlberger},
  \citenamefont {Scheel},\ and\ \citenamefont {Szilagyi}}]{Deaton:2013}%
  \BibitemOpen
  \bibfield  {author} {\bibinfo {author} {\bibfnamefont {M.~B.}\ \bibnamefont
  {Deaton}}, \bibinfo {author} {\bibfnamefont {M.~D.}\ \bibnamefont {Duez}},
  \bibinfo {author} {\bibfnamefont {F.}~\bibnamefont {Foucart}}, \bibinfo
  {author} {\bibfnamefont {E.}~\bibnamefont {O'Connor}}, \bibinfo {author}
  {\bibfnamefont {C.~D.}\ \bibnamefont {Ott}}, \bibinfo {author} {\bibfnamefont
  {L.~E.}\ \bibnamefont {Kidder}}, \bibinfo {author} {\bibfnamefont {C.~D.}\
  \bibnamefont {Muhlberger}}, \bibinfo {author} {\bibfnamefont {M.~A.}\
  \bibnamefont {Scheel}}, \ and\ \bibinfo {author} {\bibfnamefont
  {B.}~\bibnamefont {Szilagyi}},\ }\href {\doibase 10.1088/0004-637x/776/1/47}
  {\bibfield  {journal} {\bibinfo  {journal} {The Astrophysical Journal}\
  }\textbf {\bibinfo {volume} {776}},\ \bibinfo {pages} {47} (\bibinfo {year}
  {2013})}\BibitemShut {NoStop}%
\bibitem [{\citenamefont {Neilsen}\ \emph {et~al.}(2014)\citenamefont
  {Neilsen}, \citenamefont {Liebling}, \citenamefont {Anderson}, \citenamefont
  {Lehner}, \citenamefont {O'Connor},\ and\ \citenamefont
  {Palenzuela}}]{Neilson2014leakage}%
  \BibitemOpen
  \bibfield  {author} {\bibinfo {author} {\bibfnamefont {D.}~\bibnamefont
  {Neilsen}}, \bibinfo {author} {\bibfnamefont {S.~L.}\ \bibnamefont
  {Liebling}}, \bibinfo {author} {\bibfnamefont {M.}~\bibnamefont {Anderson}},
  \bibinfo {author} {\bibfnamefont {L.}~\bibnamefont {Lehner}}, \bibinfo
  {author} {\bibfnamefont {E.}~\bibnamefont {O'Connor}}, \ and\ \bibinfo
  {author} {\bibfnamefont {C.}~\bibnamefont {Palenzuela}},\ }\href {\doibase
  10.1103/PhysRevD.89.104029} {\bibfield  {journal} {\bibinfo  {journal} {Phys.
  Rev. D}\ }\textbf {\bibinfo {volume} {89}},\ \bibinfo {pages} {104029}
  (\bibinfo {year} {2014})}\BibitemShut {NoStop}%
\bibitem [{\citenamefont {Wanajo}\ \emph {et~al.}(2014)\citenamefont {Wanajo},
  \citenamefont {Sekiguchi}, \citenamefont {Nishimura}, \citenamefont {Kiuchi},
  \citenamefont {Kyutoku},\ and\ \citenamefont {Shibata}}]{Wanajo_2014}%
  \BibitemOpen
  \bibfield  {author} {\bibinfo {author} {\bibfnamefont {S.}~\bibnamefont
  {Wanajo}}, \bibinfo {author} {\bibfnamefont {Y.}~\bibnamefont {Sekiguchi}},
  \bibinfo {author} {\bibfnamefont {N.}~\bibnamefont {Nishimura}}, \bibinfo
  {author} {\bibfnamefont {K.}~\bibnamefont {Kiuchi}}, \bibinfo {author}
  {\bibfnamefont {K.}~\bibnamefont {Kyutoku}}, \ and\ \bibinfo {author}
  {\bibfnamefont {M.}~\bibnamefont {Shibata}},\ }\href {\doibase
  10.1088/2041-8205/789/2/l39} {\bibfield  {journal} {\bibinfo  {journal} {The
  Astrophysical Journal}\ }\textbf {\bibinfo {volume} {789}},\ \bibinfo {pages}
  {L39} (\bibinfo {year} {2014})}\BibitemShut {NoStop}%
\bibitem [{\citenamefont {Radice}\ \emph
  {et~al.}(2016{\natexlab{a}})\citenamefont {Radice}, \citenamefont {Galeazzi},
  \citenamefont {Lippuner}, \citenamefont {Roberts}, \citenamefont {Ott},\ and\
  \citenamefont {Rezzolla}}]{RadiceDynamicalMass}%
  \BibitemOpen
  \bibfield  {author} {\bibinfo {author} {\bibfnamefont {D.}~\bibnamefont
  {Radice}}, \bibinfo {author} {\bibfnamefont {F.}~\bibnamefont {Galeazzi}},
  \bibinfo {author} {\bibfnamefont {J.}~\bibnamefont {Lippuner}}, \bibinfo
  {author} {\bibfnamefont {L.~F.}\ \bibnamefont {Roberts}}, \bibinfo {author}
  {\bibfnamefont {C.~D.}\ \bibnamefont {Ott}}, \ and\ \bibinfo {author}
  {\bibfnamefont {L.}~\bibnamefont {Rezzolla}},\ }\href {\doibase
  10.1093/mnras/stw1227} {\bibfield  {journal} {\bibinfo  {journal} {Monthly
  Notices of the Royal Astronomical Society}\ }\textbf {\bibinfo {volume}
  {460}},\ \bibinfo {pages} {3255} (\bibinfo {year}
  {2016}{\natexlab{a}})}\BibitemShut {NoStop}%
\bibitem [{\citenamefont {Foucart}\ \emph {et~al.}(2015)\citenamefont {Foucart}
  \emph {et~al.}}]{FoucartPostMerger}%
  \BibitemOpen
  \bibfield  {author} {\bibinfo {author} {\bibfnamefont {F.}~\bibnamefont
  {Foucart}} \emph {et~al.},\ }\href {\doibase 10.1103/PhysRevD.91.124021}
  {\bibfield  {journal} {\bibinfo  {journal} {Phys. Rev. D}\ }\textbf {\bibinfo
  {volume} {91}},\ \bibinfo {pages} {124021} (\bibinfo {year}
  {2015})}\BibitemShut {NoStop}%
\bibitem [{\citenamefont {Richers}\ \emph {et~al.}(2015)\citenamefont
  {Richers}, \citenamefont {Kasen}, \citenamefont {O'Connor}, \citenamefont
  {Fern{\'{a}}ndez},\ and\ \citenamefont {Ott}}]{RichersSedonu}%
  \BibitemOpen
  \bibfield  {author} {\bibinfo {author} {\bibfnamefont {S.}~\bibnamefont
  {Richers}}, \bibinfo {author} {\bibfnamefont {D.}~\bibnamefont {Kasen}},
  \bibinfo {author} {\bibfnamefont {E.}~\bibnamefont {O'Connor}}, \bibinfo
  {author} {\bibfnamefont {R.}~\bibnamefont {Fern{\'{a}}ndez}}, \ and\ \bibinfo
  {author} {\bibfnamefont {C.~D.}\ \bibnamefont {Ott}},\ }\href {\doibase
  10.1088/0004-637x/813/1/38} {\bibfield  {journal} {\bibinfo  {journal} {The
  Astrophysical Journal}\ }\textbf {\bibinfo {volume} {813}},\ \bibinfo {pages}
  {38} (\bibinfo {year} {2015})}\BibitemShut {NoStop}%
\bibitem [{\citenamefont {Foucart}(2018)}]{FoucarMCMoment}%
  \BibitemOpen
  \bibfield  {author} {\bibinfo {author} {\bibfnamefont {F.}~\bibnamefont
  {Foucart}},\ }\href {\doibase 10.1093/mnras/sty108} {\bibfield  {journal}
  {\bibinfo  {journal} {Monthly Notices of the Royal Astronomical Society}\
  }\textbf {\bibinfo {volume} {475}},\ \bibinfo {pages} {4186} (\bibinfo {year}
  {2018})}\BibitemShut {NoStop}%
\bibitem [{\citenamefont {Hossein~Nouri}\ \emph {et~al.}(2018)\citenamefont
  {Hossein~Nouri}, \citenamefont {Duez}, \citenamefont {Foucart}, \citenamefont
  {Deaton}, \citenamefont {Haas}, \citenamefont {Haddadi}, \citenamefont
  {Kidder}, \citenamefont {Ott}, \citenamefont {Pfeiffer}, \citenamefont
  {Scheel},\ and\ \citenamefont {Szilagyi}}]{NouriPostMerger}%
  \BibitemOpen
  \bibfield  {author} {\bibinfo {author} {\bibfnamefont {F.}~\bibnamefont
  {Hossein~Nouri}}, \bibinfo {author} {\bibfnamefont {M.~D.}\ \bibnamefont
  {Duez}}, \bibinfo {author} {\bibfnamefont {F.}~\bibnamefont {Foucart}},
  \bibinfo {author} {\bibfnamefont {M.~B.}\ \bibnamefont {Deaton}}, \bibinfo
  {author} {\bibfnamefont {R.}~\bibnamefont {Haas}}, \bibinfo {author}
  {\bibfnamefont {M.}~\bibnamefont {Haddadi}}, \bibinfo {author} {\bibfnamefont
  {L.~E.}\ \bibnamefont {Kidder}}, \bibinfo {author} {\bibfnamefont {C.~D.}\
  \bibnamefont {Ott}}, \bibinfo {author} {\bibfnamefont {H.~P.}\ \bibnamefont
  {Pfeiffer}}, \bibinfo {author} {\bibfnamefont {M.~A.}\ \bibnamefont
  {Scheel}}, \ and\ \bibinfo {author} {\bibfnamefont {B.}~\bibnamefont
  {Szilagyi}},\ }\href {\doibase 10.1103/PhysRevD.97.083014} {\bibfield
  {journal} {\bibinfo  {journal} {Phys. Rev. D}\ }\textbf {\bibinfo {volume}
  {97}},\ \bibinfo {pages} {083014} (\bibinfo {year} {2018})}\BibitemShut
  {NoStop}%
\bibitem [{\citenamefont {Siegel}\ and\ \citenamefont
  {Metzger}(2018)}]{SiegelMetzger3DBNS}%
  \BibitemOpen
  \bibfield  {author} {\bibinfo {author} {\bibfnamefont {D.~M.}\ \bibnamefont
  {Siegel}}\ and\ \bibinfo {author} {\bibfnamefont {B.~D.}\ \bibnamefont
  {Metzger}},\ }\href {http://stacks.iop.org/0004-637X/858/i=1/a=52} {\bibfield
   {journal} {\bibinfo  {journal} {The Astrophysical Journal}\ }\textbf
  {\bibinfo {volume} {858}},\ \bibinfo {pages} {52} (\bibinfo {year}
  {2018})}\BibitemShut {NoStop}%
\bibitem [{\citenamefont {{Fern{\'a}ndez}}\ \emph {et~al.}(2018)\citenamefont
  {{Fern{\'a}ndez}} \emph {et~al.}}]{FernandezLongTermGRMHD}%
  \BibitemOpen
  \bibfield  {author} {\bibinfo {author} {\bibfnamefont {R.}~\bibnamefont
  {{Fern{\'a}ndez}}} \emph {et~al.},\ }\href@noop {} {\bibfield  {journal}
  {\bibinfo  {journal} {ArXiv e-prints}\ } (\bibinfo {year} {2018})},\ \Eprint
  {http://arxiv.org/abs/1808.00461} {arXiv:1808.00461 [astro-ph.HE]}
  \BibitemShut {NoStop}%
\bibitem [{\citenamefont {Baiotti}\ and\ \citenamefont
  {Rezzolla}(2017)}]{BaiottiRezzollaReview}%
  \BibitemOpen
  \bibfield  {author} {\bibinfo {author} {\bibfnamefont {L.}~\bibnamefont
  {Baiotti}}\ and\ \bibinfo {author} {\bibfnamefont {L.}~\bibnamefont
  {Rezzolla}},\ }\href {http://stacks.iop.org/0034-4885/80/i=9/a=096901}
  {\bibfield  {journal} {\bibinfo  {journal} {Reports on Progress in Physics}\
  }\textbf {\bibinfo {volume} {80}},\ \bibinfo {pages} {096901} (\bibinfo
  {year} {2017})}\BibitemShut {NoStop}%
\bibitem [{\citenamefont {Kumar}\ and\ \citenamefont
  {Zhang}(2015)}]{KUMAR20151}%
  \BibitemOpen
  \bibfield  {author} {\bibinfo {author} {\bibfnamefont {P.}~\bibnamefont
  {Kumar}}\ and\ \bibinfo {author} {\bibfnamefont {B.}~\bibnamefont {Zhang}},\
  }\href {\doibase https://doi.org/10.1016/j.physrep.2014.09.008} {\bibfield
  {journal} {\bibinfo  {journal} {Physics Reports}\ }\textbf {\bibinfo {volume}
  {561}},\ \bibinfo {pages} {1 } (\bibinfo {year} {2015})},\ \bibinfo {note}
  {the physics of gamma-ray bursts and relativistic jets}\BibitemShut {NoStop}%
\bibitem [{\citenamefont {Lippuner}\ and\ \citenamefont
  {Roberts}(2017)}]{LippunerSkynet}%
  \BibitemOpen
  \bibfield  {author} {\bibinfo {author} {\bibfnamefont {J.}~\bibnamefont
  {Lippuner}}\ and\ \bibinfo {author} {\bibfnamefont {L.~F.}\ \bibnamefont
  {Roberts}},\ }\href {http://stacks.iop.org/0067-0049/233/i=2/a=18} {\bibfield
   {journal} {\bibinfo  {journal} {The Astrophysical Journal Supplement
  Series}\ }\textbf {\bibinfo {volume} {233}},\ \bibinfo {pages} {18} (\bibinfo
  {year} {2017})}\BibitemShut {NoStop}%
\bibitem [{\citenamefont {{Kasen}}\ \emph
  {et~al.}(2013{\natexlab{a}})\citenamefont {{Kasen}}, \citenamefont
  {{Badnell}},\ and\ \citenamefont {{Barnes}}}]{KasenOpacitiesBNS}%
  \BibitemOpen
  \bibfield  {author} {\bibinfo {author} {\bibfnamefont {D.}~\bibnamefont
  {{Kasen}}}, \bibinfo {author} {\bibfnamefont {N.~R.}\ \bibnamefont
  {{Badnell}}}, \ and\ \bibinfo {author} {\bibfnamefont {J.}~\bibnamefont
  {{Barnes}}},\ }\href {\doibase 10.1088/0004-637X/774/1/25} {\bibfield
  {journal} {\bibinfo  {journal} {{ApJ}}\ }\textbf {\bibinfo {volume} {774}},\
  \bibinfo {eid} {25} (\bibinfo {year} {2013}{\natexlab{a}})},\ \Eprint
  {http://arxiv.org/abs/1303.5788} {arXiv:1303.5788 [astro-ph.HE]} \BibitemShut
  {NoStop}%
\bibitem [{\citenamefont {Tanvir}\ \emph {et~al.}(2017)\citenamefont {Tanvir}
  \emph {et~al.}}]{TanvirHST}%
  \BibitemOpen
  \bibfield  {author} {\bibinfo {author} {\bibfnamefont {N.~R.}\ \bibnamefont
  {Tanvir}} \emph {et~al.},\ }\href
  {http://stacks.iop.org/2041-8205/848/i=2/a=L27} {\bibfield  {journal}
  {\bibinfo  {journal} {The Astrophysical Journal Letters}\ }\textbf {\bibinfo
  {volume} {848}},\ \bibinfo {pages} {L27} (\bibinfo {year}
  {2017})}\BibitemShut {NoStop}%
\bibitem [{\citenamefont {Connaughton}\ \emph {et~al.}(2016)\citenamefont
  {Connaughton} \emph {et~al.}}]{ConBurGold16}%
  \BibitemOpen
  \bibfield  {author} {\bibinfo {author} {\bibfnamefont {V.}~\bibnamefont
  {Connaughton}} \emph {et~al.},\ }\href
  {http://stacks.iop.org/2041-8205/826/i=1/a=L6} {\bibfield  {journal}
  {\bibinfo  {journal} {The Astrophysical Journal Letters}\ }\textbf {\bibinfo
  {volume} {826}},\ \bibinfo {pages} {L6} (\bibinfo {year} {2016})}\BibitemShut
  {NoStop}%
\bibitem [{\citenamefont {Verrecchia}\ \emph {et~al.}(2017)\citenamefont
  {Verrecchia}, \citenamefont {Tavani}, \citenamefont {Ursi}, \citenamefont
  {Argan}, \citenamefont {Pittori}, \citenamefont {Donnarumma}, \citenamefont
  {Bulgarelli}, \citenamefont {Fuschino}, \citenamefont {Labanti},\ and\
  \citenamefont {et~al.}}]{VerTavUrs17}%
  \BibitemOpen
  \bibfield  {author} {\bibinfo {author} {\bibfnamefont {F.}~\bibnamefont
  {Verrecchia}}, \bibinfo {author} {\bibfnamefont {M.}~\bibnamefont {Tavani}},
  \bibinfo {author} {\bibfnamefont {A.}~\bibnamefont {Ursi}}, \bibinfo {author}
  {\bibfnamefont {A.}~\bibnamefont {Argan}}, \bibinfo {author} {\bibfnamefont
  {C.}~\bibnamefont {Pittori}}, \bibinfo {author} {\bibfnamefont
  {I.}~\bibnamefont {Donnarumma}}, \bibinfo {author} {\bibfnamefont
  {A.}~\bibnamefont {Bulgarelli}}, \bibinfo {author} {\bibfnamefont
  {F.}~\bibnamefont {Fuschino}}, \bibinfo {author} {\bibfnamefont
  {C.}~\bibnamefont {Labanti}}, \ and\ \bibinfo {author} {\bibnamefont
  {et~al.}},\ }\href {http://stacks.iop.org/2041-8205/847/i=2/a=L20} {\bibfield
   {journal} {\bibinfo  {journal} {The Astrophysical Journal Letters}\ }\textbf
  {\bibinfo {volume} {847}},\ \bibinfo {pages} {L20} (\bibinfo {year}
  {2017})}\BibitemShut {NoStop}%
\bibitem [{\citenamefont {{Radice}}\ \emph {et~al.}(2018)\citenamefont
  {{Radice}}, \citenamefont {{Perego}}, \citenamefont {{Hotokezaka}},
  \citenamefont {{Fromm}}, \citenamefont {{Bernuzzi}},\ and\ \citenamefont
  {{Roberts}}}]{rad:2018ApJR}%
  \BibitemOpen
  \bibfield  {author} {\bibinfo {author} {\bibfnamefont {D.}~\bibnamefont
  {{Radice}}}, \bibinfo {author} {\bibfnamefont {A.}~\bibnamefont {{Perego}}},
  \bibinfo {author} {\bibfnamefont {K.}~\bibnamefont {{Hotokezaka}}}, \bibinfo
  {author} {\bibfnamefont {S.~A.}\ \bibnamefont {{Fromm}}}, \bibinfo {author}
  {\bibfnamefont {S.}~\bibnamefont {{Bernuzzi}}}, \ and\ \bibinfo {author}
  {\bibfnamefont {L.~F.}\ \bibnamefont {{Roberts}}},\ }\href {\doibase
  10.3847/1538-4357/aaf054} {\bibfield  {journal} {\bibinfo  {journal} {\apj}\
  }\textbf {\bibinfo {volume} {869}},\ \bibinfo {eid} {130} (\bibinfo {year}
  {2018})},\ \Eprint {http://arxiv.org/abs/1809.11161} {arXiv:1809.11161
  [astro-ph.HE]} \BibitemShut {NoStop}%
\bibitem [{\citenamefont {Berger}\ and\ \citenamefont
  {Colella}(1989)}]{BERGER198964}%
  \BibitemOpen
  \bibfield  {author} {\bibinfo {author} {\bibfnamefont {M.}~\bibnamefont
  {Berger}}\ and\ \bibinfo {author} {\bibfnamefont {P.}~\bibnamefont
  {Colella}},\ }\href {\doibase https://doi.org/10.1016/0021-9991(89)90035-1}
  {\bibfield  {journal} {\bibinfo  {journal} {Journal of Computational
  Physics}\ }\textbf {\bibinfo {volume} {82}},\ \bibinfo {pages} {64 }
  (\bibinfo {year} {1989})}\BibitemShut {NoStop}%
\bibitem [{\citenamefont {{Balbus}}\ and\ \citenamefont
  {{Hawley}}(1991)}]{BH91}%
  \BibitemOpen
  \bibfield  {author} {\bibinfo {author} {\bibfnamefont {S.~A.}\ \bibnamefont
  {{Balbus}}}\ and\ \bibinfo {author} {\bibfnamefont {J.~F.}\ \bibnamefont
  {{Hawley}}},\ }\href {\doibase 10.1086/170270} {\bibfield  {journal}
  {\bibinfo  {journal} {\apj}\ }\textbf {\bibinfo {volume} {376}},\ \bibinfo
  {pages} {214} (\bibinfo {year} {1991})}\BibitemShut {NoStop}%
\bibitem [{\citenamefont {{Balbus}}\ and\ \citenamefont
  {{Hawley}}(1998)}]{BH98}%
  \BibitemOpen
  \bibfield  {author} {\bibinfo {author} {\bibfnamefont {S.~A.}\ \bibnamefont
  {{Balbus}}}\ and\ \bibinfo {author} {\bibfnamefont {J.~F.}\ \bibnamefont
  {{Hawley}}},\ }\href {\doibase 10.1103/RevModPhys.70.1} {\bibfield  {journal}
  {\bibinfo  {journal} {Reviews of Modern Physics}\ }\textbf {\bibinfo {volume}
  {70}},\ \bibinfo {pages} {1} (\bibinfo {year} {1998})}\BibitemShut {NoStop}%
\bibitem [{\citenamefont {{Shakura}}\ and\ \citenamefont
  {{Sunyaev}}(1973)}]{ShakuraSunyaevAlpha}%
  \BibitemOpen
  \bibfield  {author} {\bibinfo {author} {\bibfnamefont {N.~I.}\ \bibnamefont
  {{Shakura}}}\ and\ \bibinfo {author} {\bibfnamefont {R.~A.}\ \bibnamefont
  {{Sunyaev}}},\ }\href@noop {} {\bibfield  {journal} {\bibinfo  {journal}
  {\aap}\ }\textbf {\bibinfo {volume} {24}},\ \bibinfo {pages} {337} (\bibinfo
  {year} {1973})}\BibitemShut {NoStop}%
\bibitem [{\citenamefont {{Hawley}}\ \emph {et~al.}(1995)\citenamefont
  {{Hawley}}, \citenamefont {{Gammie}},\ and\ \citenamefont
  {{Balbus}}}]{HawleyGammieBalbusShearingBox}%
  \BibitemOpen
  \bibfield  {author} {\bibinfo {author} {\bibfnamefont {J.~F.}\ \bibnamefont
  {{Hawley}}}, \bibinfo {author} {\bibfnamefont {C.~F.}\ \bibnamefont
  {{Gammie}}}, \ and\ \bibinfo {author} {\bibfnamefont {S.~A.}\ \bibnamefont
  {{Balbus}}},\ }\href {\doibase 10.1086/175311} {\bibfield  {journal}
  {\bibinfo  {journal} {\apj}\ }\textbf {\bibinfo {volume} {440}},\ \bibinfo
  {pages} {742} (\bibinfo {year} {1995})}\BibitemShut {NoStop}%
\bibitem [{\citenamefont {Zhiyin}(2015)}]{ZHIYIN201511}%
  \BibitemOpen
  \bibfield  {author} {\bibinfo {author} {\bibfnamefont {Y.}~\bibnamefont
  {Zhiyin}},\ }\href {\doibase https://doi.org/10.1016/j.cja.2014.12.007}
  {\bibfield  {journal} {\bibinfo  {journal} {Chinese Journal of Aeronautics}\
  }\textbf {\bibinfo {volume} {28}},\ \bibinfo {pages} {11 } (\bibinfo {year}
  {2015})}\BibitemShut {NoStop}%
\bibitem [{\citenamefont {Duraisamy}\ \emph {et~al.}(2015)\citenamefont
  {Duraisamy}, \citenamefont {Zhang},\ and\ \citenamefont {Singh}}]{DZS15AIAA}%
  \BibitemOpen
  \bibfield  {author} {\bibinfo {author} {\bibfnamefont {K.}~\bibnamefont
  {Duraisamy}}, \bibinfo {author} {\bibfnamefont {Z.~J.}\ \bibnamefont
  {Zhang}}, \ and\ \bibinfo {author} {\bibfnamefont {A.~P.}\ \bibnamefont
  {Singh}},\ }in\ \href@noop {} {\emph {\bibinfo {booktitle} {53rd AIAA
  Aerospace Sciences Meeting}}}\ (\bibinfo {year} {2015})\ p.\ \bibinfo {pages}
  {1284}\BibitemShut {NoStop}%
\bibitem [{\citenamefont {Parish}\ and\ \citenamefont
  {Duraisamy}(2016)}]{PD16JCOMP}%
  \BibitemOpen
  \bibfield  {author} {\bibinfo {author} {\bibfnamefont {E.~J.}\ \bibnamefont
  {Parish}}\ and\ \bibinfo {author} {\bibfnamefont {K.}~\bibnamefont
  {Duraisamy}},\ }\href {\doibase https://doi.org/10.1016/j.jcp.2015.11.012}
  {\bibfield  {journal} {\bibinfo  {journal} {Journal of Computational
  Physics}\ }\textbf {\bibinfo {volume} {305}},\ \bibinfo {pages} {758 }
  (\bibinfo {year} {2016})}\BibitemShut {NoStop}%
\bibitem [{\citenamefont {Wang}\ \emph {et~al.}(2017)\citenamefont {Wang},
  \citenamefont {Wu},\ and\ \citenamefont {Xiao}}]{WWX17PRF}%
  \BibitemOpen
  \bibfield  {author} {\bibinfo {author} {\bibfnamefont {J.-X.}\ \bibnamefont
  {Wang}}, \bibinfo {author} {\bibfnamefont {J.-L.}\ \bibnamefont {Wu}}, \ and\
  \bibinfo {author} {\bibfnamefont {H.}~\bibnamefont {Xiao}},\ }\href {\doibase
  10.1103/PhysRevFluids.2.034603} {\bibfield  {journal} {\bibinfo  {journal}
  {Phys. Rev. Fluids}\ }\textbf {\bibinfo {volume} {2}},\ \bibinfo {pages}
  {034603} (\bibinfo {year} {2017})}\BibitemShut {NoStop}%
\bibitem [{\citenamefont {{King}}\ \emph {et~al.}(2017)\citenamefont {{King}},
  \citenamefont {{Graf}},\ and\ \citenamefont {{Chertkov}}}]{KHC17APS}%
  \BibitemOpen
  \bibfield  {author} {\bibinfo {author} {\bibfnamefont {R.}~\bibnamefont
  {{King}}}, \bibinfo {author} {\bibfnamefont {P.}~\bibnamefont {{Graf}}}, \
  and\ \bibinfo {author} {\bibfnamefont {M.}~\bibnamefont {{Chertkov}}},\ }in\
  \href@noop {} {\emph {\bibinfo {booktitle} {APS Meeting Abstracts}}}\
  (\bibinfo {year} {2017})\ p.\ \bibinfo {pages} {A31.008}\BibitemShut
  {NoStop}%
\bibitem [{\citenamefont {Maulik}\ and\ \citenamefont {San}(2017)}]{MS17JFM}%
  \BibitemOpen
  \bibfield  {author} {\bibinfo {author} {\bibfnamefont {R.}~\bibnamefont
  {Maulik}}\ and\ \bibinfo {author} {\bibfnamefont {O.}~\bibnamefont {San}},\
  }\href {\doibase 10.1017/jfm.2017.637} {\bibfield  {journal} {\bibinfo
  {journal} {Journal of Fluid Mechanics}\ }\textbf {\bibinfo {volume} {831}},\
  \bibinfo {pages} {151 to 181} (\bibinfo {year} {2017})}\BibitemShut {NoStop}%
\bibitem [{\citenamefont {{Miyanawala}}\ and\ \citenamefont
  {{Jaiman}}(2017)}]{MJ17}%
  \BibitemOpen
  \bibfield  {author} {\bibinfo {author} {\bibfnamefont {T.~P.}\ \bibnamefont
  {{Miyanawala}}}\ and\ \bibinfo {author} {\bibfnamefont {R.~K.}\ \bibnamefont
  {{Jaiman}}},\ }\href@noop {} {\bibfield  {journal} {\bibinfo  {journal}
  {arXiv e-prints}\ ,\ \bibinfo {eid} {arXiv:1710.09099}} (\bibinfo {year}
  {2017})},\ \Eprint {http://arxiv.org/abs/1710.09099} {arXiv:1710.09099
  [physics.flu-dyn]} \BibitemShut {NoStop}%
\bibitem [{\citenamefont {{Barati Farimani}}\ \emph {et~al.}(2017)\citenamefont
  {{Barati Farimani}}, \citenamefont {{Gomes}},\ and\ \citenamefont
  {{Pande}}}]{BHP17}%
  \BibitemOpen
  \bibfield  {author} {\bibinfo {author} {\bibfnamefont {A.}~\bibnamefont
  {{Barati Farimani}}}, \bibinfo {author} {\bibfnamefont {J.}~\bibnamefont
  {{Gomes}}}, \ and\ \bibinfo {author} {\bibfnamefont {V.~S.}\ \bibnamefont
  {{Pande}}},\ }\href@noop {} {\bibfield  {journal} {\bibinfo  {journal} {arXiv
  e-prints}\ ,\ \bibinfo {eid} {arXiv:1709.02432}} (\bibinfo {year} {2017})},\
  \Eprint {http://arxiv.org/abs/1709.02432} {arXiv:1709.02432 [cs.LG]}
  \BibitemShut {NoStop}%
\bibitem [{\citenamefont {{Hennigh}}(2017)}]{Hennigh17}%
  \BibitemOpen
  \bibfield  {author} {\bibinfo {author} {\bibfnamefont {O.}~\bibnamefont
  {{Hennigh}}},\ }\href@noop {} {\bibfield  {journal} {\bibinfo  {journal}
  {arXiv e-prints}\ ,\ \bibinfo {eid} {arXiv:1705.09036}} (\bibinfo {year}
  {2017})},\ \Eprint {http://arxiv.org/abs/1705.09036} {arXiv:1705.09036
  [stat.ML]} \BibitemShut {NoStop}%
\bibitem [{\citenamefont {{Xie}}\ \emph {et~al.}(2018)\citenamefont {{Xie}},
  \citenamefont {{Franz}}, \citenamefont {{Chu}},\ and\ \citenamefont
  {{Thuerey}}}]{XFCT18}%
  \BibitemOpen
  \bibfield  {author} {\bibinfo {author} {\bibfnamefont {Y.}~\bibnamefont
  {{Xie}}}, \bibinfo {author} {\bibfnamefont {E.}~\bibnamefont {{Franz}}},
  \bibinfo {author} {\bibfnamefont {M.}~\bibnamefont {{Chu}}}, \ and\ \bibinfo
  {author} {\bibfnamefont {N.}~\bibnamefont {{Thuerey}}},\ }\href@noop {}
  {\bibfield  {journal} {\bibinfo  {journal} {arXiv e-prints}\ ,\ \bibinfo
  {eid} {arXiv:1801.09710}} (\bibinfo {year} {2018})},\ \Eprint
  {http://arxiv.org/abs/1801.09710} {arXiv:1801.09710 [cs.LG]} \BibitemShut
  {NoStop}%
\bibitem [{\citenamefont {{Mohan}}\ and\ \citenamefont
  {{Gaitonde}}(2018)}]{MGD18}%
  \BibitemOpen
  \bibfield  {author} {\bibinfo {author} {\bibfnamefont {A.~T.}\ \bibnamefont
  {{Mohan}}}\ and\ \bibinfo {author} {\bibfnamefont {D.~V.}\ \bibnamefont
  {{Gaitonde}}},\ }\href@noop {} {\bibfield  {journal} {\bibinfo  {journal}
  {arXiv e-prints}\ ,\ \bibinfo {eid} {arXiv:1804.09269}} (\bibinfo {year}
  {2018})},\ \Eprint {http://arxiv.org/abs/1804.09269} {arXiv:1804.09269
  [physics.comp-ph]} \BibitemShut {NoStop}%
\bibitem [{\citenamefont {{King}}\ \emph {et~al.}(2018)\citenamefont {{King}},
  \citenamefont {{Hennigh}}, \citenamefont {{Mohan}},\ and\ \citenamefont
  {{Chertkov}}}]{KHMC18}%
  \BibitemOpen
  \bibfield  {author} {\bibinfo {author} {\bibfnamefont {R.}~\bibnamefont
  {{King}}}, \bibinfo {author} {\bibfnamefont {O.}~\bibnamefont {{Hennigh}}},
  \bibinfo {author} {\bibfnamefont {A.}~\bibnamefont {{Mohan}}}, \ and\
  \bibinfo {author} {\bibfnamefont {M.}~\bibnamefont {{Chertkov}}},\
  }\href@noop {} {\bibfield  {journal} {\bibinfo  {journal} {arXiv e-prints}\
  ,\ \bibinfo {eid} {arXiv:1810.07785}} (\bibinfo {year} {2018})},\ \Eprint
  {http://arxiv.org/abs/1810.07785} {arXiv:1810.07785 [physics.flu-dyn]}
  \BibitemShut {NoStop}%
\bibitem [{\citenamefont {{Wiewel}}\ \emph {et~al.}(2018)\citenamefont
  {{Wiewel}}, \citenamefont {{Becher}},\ and\ \citenamefont
  {{Thuerey}}}]{WBT18}%
  \BibitemOpen
  \bibfield  {author} {\bibinfo {author} {\bibfnamefont {S.}~\bibnamefont
  {{Wiewel}}}, \bibinfo {author} {\bibfnamefont {M.}~\bibnamefont {{Becher}}},
  \ and\ \bibinfo {author} {\bibfnamefont {N.}~\bibnamefont {{Thuerey}}},\
  }\href@noop {} {\bibfield  {journal} {\bibinfo  {journal} {arXiv e-prints}\
  ,\ \bibinfo {eid} {arXiv:1802.10123}} (\bibinfo {year} {2018})},\ \Eprint
  {http://arxiv.org/abs/1802.10123} {arXiv:1802.10123 [cs.LG]} \BibitemShut
  {NoStop}%
\bibitem [{\citenamefont {Weinan}\ \emph {et~al.}(2017)\citenamefont {Weinan},
  \citenamefont {Han},\ and\ \citenamefont {Jentzen}}]{weinan2017deep}%
  \BibitemOpen
  \bibfield  {author} {\bibinfo {author} {\bibfnamefont {E.}~\bibnamefont
  {Weinan}}, \bibinfo {author} {\bibfnamefont {J.}~\bibnamefont {Han}}, \ and\
  \bibinfo {author} {\bibfnamefont {A.}~\bibnamefont {Jentzen}},\ }\href@noop
  {} {\bibfield  {journal} {\bibinfo  {journal} {Communications in Mathematics
  and Statistics}\ }\textbf {\bibinfo {volume} {5}},\ \bibinfo {pages} {349}
  (\bibinfo {year} {2017})}\BibitemShut {NoStop}%
\bibitem [{\citenamefont {Berg}\ and\ \citenamefont
  {Nystr{\"o}m}(2018)}]{berg2018unified}%
  \BibitemOpen
  \bibfield  {author} {\bibinfo {author} {\bibfnamefont {J.}~\bibnamefont
  {Berg}}\ and\ \bibinfo {author} {\bibfnamefont {K.}~\bibnamefont
  {Nystr{\"o}m}},\ }\href@noop {} {\bibfield  {journal} {\bibinfo  {journal}
  {Neurocomputing}\ }\textbf {\bibinfo {volume} {317}},\ \bibinfo {pages} {28}
  (\bibinfo {year} {2018})}\BibitemShut {NoStop}%
\bibitem [{\citenamefont {Chen}\ \emph {et~al.}(2018)\citenamefont {Chen},
  \citenamefont {Rubanova}, \citenamefont {Bettencourt},\ and\ \citenamefont
  {Duvenaud}}]{chen2018neural}%
  \BibitemOpen
  \bibfield  {author} {\bibinfo {author} {\bibfnamefont {T.~Q.}\ \bibnamefont
  {Chen}}, \bibinfo {author} {\bibfnamefont {Y.}~\bibnamefont {Rubanova}},
  \bibinfo {author} {\bibfnamefont {J.}~\bibnamefont {Bettencourt}}, \ and\
  \bibinfo {author} {\bibfnamefont {D.}~\bibnamefont {Duvenaud}},\ }\href@noop
  {} {\bibfield  {journal} {\bibinfo  {journal} {arXiv preprint
  arXiv:1806.07366}\ } (\bibinfo {year} {2018})}\BibitemShut {NoStop}%
\bibitem [{\citenamefont {Mösta}\ \emph {et~al.}(2015)\citenamefont {Mösta},
  \citenamefont {Ott}, \citenamefont {Radice}, \citenamefont {Roberts},
  \citenamefont {Schnetter},\ and\ \citenamefont {Haas}}]{Mosta:2015ucs}%
  \BibitemOpen
  \bibfield  {author} {\bibinfo {author} {\bibfnamefont {P.}~\bibnamefont
  {Mösta}}, \bibinfo {author} {\bibfnamefont {C.~D.}\ \bibnamefont {Ott}},
  \bibinfo {author} {\bibfnamefont {D.}~\bibnamefont {Radice}}, \bibinfo
  {author} {\bibfnamefont {L.~F.}\ \bibnamefont {Roberts}}, \bibinfo {author}
  {\bibfnamefont {E.}~\bibnamefont {Schnetter}}, \ and\ \bibinfo {author}
  {\bibfnamefont {R.}~\bibnamefont {Haas}},\ }\href {\doibase
  10.1038/nature15755} {\bibfield  {journal} {\bibinfo  {journal} {Nature}\
  }\textbf {\bibinfo {volume} {528}},\ \bibinfo {pages} {376} (\bibinfo {year}
  {2015})},\ \Eprint {http://arxiv.org/abs/1512.00838} {arXiv:1512.00838
  [astro-ph.HE]} \BibitemShut {NoStop}%
\bibitem [{\citenamefont {Kiuchi}\ \emph {et~al.}(2018)\citenamefont {Kiuchi},
  \citenamefont {Kyutoku}, \citenamefont {Sekiguchi},\ and\ \citenamefont
  {Shibata}}]{Kiuchi:2017zzg}%
  \BibitemOpen
  \bibfield  {author} {\bibinfo {author} {\bibfnamefont {K.}~\bibnamefont
  {Kiuchi}}, \bibinfo {author} {\bibfnamefont {K.}~\bibnamefont {Kyutoku}},
  \bibinfo {author} {\bibfnamefont {Y.}~\bibnamefont {Sekiguchi}}, \ and\
  \bibinfo {author} {\bibfnamefont {M.}~\bibnamefont {Shibata}},\ }\href
  {\doibase 10.1103/PhysRevD.97.124039} {\bibfield  {journal} {\bibinfo
  {journal} {Phys. Rev.}\ }\textbf {\bibinfo {volume} {D97}},\ \bibinfo {pages}
  {124039} (\bibinfo {year} {2018})},\ \Eprint
  {http://arxiv.org/abs/1710.01311} {arXiv:1710.01311 [astro-ph.HE]}
  \BibitemShut {NoStop}%
\bibitem [{\citenamefont {Radice}\ \emph
  {et~al.}(2016{\natexlab{b}})\citenamefont {Radice}, \citenamefont {Ott},
  \citenamefont {Abdikamalov}, \citenamefont {Couch}, \citenamefont {Haas},\
  and\ \citenamefont {Schnetter}}]{Radice:2015qva}%
  \BibitemOpen
  \bibfield  {author} {\bibinfo {author} {\bibfnamefont {D.}~\bibnamefont
  {Radice}}, \bibinfo {author} {\bibfnamefont {C.~D.}\ \bibnamefont {Ott}},
  \bibinfo {author} {\bibfnamefont {E.}~\bibnamefont {Abdikamalov}}, \bibinfo
  {author} {\bibfnamefont {S.~M.}\ \bibnamefont {Couch}}, \bibinfo {author}
  {\bibfnamefont {R.}~\bibnamefont {Haas}}, \ and\ \bibinfo {author}
  {\bibfnamefont {E.}~\bibnamefont {Schnetter}},\ }\href {\doibase
  10.3847/0004-637X/820/1/76} {\bibfield  {journal} {\bibinfo  {journal}
  {Astrophys. J.}\ }\textbf {\bibinfo {volume} {820}},\ \bibinfo {pages} {76}
  (\bibinfo {year} {2016}{\natexlab{b}})},\ \Eprint
  {http://arxiv.org/abs/1510.05022} {arXiv:1510.05022 [astro-ph.HE]}
  \BibitemShut {NoStop}%
\bibitem [{\citenamefont {Giacomazzo}\ \emph {et~al.}(2015)\citenamefont
  {Giacomazzo}, \citenamefont {Zrake}, \citenamefont {Duffell}, \citenamefont
  {MacFadyen},\ and\ \citenamefont {Perna}}]{Giacomazzo:2014qba}%
  \BibitemOpen
  \bibfield  {author} {\bibinfo {author} {\bibfnamefont {B.}~\bibnamefont
  {Giacomazzo}}, \bibinfo {author} {\bibfnamefont {J.}~\bibnamefont {Zrake}},
  \bibinfo {author} {\bibfnamefont {P.}~\bibnamefont {Duffell}}, \bibinfo
  {author} {\bibfnamefont {A.~I.}\ \bibnamefont {MacFadyen}}, \ and\ \bibinfo
  {author} {\bibfnamefont {R.}~\bibnamefont {Perna}},\ }\href {\doibase
  10.1088/0004-637X/809/1/39} {\bibfield  {journal} {\bibinfo  {journal}
  {Astrophys. J.}\ }\textbf {\bibinfo {volume} {809}},\ \bibinfo {pages} {39}
  (\bibinfo {year} {2015})},\ \Eprint {http://arxiv.org/abs/1410.0013}
  {arXiv:1410.0013 [astro-ph.HE]} \BibitemShut {NoStop}%
\bibitem [{\citenamefont {{Metzger}}\ and\ \citenamefont
  {{Berger}}(2012)}]{Metzger:2012}%
  \BibitemOpen
  \bibfield  {author} {\bibinfo {author} {\bibfnamefont {B.~D.}\ \bibnamefont
  {{Metzger}}}\ and\ \bibinfo {author} {\bibfnamefont {E.}~\bibnamefont
  {{Berger}}},\ }\href {\doibase 10.1088/0004-637X/746/1/48} {\bibfield
  {journal} {\bibinfo  {journal} {\apj}\ }\textbf {\bibinfo {volume} {746}},\
  \bibinfo {eid} {48} (\bibinfo {year} {2012})},\ \Eprint
  {http://arxiv.org/abs/1108.6056} {arXiv:1108.6056 [astro-ph.HE]} \BibitemShut
  {NoStop}%
\bibitem [{\citenamefont {{Barnes}}\ and\ \citenamefont
  {{Kasen}}(2013)}]{Barnes2013}%
  \BibitemOpen
  \bibfield  {author} {\bibinfo {author} {\bibfnamefont {J.}~\bibnamefont
  {{Barnes}}}\ and\ \bibinfo {author} {\bibfnamefont {D.}~\bibnamefont
  {{Kasen}}},\ }\href {\doibase 10.1088/0004-637X/775/1/18} {\bibfield
  {journal} {\bibinfo  {journal} {\apj}\ }\textbf {\bibinfo {volume} {775}},\
  \bibinfo {eid} {18} (\bibinfo {year} {2013})},\ \Eprint
  {http://arxiv.org/abs/1303.5787} {arXiv:1303.5787 [astro-ph.HE]} \BibitemShut
  {NoStop}%
\bibitem [{\citenamefont {{Kasen}}\ \emph
  {et~al.}(2013{\natexlab{b}})\citenamefont {{Kasen}}, \citenamefont
  {{Badnell}},\ and\ \citenamefont {{Barnes}}}]{Kasen2013}%
  \BibitemOpen
  \bibfield  {author} {\bibinfo {author} {\bibfnamefont {D.}~\bibnamefont
  {{Kasen}}}, \bibinfo {author} {\bibfnamefont {N.~R.}\ \bibnamefont
  {{Badnell}}}, \ and\ \bibinfo {author} {\bibfnamefont {J.}~\bibnamefont
  {{Barnes}}},\ }\href {\doibase 10.1088/0004-637X/774/1/25} {\bibfield
  {journal} {\bibinfo  {journal} {\apj}\ }\textbf {\bibinfo {volume} {774}},\
  \bibinfo {eid} {25} (\bibinfo {year} {2013}{\natexlab{b}})},\ \Eprint
  {http://arxiv.org/abs/1303.5788} {arXiv:1303.5788 [astro-ph.HE]} \BibitemShut
  {NoStop}%
\bibitem [{\citenamefont {{Rosswog}}(2015)}]{2015IJMPD..2430012R}%
  \BibitemOpen
  \bibfield  {author} {\bibinfo {author} {\bibfnamefont {S.}~\bibnamefont
  {{Rosswog}}},\ }\href {\doibase 10.1142/S0218271815300128} {\bibfield
  {journal} {\bibinfo  {journal} {International Journal of Modern Physics D}\
  }\textbf {\bibinfo {volume} {24}},\ \bibinfo {eid} {1530012-52} (\bibinfo
  {year} {2015})},\ \Eprint {http://arxiv.org/abs/1501.02081} {arXiv:1501.02081
  [astro-ph.HE]} \BibitemShut {NoStop}%
\bibitem [{\citenamefont {{Villar}}\ \emph {et~al.}(2017)\citenamefont
  {{Villar}}, \citenamefont {{Guillochon}}, \citenamefont {{Berger}},
  \citenamefont {{Metzger}}, \citenamefont {{Cowperthwaite}}, \citenamefont
  {{Nicholl}}, \citenamefont {{Alexander}}, \citenamefont {{Blanchard}},
  \citenamefont {{Chornock}}, \citenamefont {{Eftekhari}}, \citenamefont
  {{Fong}}, \citenamefont {{Margutti}},\ and\ \citenamefont
  {{Williams}}}]{villar:2017ApJ}%
  \BibitemOpen
  \bibfield  {author} {\bibinfo {author} {\bibfnamefont {V.~A.}\ \bibnamefont
  {{Villar}}}, \bibinfo {author} {\bibfnamefont {J.}~\bibnamefont
  {{Guillochon}}}, \bibinfo {author} {\bibfnamefont {E.}~\bibnamefont
  {{Berger}}}, \bibinfo {author} {\bibfnamefont {B.~D.}\ \bibnamefont
  {{Metzger}}}, \bibinfo {author} {\bibfnamefont {P.~S.}\ \bibnamefont
  {{Cowperthwaite}}}, \bibinfo {author} {\bibfnamefont {M.}~\bibnamefont
  {{Nicholl}}}, \bibinfo {author} {\bibfnamefont {K.~D.}\ \bibnamefont
  {{Alexander}}}, \bibinfo {author} {\bibfnamefont {P.~K.}\ \bibnamefont
  {{Blanchard}}}, \bibinfo {author} {\bibfnamefont {R.}~\bibnamefont
  {{Chornock}}}, \bibinfo {author} {\bibfnamefont {T.}~\bibnamefont
  {{Eftekhari}}}, \bibinfo {author} {\bibfnamefont {W.}~\bibnamefont {{Fong}}},
  \bibinfo {author} {\bibfnamefont {R.}~\bibnamefont {{Margutti}}}, \ and\
  \bibinfo {author} {\bibfnamefont {P.~K.~G.}\ \bibnamefont {{Williams}}},\
  }\href {\doibase 10.3847/2041-8213/aa9c84} {\bibfield  {journal} {\bibinfo
  {journal} {\apj}\ }\textbf {\bibinfo {volume} {851}},\ \bibinfo {eid} {L21}
  (\bibinfo {year} {2017})},\ \Eprint {http://arxiv.org/abs/1710.11576}
  {arXiv:1710.11576 [astro-ph.HE]} \BibitemShut {NoStop}%
\bibitem [{\citenamefont {{Metzger}}(2017)}]{2017arXiv171005931M}%
  \BibitemOpen
  \bibfield  {author} {\bibinfo {author} {\bibfnamefont {B.~D.}\ \bibnamefont
  {{Metzger}}},\ }\href@noop {} {\bibfield  {journal} {\bibinfo  {journal}
  {arXiv e-prints}\ } (\bibinfo {year} {2017})},\ \Eprint
  {http://arxiv.org/abs/1710.05931} {arXiv:1710.05931 [astro-ph.HE]}
  \BibitemShut {NoStop}%
\bibitem [{\citenamefont {Tanaka}\ \emph {et~al.}(2018)\citenamefont {Tanaka},
  \citenamefont {Kato}, \citenamefont {Gaigalas}, \citenamefont {Rynkun},
  \citenamefont {Rad{\v{z}}i{\={u}}t{\.{e}}}, \citenamefont {Wanajo},
  \citenamefont {Sekiguchi}, \citenamefont {Nakamura}, \citenamefont {Tanuma},
  \citenamefont {Murakami},\ and\ \citenamefont {Sakaue}}]{Tanaka_2018}%
  \BibitemOpen
  \bibfield  {author} {\bibinfo {author} {\bibfnamefont {M.}~\bibnamefont
  {Tanaka}}, \bibinfo {author} {\bibfnamefont {D.}~\bibnamefont {Kato}},
  \bibinfo {author} {\bibfnamefont {G.}~\bibnamefont {Gaigalas}}, \bibinfo
  {author} {\bibfnamefont {P.}~\bibnamefont {Rynkun}}, \bibinfo {author}
  {\bibfnamefont {L.}~\bibnamefont {Rad{\v{z}}i{\={u}}t{\.{e}}}}, \bibinfo
  {author} {\bibfnamefont {S.}~\bibnamefont {Wanajo}}, \bibinfo {author}
  {\bibfnamefont {Y.}~\bibnamefont {Sekiguchi}}, \bibinfo {author}
  {\bibfnamefont {N.}~\bibnamefont {Nakamura}}, \bibinfo {author}
  {\bibfnamefont {H.}~\bibnamefont {Tanuma}}, \bibinfo {author} {\bibfnamefont
  {I.}~\bibnamefont {Murakami}}, \ and\ \bibinfo {author} {\bibfnamefont
  {H.~A.}\ \bibnamefont {Sakaue}},\ }\href {\doibase 10.3847/1538-4357/aaa0cb}
  {\bibfield  {journal} {\bibinfo  {journal} {The Astrophysical Journal}\
  }\textbf {\bibinfo {volume} {852}},\ \bibinfo {pages} {109} (\bibinfo {year}
  {2018})}\BibitemShut {NoStop}%
\bibitem [{\citenamefont {Cowperthwaite}\ \emph {et~al.}(2017)\citenamefont
  {Cowperthwaite} \emph {et~al.}}]{2041-8205-848-2-L17}%
  \BibitemOpen
  \bibfield  {author} {\bibinfo {author} {\bibfnamefont {P.~S.}\ \bibnamefont
  {Cowperthwaite}} \emph {et~al.},\ }\href
  {http://stacks.iop.org/2041-8205/848/i=2/a=L17} {\bibfield  {journal}
  {\bibinfo  {journal} {The Astrophysical Journal Letters}\ }\textbf {\bibinfo
  {volume} {848}},\ \bibinfo {pages} {L17} (\bibinfo {year}
  {2017})}\BibitemShut {NoStop}%
\bibitem [{\citenamefont {{Kessler}}\ \emph {et~al.}(2015)\citenamefont
  {{Kessler}}, \citenamefont {{Marriner}}, \citenamefont {{Childress}},
  \citenamefont {{Covarrubias}}, \citenamefont {{D'Andrea}}, \citenamefont
  {{Finley}}, \citenamefont {{Fischer}}, \citenamefont {{Foley}}, \citenamefont
  {{Goldstein}}, \citenamefont {{Gupta}}, \citenamefont {{Kuehn}},
  \citenamefont {{Marcha}}, \citenamefont {{Nichol}}, \citenamefont
  {{Papadopoulos}}, \citenamefont {{Sako}}, \citenamefont {{Scolnic}},
  \citenamefont {{Smith}}, \citenamefont {{Sullivan}}, \citenamefont
  {{Wester}}, \citenamefont {{Yuan}}, \citenamefont {{Abbott}}, \citenamefont
  {{Abdalla}}, \citenamefont {{Allam}}, \citenamefont {{Benoit-L{\'e}vy}},
  \citenamefont {{Bernstein}}, \citenamefont {{Bertin}}, \citenamefont
  {{Brooks}}, \citenamefont {{Carnero Rosell}}, \citenamefont {{Carrasco
  Kind}}, \citenamefont {{Castander}}, \citenamefont {{Crocce}}, \citenamefont
  {{da Costa}}, \citenamefont {{Desai}}, \citenamefont {{Diehl}}, \citenamefont
  {{Eifler}}, \citenamefont {{Fausti Neto}}, \citenamefont {{Flaugher}},
  \citenamefont {{Frieman}}, \citenamefont {{Gerdes}}, \citenamefont {{Gruen}},
  \citenamefont {{Gruendl}}, \citenamefont {{Honscheid}}, \citenamefont
  {{James}}, \citenamefont {{Kuropatkin}}, \citenamefont {{Li}}, \citenamefont
  {{Maia}}, \citenamefont {{Marshall}}, \citenamefont {{Martini}},
  \citenamefont {{Miller}}, \citenamefont {{Miquel}}, \citenamefont {{Nord}},
  \citenamefont {{Ogando}}, \citenamefont {{Plazas}}, \citenamefont {{Reil}},
  \citenamefont {{Romer}}, \citenamefont {{Roodman}}, \citenamefont
  {{Sanchez}}, \citenamefont {{Sevilla- Noarbe}}, \citenamefont {{Smith}},
  \citenamefont {{Soares-Santos}}, \citenamefont {{Sobreira}}, \citenamefont
  {{Tarle}}, \citenamefont {{Thaler}}, \citenamefont {{Thomas}}, \citenamefont
  {{Tucker}}, \citenamefont {{Walker}},\ and\ \citenamefont {{DES
  Collaboration}}}]{2015AJ....150..172K}%
  \BibitemOpen
  \bibfield  {author} {\bibinfo {author} {\bibfnamefont {R.}~\bibnamefont
  {{Kessler}}}, \bibinfo {author} {\bibfnamefont {J.}~\bibnamefont
  {{Marriner}}}, \bibinfo {author} {\bibfnamefont {M.}~\bibnamefont
  {{Childress}}}, \bibinfo {author} {\bibfnamefont {R.}~\bibnamefont
  {{Covarrubias}}}, \bibinfo {author} {\bibfnamefont {C.~B.}\ \bibnamefont
  {{D'Andrea}}}, \bibinfo {author} {\bibfnamefont {D.~A.}\ \bibnamefont
  {{Finley}}}, \bibinfo {author} {\bibfnamefont {J.}~\bibnamefont {{Fischer}}},
  \bibinfo {author} {\bibfnamefont {R.~J.}\ \bibnamefont {{Foley}}}, \bibinfo
  {author} {\bibfnamefont {D.}~\bibnamefont {{Goldstein}}}, \bibinfo {author}
  {\bibfnamefont {R.~R.}\ \bibnamefont {{Gupta}}}, \bibinfo {author}
  {\bibfnamefont {K.}~\bibnamefont {{Kuehn}}}, \bibinfo {author} {\bibfnamefont
  {M.}~\bibnamefont {{Marcha}}}, \bibinfo {author} {\bibfnamefont {R.~C.}\
  \bibnamefont {{Nichol}}}, \bibinfo {author} {\bibfnamefont {A.}~\bibnamefont
  {{Papadopoulos}}}, \bibinfo {author} {\bibfnamefont {M.}~\bibnamefont
  {{Sako}}}, \bibinfo {author} {\bibfnamefont {D.}~\bibnamefont {{Scolnic}}},
  \bibinfo {author} {\bibfnamefont {M.}~\bibnamefont {{Smith}}}, \bibinfo
  {author} {\bibfnamefont {M.}~\bibnamefont {{Sullivan}}}, \bibinfo {author}
  {\bibfnamefont {W.}~\bibnamefont {{Wester}}}, \bibinfo {author}
  {\bibfnamefont {F.}~\bibnamefont {{Yuan}}}, \bibinfo {author} {\bibfnamefont
  {T.}~\bibnamefont {{Abbott}}}, \bibinfo {author} {\bibfnamefont {F.~B.}\
  \bibnamefont {{Abdalla}}}, \bibinfo {author} {\bibfnamefont {S.}~\bibnamefont
  {{Allam}}}, \bibinfo {author} {\bibfnamefont {A.}~\bibnamefont
  {{Benoit-L{\'e}vy}}}, \bibinfo {author} {\bibfnamefont {G.~M.}\ \bibnamefont
  {{Bernstein}}}, \bibinfo {author} {\bibfnamefont {E.}~\bibnamefont
  {{Bertin}}}, \bibinfo {author} {\bibfnamefont {D.}~\bibnamefont {{Brooks}}},
  \bibinfo {author} {\bibfnamefont {A.}~\bibnamefont {{Carnero Rosell}}},
  \bibinfo {author} {\bibfnamefont {M.}~\bibnamefont {{Carrasco Kind}}},
  \bibinfo {author} {\bibfnamefont {F.~J.}\ \bibnamefont {{Castander}}},
  \bibinfo {author} {\bibfnamefont {M.}~\bibnamefont {{Crocce}}}, \bibinfo
  {author} {\bibfnamefont {L.~N.}\ \bibnamefont {{da Costa}}}, \bibinfo
  {author} {\bibfnamefont {S.}~\bibnamefont {{Desai}}}, \bibinfo {author}
  {\bibfnamefont {H.~T.}\ \bibnamefont {{Diehl}}}, \bibinfo {author}
  {\bibfnamefont {T.~F.}\ \bibnamefont {{Eifler}}}, \bibinfo {author}
  {\bibfnamefont {A.}~\bibnamefont {{Fausti Neto}}}, \bibinfo {author}
  {\bibfnamefont {B.}~\bibnamefont {{Flaugher}}}, \bibinfo {author}
  {\bibfnamefont {J.}~\bibnamefont {{Frieman}}}, \bibinfo {author}
  {\bibfnamefont {D.~W.}\ \bibnamefont {{Gerdes}}}, \bibinfo {author}
  {\bibfnamefont {D.}~\bibnamefont {{Gruen}}}, \bibinfo {author} {\bibfnamefont
  {R.~A.}\ \bibnamefont {{Gruendl}}}, \bibinfo {author} {\bibfnamefont
  {K.}~\bibnamefont {{Honscheid}}}, \bibinfo {author} {\bibfnamefont {D.~J.}\
  \bibnamefont {{James}}}, \bibinfo {author} {\bibfnamefont {N.}~\bibnamefont
  {{Kuropatkin}}}, \bibinfo {author} {\bibfnamefont {T.~S.}\ \bibnamefont
  {{Li}}}, \bibinfo {author} {\bibfnamefont {M.~A.~G.}\ \bibnamefont {{Maia}}},
  \bibinfo {author} {\bibfnamefont {J.~L.}\ \bibnamefont {{Marshall}}},
  \bibinfo {author} {\bibfnamefont {P.}~\bibnamefont {{Martini}}}, \bibinfo
  {author} {\bibfnamefont {C.~J.}\ \bibnamefont {{Miller}}}, \bibinfo {author}
  {\bibfnamefont {R.}~\bibnamefont {{Miquel}}}, \bibinfo {author}
  {\bibfnamefont {B.}~\bibnamefont {{Nord}}}, \bibinfo {author} {\bibfnamefont
  {R.}~\bibnamefont {{Ogando}}}, \bibinfo {author} {\bibfnamefont {A.~A.}\
  \bibnamefont {{Plazas}}}, \bibinfo {author} {\bibfnamefont {K.}~\bibnamefont
  {{Reil}}}, \bibinfo {author} {\bibfnamefont {A.~K.}\ \bibnamefont {{Romer}}},
  \bibinfo {author} {\bibfnamefont {A.}~\bibnamefont {{Roodman}}}, \bibinfo
  {author} {\bibfnamefont {E.}~\bibnamefont {{Sanchez}}}, \bibinfo {author}
  {\bibfnamefont {I.}~\bibnamefont {{Sevilla- Noarbe}}}, \bibinfo {author}
  {\bibfnamefont {R.~C.}\ \bibnamefont {{Smith}}}, \bibinfo {author}
  {\bibfnamefont {M.}~\bibnamefont {{Soares-Santos}}}, \bibinfo {author}
  {\bibfnamefont {F.}~\bibnamefont {{Sobreira}}}, \bibinfo {author}
  {\bibfnamefont {G.}~\bibnamefont {{Tarle}}}, \bibinfo {author} {\bibfnamefont
  {J.}~\bibnamefont {{Thaler}}}, \bibinfo {author} {\bibfnamefont {R.~C.}\
  \bibnamefont {{Thomas}}}, \bibinfo {author} {\bibfnamefont {D.}~\bibnamefont
  {{Tucker}}}, \bibinfo {author} {\bibfnamefont {A.~R.}\ \bibnamefont
  {{Walker}}}, \ and\ \bibinfo {author} {\bibnamefont {{DES Collaboration}}},\
  }\href {\doibase 10.1088/0004-6256/150/6/172} {\bibfield  {journal} {\bibinfo
   {journal} {\aj}\ }\textbf {\bibinfo {volume} {150}},\ \bibinfo {eid} {172}
  (\bibinfo {year} {2015})},\ \Eprint {http://arxiv.org/abs/1507.05137}
  {arXiv:1507.05137 [astro-ph.IM]} \BibitemShut {NoStop}%
\bibitem [{\citenamefont {{Goldstein}}\ \emph
  {et~al.}(2015{\natexlab{a}})\citenamefont {{Goldstein}}, \citenamefont
  {{D'Andrea}}, \citenamefont {{Fischer}}, \citenamefont {{Foley}},
  \citenamefont {{Gupta}}, \citenamefont {{Kessler}}, \citenamefont {{Kim}},
  \citenamefont {{Nichol}}, \citenamefont {{Nugent}}, \citenamefont
  {{Papadopoulos}}, \citenamefont {{Sako}}, \citenamefont {{Smith}},
  \citenamefont {{Sullivan}}, \citenamefont {{Thomas}}, \citenamefont
  {{Wester}}, \citenamefont {{Wolf}}, \citenamefont {{Abdalla}}, \citenamefont
  {{Banerji}}, \citenamefont {{Benoit-L{\'e}vy}}, \citenamefont {{Bertin}},
  \citenamefont {{Brooks}}, \citenamefont {{Carnero Rosell}}, \citenamefont
  {{Castander}}, \citenamefont {{da Costa}}, \citenamefont {{Covarrubias}},
  \citenamefont {{DePoy}}, \citenamefont {{Desai}}, \citenamefont {{Diehl}},
  \citenamefont {{Doel}}, \citenamefont {{Eifler}}, \citenamefont {{Fausti
  Neto}}, \citenamefont {{Finley}}, \citenamefont {{Flaugher}}, \citenamefont
  {{Fosalba}}, \citenamefont {{Frieman}}, \citenamefont {{Gerdes}},
  \citenamefont {{Gruen}}, \citenamefont {{Gruendl}}, \citenamefont {{James}},
  \citenamefont {{Kuehn}}, \citenamefont {{Kuropatkin}}, \citenamefont
  {{Lahav}}, \citenamefont {{Li}}, \citenamefont {{Maia}}, \citenamefont
  {{Makler}}, \citenamefont {{March}}, \citenamefont {{Marshall}},
  \citenamefont {{Martini}}, \citenamefont {{Merritt}}, \citenamefont
  {{Miquel}}, \citenamefont {{Nord}}, \citenamefont {{Ogando}}, \citenamefont
  {{Plazas}}, \citenamefont {{Romer}}, \citenamefont {{Roodman}}, \citenamefont
  {{Sanchez}}, \citenamefont {{Scarpine}}, \citenamefont {{Schubnell}},
  \citenamefont {{Sevilla-Noarbe}}, \citenamefont {{Smith}}, \citenamefont
  {{Soares-Santos}}, \citenamefont {{Sobreira}}, \citenamefont {{Suchyta}},
  \citenamefont {{Swanson}}, \citenamefont {{Tarle}}, \citenamefont
  {{Thaler}},\ and\ \citenamefont {{Walker}}}]{2015AJ....150...82G}%
  \BibitemOpen
  \bibfield  {author} {\bibinfo {author} {\bibfnamefont {D.~A.}\ \bibnamefont
  {{Goldstein}}}, \bibinfo {author} {\bibfnamefont {C.~B.}\ \bibnamefont
  {{D'Andrea}}}, \bibinfo {author} {\bibfnamefont {J.~A.}\ \bibnamefont
  {{Fischer}}}, \bibinfo {author} {\bibfnamefont {R.~J.}\ \bibnamefont
  {{Foley}}}, \bibinfo {author} {\bibfnamefont {R.~R.}\ \bibnamefont
  {{Gupta}}}, \bibinfo {author} {\bibfnamefont {R.}~\bibnamefont {{Kessler}}},
  \bibinfo {author} {\bibfnamefont {A.~G.}\ \bibnamefont {{Kim}}}, \bibinfo
  {author} {\bibfnamefont {R.~C.}\ \bibnamefont {{Nichol}}}, \bibinfo {author}
  {\bibfnamefont {P.~E.}\ \bibnamefont {{Nugent}}}, \bibinfo {author}
  {\bibfnamefont {A.}~\bibnamefont {{Papadopoulos}}}, \bibinfo {author}
  {\bibfnamefont {M.}~\bibnamefont {{Sako}}}, \bibinfo {author} {\bibfnamefont
  {M.}~\bibnamefont {{Smith}}}, \bibinfo {author} {\bibfnamefont
  {M.}~\bibnamefont {{Sullivan}}}, \bibinfo {author} {\bibfnamefont {R.~C.}\
  \bibnamefont {{Thomas}}}, \bibinfo {author} {\bibfnamefont {W.}~\bibnamefont
  {{Wester}}}, \bibinfo {author} {\bibfnamefont {R.~C.}\ \bibnamefont
  {{Wolf}}}, \bibinfo {author} {\bibfnamefont {F.~B.}\ \bibnamefont
  {{Abdalla}}}, \bibinfo {author} {\bibfnamefont {M.}~\bibnamefont
  {{Banerji}}}, \bibinfo {author} {\bibfnamefont {A.}~\bibnamefont
  {{Benoit-L{\'e}vy}}}, \bibinfo {author} {\bibfnamefont {E.}~\bibnamefont
  {{Bertin}}}, \bibinfo {author} {\bibfnamefont {D.}~\bibnamefont {{Brooks}}},
  \bibinfo {author} {\bibfnamefont {A.}~\bibnamefont {{Carnero Rosell}}},
  \bibinfo {author} {\bibfnamefont {F.~J.}\ \bibnamefont {{Castander}}},
  \bibinfo {author} {\bibfnamefont {L.~N.}\ \bibnamefont {{da Costa}}},
  \bibinfo {author} {\bibfnamefont {R.}~\bibnamefont {{Covarrubias}}}, \bibinfo
  {author} {\bibfnamefont {D.~L.}\ \bibnamefont {{DePoy}}}, \bibinfo {author}
  {\bibfnamefont {S.}~\bibnamefont {{Desai}}}, \bibinfo {author} {\bibfnamefont
  {H.~T.}\ \bibnamefont {{Diehl}}}, \bibinfo {author} {\bibfnamefont
  {P.}~\bibnamefont {{Doel}}}, \bibinfo {author} {\bibfnamefont {T.~F.}\
  \bibnamefont {{Eifler}}}, \bibinfo {author} {\bibfnamefont {A.}~\bibnamefont
  {{Fausti Neto}}}, \bibinfo {author} {\bibfnamefont {D.~A.}\ \bibnamefont
  {{Finley}}}, \bibinfo {author} {\bibfnamefont {B.}~\bibnamefont
  {{Flaugher}}}, \bibinfo {author} {\bibfnamefont {P.}~\bibnamefont
  {{Fosalba}}}, \bibinfo {author} {\bibfnamefont {J.}~\bibnamefont
  {{Frieman}}}, \bibinfo {author} {\bibfnamefont {D.}~\bibnamefont {{Gerdes}}},
  \bibinfo {author} {\bibfnamefont {D.}~\bibnamefont {{Gruen}}}, \bibinfo
  {author} {\bibfnamefont {R.~A.}\ \bibnamefont {{Gruendl}}}, \bibinfo {author}
  {\bibfnamefont {D.}~\bibnamefont {{James}}}, \bibinfo {author} {\bibfnamefont
  {K.}~\bibnamefont {{Kuehn}}}, \bibinfo {author} {\bibfnamefont
  {N.}~\bibnamefont {{Kuropatkin}}}, \bibinfo {author} {\bibfnamefont
  {O.}~\bibnamefont {{Lahav}}}, \bibinfo {author} {\bibfnamefont {T.~S.}\
  \bibnamefont {{Li}}}, \bibinfo {author} {\bibfnamefont {M.~A.~G.}\
  \bibnamefont {{Maia}}}, \bibinfo {author} {\bibfnamefont {M.}~\bibnamefont
  {{Makler}}}, \bibinfo {author} {\bibfnamefont {M.}~\bibnamefont {{March}}},
  \bibinfo {author} {\bibfnamefont {J.~L.}\ \bibnamefont {{Marshall}}},
  \bibinfo {author} {\bibfnamefont {P.}~\bibnamefont {{Martini}}}, \bibinfo
  {author} {\bibfnamefont {K.~W.}\ \bibnamefont {{Merritt}}}, \bibinfo {author}
  {\bibfnamefont {R.}~\bibnamefont {{Miquel}}}, \bibinfo {author}
  {\bibfnamefont {B.}~\bibnamefont {{Nord}}}, \bibinfo {author} {\bibfnamefont
  {R.}~\bibnamefont {{Ogando}}}, \bibinfo {author} {\bibfnamefont {A.~A.}\
  \bibnamefont {{Plazas}}}, \bibinfo {author} {\bibfnamefont {A.~K.}\
  \bibnamefont {{Romer}}}, \bibinfo {author} {\bibfnamefont {A.}~\bibnamefont
  {{Roodman}}}, \bibinfo {author} {\bibfnamefont {E.}~\bibnamefont
  {{Sanchez}}}, \bibinfo {author} {\bibfnamefont {V.}~\bibnamefont
  {{Scarpine}}}, \bibinfo {author} {\bibfnamefont {M.}~\bibnamefont
  {{Schubnell}}}, \bibinfo {author} {\bibfnamefont {I.}~\bibnamefont
  {{Sevilla-Noarbe}}}, \bibinfo {author} {\bibfnamefont {R.~C.}\ \bibnamefont
  {{Smith}}}, \bibinfo {author} {\bibfnamefont {M.}~\bibnamefont
  {{Soares-Santos}}}, \bibinfo {author} {\bibfnamefont {F.}~\bibnamefont
  {{Sobreira}}}, \bibinfo {author} {\bibfnamefont {E.}~\bibnamefont
  {{Suchyta}}}, \bibinfo {author} {\bibfnamefont {M.~E.~C.}\ \bibnamefont
  {{Swanson}}}, \bibinfo {author} {\bibfnamefont {G.}~\bibnamefont {{Tarle}}},
  \bibinfo {author} {\bibfnamefont {J.}~\bibnamefont {{Thaler}}}, \ and\
  \bibinfo {author} {\bibfnamefont {A.~R.}\ \bibnamefont {{Walker}}},\ }\href
  {\doibase 10.1088/0004-6256/150/3/82} {\bibfield  {journal} {\bibinfo
  {journal} {\aj}\ }\textbf {\bibinfo {volume} {150}},\ \bibinfo {eid} {82}
  (\bibinfo {year} {2015}{\natexlab{a}})},\ \Eprint
  {http://arxiv.org/abs/1504.02936} {arXiv:1504.02936 [astro-ph.IM]}
  \BibitemShut {NoStop}%
\bibitem [{\citenamefont {{Hughes}}\ and\ \citenamefont
  {{Holz}}(2003)}]{2003CQGra..20S..65H}%
  \BibitemOpen
  \bibfield  {author} {\bibinfo {author} {\bibfnamefont {S.~A.}\ \bibnamefont
  {{Hughes}}}\ and\ \bibinfo {author} {\bibfnamefont {D.~E.}\ \bibnamefont
  {{Holz}}},\ }\href {\doibase 10.1088/0264-9381/20/10/308} {\bibfield
  {journal} {\bibinfo  {journal} {Classical and Quantum Gravity}\ }\textbf
  {\bibinfo {volume} {20}},\ \bibinfo {pages} {S65} (\bibinfo {year} {2003})},\
  \Eprint {http://arxiv.org/abs/astro-ph/0212218} {arXiv:astro-ph/0212218
  [astro-ph]} \BibitemShut {NoStop}%
\bibitem [{\citenamefont {{Feeney}}\ \emph {et~al.}(2018)\citenamefont
  {{Feeney}}, \citenamefont {{Peiris}}, \citenamefont {{Williamson}},
  \citenamefont {{Nissanke}}, \citenamefont {{Mortlock}}, \citenamefont
  {{Alsing}},\ and\ \citenamefont {{Scolnic}}}]{2018arXiv180203404F}%
  \BibitemOpen
  \bibfield  {author} {\bibinfo {author} {\bibfnamefont {S.~M.}\ \bibnamefont
  {{Feeney}}}, \bibinfo {author} {\bibfnamefont {H.~V.}\ \bibnamefont
  {{Peiris}}}, \bibinfo {author} {\bibfnamefont {A.~R.}\ \bibnamefont
  {{Williamson}}}, \bibinfo {author} {\bibfnamefont {S.~M.}\ \bibnamefont
  {{Nissanke}}}, \bibinfo {author} {\bibfnamefont {D.~J.}\ \bibnamefont
  {{Mortlock}}}, \bibinfo {author} {\bibfnamefont {J.}~\bibnamefont
  {{Alsing}}}, \ and\ \bibinfo {author} {\bibfnamefont {D.}~\bibnamefont
  {{Scolnic}}},\ }\href@noop {} {\bibfield  {journal} {\bibinfo  {journal}
  {arXiv e-prints}\ } (\bibinfo {year} {2018})},\ \Eprint
  {http://arxiv.org/abs/1802.03404} {arXiv:1802.03404} \BibitemShut {NoStop}%
\bibitem [{\citenamefont {{Chen}}\ \emph {et~al.}(2018)\citenamefont {{Chen}},
  \citenamefont {{Fishbach}},\ and\ \citenamefont
  {{Holz}}}]{2018Natur.562..545C}%
  \BibitemOpen
  \bibfield  {author} {\bibinfo {author} {\bibfnamefont {H.-Y.}\ \bibnamefont
  {{Chen}}}, \bibinfo {author} {\bibfnamefont {M.}~\bibnamefont {{Fishbach}}},
  \ and\ \bibinfo {author} {\bibfnamefont {D.~E.}\ \bibnamefont {{Holz}}},\
  }\href {\doibase 10.1038/s41586-018-0606-0} {\bibfield  {journal} {\bibinfo
  {journal} {\nat}\ }\textbf {\bibinfo {volume} {562}},\ \bibinfo {pages} {545}
  (\bibinfo {year} {2018})},\ \Eprint {http://arxiv.org/abs/1712.06531}
  {arXiv:1712.06531 [astro-ph.CO]} \BibitemShut {NoStop}%
\bibitem [{\citenamefont {{Mortlock}}\ \emph {et~al.}(2018)\citenamefont
  {{Mortlock}}, \citenamefont {{Feeney}}, \citenamefont {{Peiris}},
  \citenamefont {{Williamson}},\ and\ \citenamefont
  {{Nissanke}}}]{2018arXiv181111723M}%
  \BibitemOpen
  \bibfield  {author} {\bibinfo {author} {\bibfnamefont {D.~J.}\ \bibnamefont
  {{Mortlock}}}, \bibinfo {author} {\bibfnamefont {S.~M.}\ \bibnamefont
  {{Feeney}}}, \bibinfo {author} {\bibfnamefont {H.~V.}\ \bibnamefont
  {{Peiris}}}, \bibinfo {author} {\bibfnamefont {A.~R.}\ \bibnamefont
  {{Williamson}}}, \ and\ \bibinfo {author} {\bibfnamefont {S.~M.}\
  \bibnamefont {{Nissanke}}},\ }\href@noop {} {\bibfield  {journal} {\bibinfo
  {journal} {arXiv e-prints}\ ,\ \bibinfo {eid} {arXiv:1811.11723}} (\bibinfo
  {year} {2018})},\ \Eprint {http://arxiv.org/abs/1811.11723} {arXiv:1811.11723
  [astro-ph.CO]} \BibitemShut {NoStop}%
\bibitem [{\citenamefont {{Khan}}\ \emph {et~al.}(2018)\citenamefont {{Khan}},
  \citenamefont {{Huerta}}, \citenamefont {{Wang}},\ and\ \citenamefont
  {{Gruendl}}}]{asad:2018K}%
  \BibitemOpen
  \bibfield  {author} {\bibinfo {author} {\bibfnamefont {A.}~\bibnamefont
  {{Khan}}}, \bibinfo {author} {\bibfnamefont {E.~A.}\ \bibnamefont
  {{Huerta}}}, \bibinfo {author} {\bibfnamefont {S.}~\bibnamefont {{Wang}}}, \
  and\ \bibinfo {author} {\bibfnamefont {R.}~\bibnamefont {{Gruendl}}},\
  }\href@noop {} {\bibfield  {journal} {\bibinfo  {journal} {arXiv e-prints}\
  ,\ \bibinfo {eid} {arXiv:1812.02183}} (\bibinfo {year} {2018})},\ \Eprint
  {http://arxiv.org/abs/1812.02183} {arXiv:1812.02183 [astro-ph.IM]}
  \BibitemShut {NoStop}%
\bibitem [{\citenamefont {{Dom{\'\i}nguez S{\'a}nchez}}\ \emph
  {et~al.}(2018{\natexlab{a}})\citenamefont {{Dom{\'\i}nguez S{\'a}nchez}},
  \citenamefont {{Huertas-Company}}, \citenamefont {{Bernardi}}, \citenamefont
  {{Tuccillo}},\ and\ \citenamefont {{Fischer}}}]{Dom:2018MNRAS}%
  \BibitemOpen
  \bibfield  {author} {\bibinfo {author} {\bibfnamefont {H.}~\bibnamefont
  {{Dom{\'\i}nguez S{\'a}nchez}}}, \bibinfo {author} {\bibfnamefont
  {M.}~\bibnamefont {{Huertas-Company}}}, \bibinfo {author} {\bibfnamefont
  {M.}~\bibnamefont {{Bernardi}}}, \bibinfo {author} {\bibfnamefont
  {D.}~\bibnamefont {{Tuccillo}}}, \ and\ \bibinfo {author} {\bibfnamefont
  {J.~L.}\ \bibnamefont {{Fischer}}},\ }\href {\doibase 10.1093/mnras/sty338}
  {\bibfield  {journal} {\bibinfo  {journal} {\mnras}\ }\textbf {\bibinfo
  {volume} {476}},\ \bibinfo {pages} {3661} (\bibinfo {year}
  {2018}{\natexlab{a}})}\BibitemShut {NoStop}%
\bibitem [{\citenamefont {{Dom{\'\i}nguez S{\'a}nchez}}\ \emph
  {et~al.}(2018{\natexlab{b}})\citenamefont {{Dom{\'\i}nguez S{\'a}nchez}},
  \citenamefont {{Huertas-Company}} \emph {et~al.}}]{Dom:2018D}%
  \BibitemOpen
  \bibfield  {author} {\bibinfo {author} {\bibfnamefont {H.}~\bibnamefont
  {{Dom{\'\i}nguez S{\'a}nchez}}}, \bibinfo {author} {\bibnamefont
  {{Huertas-Company}}},  \emph {et~al.},\ }\href@noop {} {\bibfield  {journal}
  {\bibinfo  {journal} {ArXiv e-prints}\ ,\ \bibinfo {eid} {arXiv:1807.00807}}
  (\bibinfo {year} {2018}{\natexlab{b}})},\ \Eprint
  {http://arxiv.org/abs/1807.00807} {arXiv:1807.00807 [astro-ph.GA]}
  \BibitemShut {NoStop}%
\bibitem [{\citenamefont {Levan}\ \emph {et~al.}(2017)\citenamefont {Levan}
  \emph {et~al.}}]{2041-8205-848-2-L28}%
  \BibitemOpen
  \bibfield  {author} {\bibinfo {author} {\bibfnamefont {A.~J.}\ \bibnamefont
  {Levan}} \emph {et~al.},\ }\href
  {http://stacks.iop.org/2041-8205/848/i=2/a=L28} {\bibfield  {journal}
  {\bibinfo  {journal} {The Astrophysical Journal Letters}\ }\textbf {\bibinfo
  {volume} {848}},\ \bibinfo {pages} {L28} (\bibinfo {year}
  {2017})}\BibitemShut {NoStop}%
\bibitem [{\citenamefont {{Mooley}}\ \emph {et~al.}(2018)\citenamefont
  {{Mooley}}, \citenamefont {{Nakar}}, \citenamefont {{Hotokezaka}},
  \citenamefont {{Hallinan}}, \citenamefont {{Corsi}}, \citenamefont {{Frail}},
  \citenamefont {{Horesh}}, \citenamefont {{Murphy}}, \citenamefont {{Lenc}},
  \citenamefont {{Kaplan}}, \citenamefont {{de}}, \citenamefont {{Dobie}},
  \citenamefont {{Chandra}}, \citenamefont {{Deller}}, \citenamefont
  {{Gottlieb}}, \citenamefont {{Kasliwal}}, \citenamefont {{Kulkarni}},
  \citenamefont {{Myers}}, \citenamefont {{Nissanke}}, \citenamefont {{Piran}},
  \citenamefont {{Lynch}}, \citenamefont {{Bhalerao}}, \citenamefont
  {{Bourke}}, \citenamefont {{Bannister}},\ and\ \citenamefont
  {{Singer}}}]{Molley:2018Natur}%
  \BibitemOpen
  \bibfield  {author} {\bibinfo {author} {\bibfnamefont {K.~P.}\ \bibnamefont
  {{Mooley}}}, \bibinfo {author} {\bibfnamefont {E.}~\bibnamefont {{Nakar}}},
  \bibinfo {author} {\bibfnamefont {K.}~\bibnamefont {{Hotokezaka}}}, \bibinfo
  {author} {\bibfnamefont {G.}~\bibnamefont {{Hallinan}}}, \bibinfo {author}
  {\bibfnamefont {A.}~\bibnamefont {{Corsi}}}, \bibinfo {author} {\bibfnamefont
  {D.~A.}\ \bibnamefont {{Frail}}}, \bibinfo {author} {\bibfnamefont
  {A.}~\bibnamefont {{Horesh}}}, \bibinfo {author} {\bibfnamefont
  {T.}~\bibnamefont {{Murphy}}}, \bibinfo {author} {\bibfnamefont
  {E.}~\bibnamefont {{Lenc}}}, \bibinfo {author} {\bibfnamefont {D.~L.}\
  \bibnamefont {{Kaplan}}}, \bibinfo {author} {\bibfnamefont {K.}~\bibnamefont
  {{de}}}, \bibinfo {author} {\bibfnamefont {D.}~\bibnamefont {{Dobie}}},
  \bibinfo {author} {\bibfnamefont {P.}~\bibnamefont {{Chandra}}}, \bibinfo
  {author} {\bibfnamefont {A.}~\bibnamefont {{Deller}}}, \bibinfo {author}
  {\bibfnamefont {O.}~\bibnamefont {{Gottlieb}}}, \bibinfo {author}
  {\bibfnamefont {M.~M.}\ \bibnamefont {{Kasliwal}}}, \bibinfo {author}
  {\bibfnamefont {S.~R.}\ \bibnamefont {{Kulkarni}}}, \bibinfo {author}
  {\bibfnamefont {S.~T.}\ \bibnamefont {{Myers}}}, \bibinfo {author}
  {\bibfnamefont {S.}~\bibnamefont {{Nissanke}}}, \bibinfo {author}
  {\bibfnamefont {T.}~\bibnamefont {{Piran}}}, \bibinfo {author} {\bibfnamefont
  {C.}~\bibnamefont {{Lynch}}}, \bibinfo {author} {\bibfnamefont
  {V.}~\bibnamefont {{Bhalerao}}}, \bibinfo {author} {\bibfnamefont
  {S.}~\bibnamefont {{Bourke}}}, \bibinfo {author} {\bibfnamefont {K.~W.}\
  \bibnamefont {{Bannister}}}, \ and\ \bibinfo {author} {\bibfnamefont {L.~P.}\
  \bibnamefont {{Singer}}},\ }\href {\doibase 10.1038/nature25452} {\bibfield
  {journal} {\bibinfo  {journal} {\nat}\ }\textbf {\bibinfo {volume} {554}},\
  \bibinfo {pages} {207} (\bibinfo {year} {2018})},\ \Eprint
  {http://arxiv.org/abs/1711.11573} {arXiv:1711.11573 [astro-ph.HE]}
  \BibitemShut {NoStop}%
\bibitem [{\citenamefont {{Drout}}\ \emph {et~al.}(2017)\citenamefont
  {{Drout}}, \citenamefont {{Piro}}, \citenamefont {{Shappee}}, \citenamefont
  {{Kilpatrick}}, \citenamefont {{Simon}}, \citenamefont {{Contreras}},
  \citenamefont {{Coulter}}, \citenamefont {{Foley}}, \citenamefont
  {{Siebert}}, \citenamefont {{Morrell}}, \citenamefont {{Boutsia}},
  \citenamefont {{Di Mille}}, \citenamefont {{Holoien}}, \citenamefont
  {{Kasen}}, \citenamefont {{Kollmeier}}, \citenamefont {{Madore}},
  \citenamefont {{Monson}}, \citenamefont {{Murguia-Berthier}}, \citenamefont
  {{Pan}}, \citenamefont {{Prochaska}}, \citenamefont {{Ramirez-Ruiz}},
  \citenamefont {{Rest}}, \citenamefont {{Adams}}, \citenamefont {{Alatalo}},
  \citenamefont {{Ba{\~n}ados}}, \citenamefont {{Baughman}}, \citenamefont
  {{Beers}}, \citenamefont {{Bernstein}}, \citenamefont {{Bitsakis}},
  \citenamefont {{Campillay}}, \citenamefont {{Hansen}}, \citenamefont
  {{Higgs}}, \citenamefont {{Ji}}, \citenamefont {{Maravelias}}, \citenamefont
  {{Marshall}}, \citenamefont {{Moni Bidin}}, \citenamefont {{Prieto}},
  \citenamefont {{Rasmussen}}, \citenamefont {{Rojas-Bravo}}, \citenamefont
  {{Strom}}, \citenamefont {{Ulloa}}, \citenamefont {{Vargas-Gonz{\'a}lez}},
  \citenamefont {{Wan}},\ and\ \citenamefont {{Whitten}}}]{drout:2017SCIENCE}%
  \BibitemOpen
  \bibfield  {author} {\bibinfo {author} {\bibfnamefont {M.~R.}\ \bibnamefont
  {{Drout}}}, \bibinfo {author} {\bibfnamefont {A.~L.}\ \bibnamefont {{Piro}}},
  \bibinfo {author} {\bibfnamefont {B.~J.}\ \bibnamefont {{Shappee}}}, \bibinfo
  {author} {\bibfnamefont {C.~D.}\ \bibnamefont {{Kilpatrick}}}, \bibinfo
  {author} {\bibfnamefont {J.~D.}\ \bibnamefont {{Simon}}}, \bibinfo {author}
  {\bibfnamefont {C.}~\bibnamefont {{Contreras}}}, \bibinfo {author}
  {\bibfnamefont {D.~A.}\ \bibnamefont {{Coulter}}}, \bibinfo {author}
  {\bibfnamefont {R.~J.}\ \bibnamefont {{Foley}}}, \bibinfo {author}
  {\bibfnamefont {M.~R.}\ \bibnamefont {{Siebert}}}, \bibinfo {author}
  {\bibfnamefont {N.}~\bibnamefont {{Morrell}}}, \bibinfo {author}
  {\bibfnamefont {K.}~\bibnamefont {{Boutsia}}}, \bibinfo {author}
  {\bibfnamefont {F.}~\bibnamefont {{Di Mille}}}, \bibinfo {author}
  {\bibfnamefont {T.~W.-S.}\ \bibnamefont {{Holoien}}}, \bibinfo {author}
  {\bibfnamefont {D.}~\bibnamefont {{Kasen}}}, \bibinfo {author} {\bibfnamefont
  {J.~A.}\ \bibnamefont {{Kollmeier}}}, \bibinfo {author} {\bibfnamefont
  {B.~F.}\ \bibnamefont {{Madore}}}, \bibinfo {author} {\bibfnamefont {A.~J.}\
  \bibnamefont {{Monson}}}, \bibinfo {author} {\bibfnamefont {A.}~\bibnamefont
  {{Murguia-Berthier}}}, \bibinfo {author} {\bibfnamefont {Y.-C.}\ \bibnamefont
  {{Pan}}}, \bibinfo {author} {\bibfnamefont {J.~X.}\ \bibnamefont
  {{Prochaska}}}, \bibinfo {author} {\bibfnamefont {E.}~\bibnamefont
  {{Ramirez-Ruiz}}}, \bibinfo {author} {\bibfnamefont {A.}~\bibnamefont
  {{Rest}}}, \bibinfo {author} {\bibfnamefont {C.}~\bibnamefont {{Adams}}},
  \bibinfo {author} {\bibfnamefont {K.}~\bibnamefont {{Alatalo}}}, \bibinfo
  {author} {\bibfnamefont {E.}~\bibnamefont {{Ba{\~n}ados}}}, \bibinfo {author}
  {\bibfnamefont {J.}~\bibnamefont {{Baughman}}}, \bibinfo {author}
  {\bibfnamefont {T.~C.}\ \bibnamefont {{Beers}}}, \bibinfo {author}
  {\bibfnamefont {R.~A.}\ \bibnamefont {{Bernstein}}}, \bibinfo {author}
  {\bibfnamefont {T.}~\bibnamefont {{Bitsakis}}}, \bibinfo {author}
  {\bibfnamefont {A.}~\bibnamefont {{Campillay}}}, \bibinfo {author}
  {\bibfnamefont {T.~T.}\ \bibnamefont {{Hansen}}}, \bibinfo {author}
  {\bibfnamefont {C.~R.}\ \bibnamefont {{Higgs}}}, \bibinfo {author}
  {\bibfnamefont {A.~P.}\ \bibnamefont {{Ji}}}, \bibinfo {author}
  {\bibfnamefont {G.}~\bibnamefont {{Maravelias}}}, \bibinfo {author}
  {\bibfnamefont {J.~L.}\ \bibnamefont {{Marshall}}}, \bibinfo {author}
  {\bibfnamefont {C.}~\bibnamefont {{Moni Bidin}}}, \bibinfo {author}
  {\bibfnamefont {J.~L.}\ \bibnamefont {{Prieto}}}, \bibinfo {author}
  {\bibfnamefont {K.~C.}\ \bibnamefont {{Rasmussen}}}, \bibinfo {author}
  {\bibfnamefont {C.}~\bibnamefont {{Rojas-Bravo}}}, \bibinfo {author}
  {\bibfnamefont {A.~L.}\ \bibnamefont {{Strom}}}, \bibinfo {author}
  {\bibfnamefont {N.}~\bibnamefont {{Ulloa}}}, \bibinfo {author} {\bibfnamefont
  {J.}~\bibnamefont {{Vargas-Gonz{\'a}lez}}}, \bibinfo {author} {\bibfnamefont
  {Z.}~\bibnamefont {{Wan}}}, \ and\ \bibinfo {author} {\bibfnamefont {D.~D.}\
  \bibnamefont {{Whitten}}},\ }\href {\doibase 10.1126/science.aaq0049}
  {\bibfield  {journal} {\bibinfo  {journal} {Science}\ }\textbf {\bibinfo
  {volume} {358}},\ \bibinfo {pages} {1570} (\bibinfo {year} {2017})},\ \Eprint
  {http://arxiv.org/abs/1710.05443} {arXiv:1710.05443 [astro-ph.HE]}
  \BibitemShut {NoStop}%
\bibitem [{\citenamefont {{Abbott}}\ \emph
  {et~al.}(2017{\natexlab{f}})\citenamefont {{Abbott}}, \citenamefont
  {{Abbott}}, \citenamefont {{Abbott}}, \citenamefont {{Acernese}},
  \citenamefont {{Ackley}}, \citenamefont {{Adams}}, \citenamefont {{Adams}},
  \citenamefont {{Addesso}}, \citenamefont {{Adhikari}}, \citenamefont
  {{Adya}},\ and\ \citenamefont {et~al.}}]{mma:2017arXiv}%
  \BibitemOpen
  \bibfield  {author} {\bibinfo {author} {\bibfnamefont {B.~P.}\ \bibnamefont
  {{Abbott}}}, \bibinfo {author} {\bibfnamefont {R.}~\bibnamefont {{Abbott}}},
  \bibinfo {author} {\bibfnamefont {T.~D.}\ \bibnamefont {{Abbott}}}, \bibinfo
  {author} {\bibfnamefont {F.}~\bibnamefont {{Acernese}}}, \bibinfo {author}
  {\bibfnamefont {K.}~\bibnamefont {{Ackley}}}, \bibinfo {author}
  {\bibfnamefont {C.}~\bibnamefont {{Adams}}}, \bibinfo {author} {\bibfnamefont
  {T.}~\bibnamefont {{Adams}}}, \bibinfo {author} {\bibfnamefont
  {P.}~\bibnamefont {{Addesso}}}, \bibinfo {author} {\bibfnamefont {R.~X.}\
  \bibnamefont {{Adhikari}}}, \bibinfo {author} {\bibfnamefont {V.~B.}\
  \bibnamefont {{Adya}}}, \ and\ \bibinfo {author} {\bibnamefont {et~al.}},\
  }\href {\doibase 10.3847/2041-8213/aa91c9} {\bibfield  {journal} {\bibinfo
  {journal} {\apjl}\ }\textbf {\bibinfo {volume} {848}},\ \bibinfo {eid} {L12}
  (\bibinfo {year} {2017}{\natexlab{f}})},\ \Eprint
  {http://arxiv.org/abs/1710.05833} {arXiv:1710.05833 [astro-ph.HE]}
  \BibitemShut {NoStop}%
\bibitem [{\citenamefont {{Andreoni}}\ \emph {et~al.}(2017)\citenamefont
  {{Andreoni}}, \citenamefont {{Jacobs}}, \citenamefont {{Hegarty}},
  \citenamefont {{Pritchard}}, \citenamefont {{Cooke}},\ and\ \citenamefont
  {{Ryder}}}]{igor:2017PASA}%
  \BibitemOpen
  \bibfield  {author} {\bibinfo {author} {\bibfnamefont {I.}~\bibnamefont
  {{Andreoni}}}, \bibinfo {author} {\bibfnamefont {C.}~\bibnamefont
  {{Jacobs}}}, \bibinfo {author} {\bibfnamefont {S.}~\bibnamefont {{Hegarty}}},
  \bibinfo {author} {\bibfnamefont {T.}~\bibnamefont {{Pritchard}}}, \bibinfo
  {author} {\bibfnamefont {J.}~\bibnamefont {{Cooke}}}, \ and\ \bibinfo
  {author} {\bibfnamefont {S.}~\bibnamefont {{Ryder}}},\ }\href {\doibase
  10.1017/pasa.2017.33} {\bibfield  {journal} {\bibinfo  {journal}
  {Publications of the Astronomical Society of Australia}\ }\textbf {\bibinfo
  {volume} {34}},\ \bibinfo {eid} {e037} (\bibinfo {year} {2017})},\ \Eprint
  {http://arxiv.org/abs/1708.04629} {arXiv:1708.04629 [astro-ph.IM]}
  \BibitemShut {NoStop}%
\bibitem [{\citenamefont {{Goldstein}}\ \emph
  {et~al.}(2015{\natexlab{b}})\citenamefont {{Goldstein}}, \citenamefont
  {{D'Andrea}}, \citenamefont {{Fischer}}, \citenamefont {{Foley}},
  \citenamefont {{Gupta}}, \citenamefont {{Kessler}}, \citenamefont {{Kim}},
  \citenamefont {{Nichol}}, \citenamefont {{Nugent}}, \citenamefont
  {{Papadopoulos}}, \citenamefont {{Sako}}, \citenamefont {{Smith}},
  \citenamefont {{Sullivan}}, \citenamefont {{Thomas}}, \citenamefont
  {{Wester}}, \citenamefont {{Wolf}}, \citenamefont {{Abdalla}}, \citenamefont
  {{Banerji}}, \citenamefont {{Benoit-L{\'e}vy}}, \citenamefont {{Bertin}},
  \citenamefont {{Brooks}}, \citenamefont {{Carnero Rosell}}, \citenamefont
  {{Castander}}, \citenamefont {{da Costa}}, \citenamefont {{Covarrubias}},
  \citenamefont {{DePoy}}, \citenamefont {{Desai}}, \citenamefont {{Diehl}},
  \citenamefont {{Doel}}, \citenamefont {{Eifler}}, \citenamefont {{Fausti
  Neto}}, \citenamefont {{Finley}}, \citenamefont {{Flaugher}}, \citenamefont
  {{Fosalba}}, \citenamefont {{Frieman}}, \citenamefont {{Gerdes}},
  \citenamefont {{Gruen}}, \citenamefont {{Gruendl}}, \citenamefont {{James}},
  \citenamefont {{Kuehn}}, \citenamefont {{Kuropatkin}}, \citenamefont
  {{Lahav}}, \citenamefont {{Li}}, \citenamefont {{Maia}}, \citenamefont
  {{Makler}}, \citenamefont {{March}}, \citenamefont {{Marshall}},
  \citenamefont {{Martini}}, \citenamefont {{Merritt}}, \citenamefont
  {{Miquel}}, \citenamefont {{Nord}}, \citenamefont {{Ogando}}, \citenamefont
  {{Plazas}}, \citenamefont {{Romer}}, \citenamefont {{Roodman}}, \citenamefont
  {{Sanchez}}, \citenamefont {{Scarpine}}, \citenamefont {{Schubnell}},
  \citenamefont {{Sevilla-Noarbe}}, \citenamefont {{Smith}}, \citenamefont
  {{Soares-Santos}}, \citenamefont {{Sobreira}}, \citenamefont {{Suchyta}},
  \citenamefont {{Swanson}}, \citenamefont {{Tarle}}, \citenamefont
  {{Thaler}},\ and\ \citenamefont {{Walker}}}]{gold:2015AJG}%
  \BibitemOpen
  \bibfield  {author} {\bibinfo {author} {\bibfnamefont {D.~A.}\ \bibnamefont
  {{Goldstein}}}, \bibinfo {author} {\bibfnamefont {C.~B.}\ \bibnamefont
  {{D'Andrea}}}, \bibinfo {author} {\bibfnamefont {J.~A.}\ \bibnamefont
  {{Fischer}}}, \bibinfo {author} {\bibfnamefont {R.~J.}\ \bibnamefont
  {{Foley}}}, \bibinfo {author} {\bibfnamefont {R.~R.}\ \bibnamefont
  {{Gupta}}}, \bibinfo {author} {\bibfnamefont {R.}~\bibnamefont {{Kessler}}},
  \bibinfo {author} {\bibfnamefont {A.~G.}\ \bibnamefont {{Kim}}}, \bibinfo
  {author} {\bibfnamefont {R.~C.}\ \bibnamefont {{Nichol}}}, \bibinfo {author}
  {\bibfnamefont {P.~E.}\ \bibnamefont {{Nugent}}}, \bibinfo {author}
  {\bibfnamefont {A.}~\bibnamefont {{Papadopoulos}}}, \bibinfo {author}
  {\bibfnamefont {M.}~\bibnamefont {{Sako}}}, \bibinfo {author} {\bibfnamefont
  {M.}~\bibnamefont {{Smith}}}, \bibinfo {author} {\bibfnamefont
  {M.}~\bibnamefont {{Sullivan}}}, \bibinfo {author} {\bibfnamefont {R.~C.}\
  \bibnamefont {{Thomas}}}, \bibinfo {author} {\bibfnamefont {W.}~\bibnamefont
  {{Wester}}}, \bibinfo {author} {\bibfnamefont {R.~C.}\ \bibnamefont
  {{Wolf}}}, \bibinfo {author} {\bibfnamefont {F.~B.}\ \bibnamefont
  {{Abdalla}}}, \bibinfo {author} {\bibfnamefont {M.}~\bibnamefont
  {{Banerji}}}, \bibinfo {author} {\bibfnamefont {A.}~\bibnamefont
  {{Benoit-L{\'e}vy}}}, \bibinfo {author} {\bibfnamefont {E.}~\bibnamefont
  {{Bertin}}}, \bibinfo {author} {\bibfnamefont {D.}~\bibnamefont {{Brooks}}},
  \bibinfo {author} {\bibfnamefont {A.}~\bibnamefont {{Carnero Rosell}}},
  \bibinfo {author} {\bibfnamefont {F.~J.}\ \bibnamefont {{Castander}}},
  \bibinfo {author} {\bibfnamefont {L.~N.}\ \bibnamefont {{da Costa}}},
  \bibinfo {author} {\bibfnamefont {R.}~\bibnamefont {{Covarrubias}}}, \bibinfo
  {author} {\bibfnamefont {D.~L.}\ \bibnamefont {{DePoy}}}, \bibinfo {author}
  {\bibfnamefont {S.}~\bibnamefont {{Desai}}}, \bibinfo {author} {\bibfnamefont
  {H.~T.}\ \bibnamefont {{Diehl}}}, \bibinfo {author} {\bibfnamefont
  {P.}~\bibnamefont {{Doel}}}, \bibinfo {author} {\bibfnamefont {T.~F.}\
  \bibnamefont {{Eifler}}}, \bibinfo {author} {\bibfnamefont {A.}~\bibnamefont
  {{Fausti Neto}}}, \bibinfo {author} {\bibfnamefont {D.~A.}\ \bibnamefont
  {{Finley}}}, \bibinfo {author} {\bibfnamefont {B.}~\bibnamefont
  {{Flaugher}}}, \bibinfo {author} {\bibfnamefont {P.}~\bibnamefont
  {{Fosalba}}}, \bibinfo {author} {\bibfnamefont {J.}~\bibnamefont
  {{Frieman}}}, \bibinfo {author} {\bibfnamefont {D.}~\bibnamefont {{Gerdes}}},
  \bibinfo {author} {\bibfnamefont {D.}~\bibnamefont {{Gruen}}}, \bibinfo
  {author} {\bibfnamefont {R.~A.}\ \bibnamefont {{Gruendl}}}, \bibinfo {author}
  {\bibfnamefont {D.}~\bibnamefont {{James}}}, \bibinfo {author} {\bibfnamefont
  {K.}~\bibnamefont {{Kuehn}}}, \bibinfo {author} {\bibfnamefont
  {N.}~\bibnamefont {{Kuropatkin}}}, \bibinfo {author} {\bibfnamefont
  {O.}~\bibnamefont {{Lahav}}}, \bibinfo {author} {\bibfnamefont {T.~S.}\
  \bibnamefont {{Li}}}, \bibinfo {author} {\bibfnamefont {M.~A.~G.}\
  \bibnamefont {{Maia}}}, \bibinfo {author} {\bibfnamefont {M.}~\bibnamefont
  {{Makler}}}, \bibinfo {author} {\bibfnamefont {M.}~\bibnamefont {{March}}},
  \bibinfo {author} {\bibfnamefont {J.~L.}\ \bibnamefont {{Marshall}}},
  \bibinfo {author} {\bibfnamefont {P.}~\bibnamefont {{Martini}}}, \bibinfo
  {author} {\bibfnamefont {K.~W.}\ \bibnamefont {{Merritt}}}, \bibinfo {author}
  {\bibfnamefont {R.}~\bibnamefont {{Miquel}}}, \bibinfo {author}
  {\bibfnamefont {B.}~\bibnamefont {{Nord}}}, \bibinfo {author} {\bibfnamefont
  {R.}~\bibnamefont {{Ogando}}}, \bibinfo {author} {\bibfnamefont {A.~A.}\
  \bibnamefont {{Plazas}}}, \bibinfo {author} {\bibfnamefont {A.~K.}\
  \bibnamefont {{Romer}}}, \bibinfo {author} {\bibfnamefont {A.}~\bibnamefont
  {{Roodman}}}, \bibinfo {author} {\bibfnamefont {E.}~\bibnamefont
  {{Sanchez}}}, \bibinfo {author} {\bibfnamefont {V.}~\bibnamefont
  {{Scarpine}}}, \bibinfo {author} {\bibfnamefont {M.}~\bibnamefont
  {{Schubnell}}}, \bibinfo {author} {\bibfnamefont {I.}~\bibnamefont
  {{Sevilla-Noarbe}}}, \bibinfo {author} {\bibfnamefont {R.~C.}\ \bibnamefont
  {{Smith}}}, \bibinfo {author} {\bibfnamefont {M.}~\bibnamefont
  {{Soares-Santos}}}, \bibinfo {author} {\bibfnamefont {F.}~\bibnamefont
  {{Sobreira}}}, \bibinfo {author} {\bibfnamefont {E.}~\bibnamefont
  {{Suchyta}}}, \bibinfo {author} {\bibfnamefont {M.~E.~C.}\ \bibnamefont
  {{Swanson}}}, \bibinfo {author} {\bibfnamefont {G.}~\bibnamefont {{Tarle}}},
  \bibinfo {author} {\bibfnamefont {J.}~\bibnamefont {{Thaler}}}, \ and\
  \bibinfo {author} {\bibfnamefont {A.~R.}\ \bibnamefont {{Walker}}},\ }\href
  {\doibase 10.1088/0004-6256/150/3/82} {\bibfield  {journal} {\bibinfo
  {journal} {The Astronomical Journal}\ }\textbf {\bibinfo {volume} {150}},\
  \bibinfo {eid} {82} (\bibinfo {year} {2015}{\natexlab{b}})},\ \Eprint
  {http://arxiv.org/abs/1504.02936} {arXiv:1504.02936 [astro-ph.IM]}
  \BibitemShut {NoStop}%
\bibitem [{\citenamefont {{Sedaghat}}\ and\ \citenamefont
  {{Mahabal}}(2018)}]{seda:2018MNRAS}%
  \BibitemOpen
  \bibfield  {author} {\bibinfo {author} {\bibfnamefont {N.}~\bibnamefont
  {{Sedaghat}}}\ and\ \bibinfo {author} {\bibfnamefont {A.}~\bibnamefont
  {{Mahabal}}},\ }\href {\doibase 10.1093/mnras/sty613} {\bibfield  {journal}
  {\bibinfo  {journal} {\mnras}\ }\textbf {\bibinfo {volume} {476}},\ \bibinfo
  {pages} {5365} (\bibinfo {year} {2018})},\ \Eprint
  {http://arxiv.org/abs/1710.01422} {arXiv:1710.01422 [astro-ph.IM]}
  \BibitemShut {NoStop}%
\bibitem [{\citenamefont {{Kessler}}\ \emph {et~al.}(2010)\citenamefont
  {{Kessler}}, \citenamefont {{Bassett}}, \citenamefont {{Belov}},
  \citenamefont {{Bhatnagar}}, \citenamefont {{Campbell}}, \citenamefont
  {{Conley}}, \citenamefont {{Frieman}}, \citenamefont {{Glazov}},
  \citenamefont {{Gonz{\'a}lez-Gait{\'a}n}}, \citenamefont {{Hlozek}},
  \citenamefont {{Jha}}, \citenamefont {{Kuhlmann}}, \citenamefont {{Kunz}},
  \citenamefont {{Lampeitl}}, \citenamefont {{Mahabal}}, \citenamefont
  {{Newling}}, \citenamefont {{Nichol}}, \citenamefont {{Parkinson}},
  \citenamefont {{Sajeeth Philip}}, \citenamefont {{Poznanski}}, \citenamefont
  {{Richards}}, \citenamefont {{Rodney}}, \citenamefont {{Sako}}, \citenamefont
  {{Schneider}}, \citenamefont {{Smith}}, \citenamefont {{Stritzinger}},\ and\
  \citenamefont {{Varughese}}}]{2010PASP..122.1415K}%
  \BibitemOpen
  \bibfield  {author} {\bibinfo {author} {\bibfnamefont {R.}~\bibnamefont
  {{Kessler}}}, \bibinfo {author} {\bibfnamefont {B.}~\bibnamefont
  {{Bassett}}}, \bibinfo {author} {\bibfnamefont {P.}~\bibnamefont {{Belov}}},
  \bibinfo {author} {\bibfnamefont {V.}~\bibnamefont {{Bhatnagar}}}, \bibinfo
  {author} {\bibfnamefont {H.}~\bibnamefont {{Campbell}}}, \bibinfo {author}
  {\bibfnamefont {A.}~\bibnamefont {{Conley}}}, \bibinfo {author}
  {\bibfnamefont {J.~A.}\ \bibnamefont {{Frieman}}}, \bibinfo {author}
  {\bibfnamefont {A.}~\bibnamefont {{Glazov}}}, \bibinfo {author}
  {\bibfnamefont {S.}~\bibnamefont {{Gonz{\'a}lez-Gait{\'a}n}}}, \bibinfo
  {author} {\bibfnamefont {R.}~\bibnamefont {{Hlozek}}}, \bibinfo {author}
  {\bibfnamefont {S.}~\bibnamefont {{Jha}}}, \bibinfo {author} {\bibfnamefont
  {S.}~\bibnamefont {{Kuhlmann}}}, \bibinfo {author} {\bibfnamefont
  {M.}~\bibnamefont {{Kunz}}}, \bibinfo {author} {\bibfnamefont
  {H.}~\bibnamefont {{Lampeitl}}}, \bibinfo {author} {\bibfnamefont
  {A.}~\bibnamefont {{Mahabal}}}, \bibinfo {author} {\bibfnamefont
  {J.}~\bibnamefont {{Newling}}}, \bibinfo {author} {\bibfnamefont {R.~C.}\
  \bibnamefont {{Nichol}}}, \bibinfo {author} {\bibfnamefont {D.}~\bibnamefont
  {{Parkinson}}}, \bibinfo {author} {\bibfnamefont {N.}~\bibnamefont {{Sajeeth
  Philip}}}, \bibinfo {author} {\bibfnamefont {D.}~\bibnamefont {{Poznanski}}},
  \bibinfo {author} {\bibfnamefont {J.~W.}\ \bibnamefont {{Richards}}},
  \bibinfo {author} {\bibfnamefont {S.~A.}\ \bibnamefont {{Rodney}}}, \bibinfo
  {author} {\bibfnamefont {M.}~\bibnamefont {{Sako}}}, \bibinfo {author}
  {\bibfnamefont {D.~P.}\ \bibnamefont {{Schneider}}}, \bibinfo {author}
  {\bibfnamefont {M.}~\bibnamefont {{Smith}}}, \bibinfo {author} {\bibfnamefont
  {M.}~\bibnamefont {{Stritzinger}}}, \ and\ \bibinfo {author} {\bibfnamefont
  {M.}~\bibnamefont {{Varughese}}},\ }\href {\doibase 10.1086/657607}
  {\bibfield  {journal} {\bibinfo  {journal} {\pasp}\ }\textbf {\bibinfo
  {volume} {122}},\ \bibinfo {pages} {1415} (\bibinfo {year} {2010})},\ \Eprint
  {http://arxiv.org/abs/1008.1024} {arXiv:1008.1024 [astro-ph.CO]} \BibitemShut
  {NoStop}%
\bibitem [{\citenamefont {{Lochner}}\ \emph {et~al.}(2016)\citenamefont
  {{Lochner}}, \citenamefont {{McEwen}}, \citenamefont {{Peiris}},
  \citenamefont {{Lahav}},\ and\ \citenamefont
  {{Winter}}}]{2016ApJS..225...31L}%
  \BibitemOpen
  \bibfield  {author} {\bibinfo {author} {\bibfnamefont {M.}~\bibnamefont
  {{Lochner}}}, \bibinfo {author} {\bibfnamefont {J.~D.}\ \bibnamefont
  {{McEwen}}}, \bibinfo {author} {\bibfnamefont {H.~V.}\ \bibnamefont
  {{Peiris}}}, \bibinfo {author} {\bibfnamefont {O.}~\bibnamefont {{Lahav}}}, \
  and\ \bibinfo {author} {\bibfnamefont {M.~K.}\ \bibnamefont {{Winter}}},\
  }\href {\doibase 10.3847/0067-0049/225/2/31} {\bibfield  {journal} {\bibinfo
  {journal} {The Astrophysical Journal Supplement Series}\ }\textbf {\bibinfo
  {volume} {225}},\ \bibinfo {eid} {31} (\bibinfo {year} {2016})},\ \Eprint
  {http://arxiv.org/abs/1603.00882} {arXiv:1603.00882 [astro-ph.IM]}
  \BibitemShut {NoStop}%
\bibitem [{\citenamefont {{Charnock}}\ and\ \citenamefont
  {{Moss}}(2017)}]{2017ApJ...837L..28C}%
  \BibitemOpen
  \bibfield  {author} {\bibinfo {author} {\bibfnamefont {T.}~\bibnamefont
  {{Charnock}}}\ and\ \bibinfo {author} {\bibfnamefont {A.}~\bibnamefont
  {{Moss}}},\ }\href {\doibase 10.3847/2041-8213/aa603d} {\bibfield  {journal}
  {\bibinfo  {journal} {\apjl}\ }\textbf {\bibinfo {volume} {837}},\ \bibinfo
  {eid} {L28} (\bibinfo {year} {2017})},\ \Eprint
  {http://arxiv.org/abs/1606.07442} {arXiv:1606.07442 [astro-ph.IM]}
  \BibitemShut {NoStop}%
\bibitem [{\citenamefont {{Campbell}}\ \emph {et~al.}(2013)\citenamefont
  {{Campbell}}, \citenamefont {{D'Andrea}}, \citenamefont {{Nichol}},
  \citenamefont {{Sako}}, \citenamefont {{Smith}}, \citenamefont {{Lampeitl}},
  \citenamefont {{Olmstead}}, \citenamefont {{Bassett}}, \citenamefont
  {{Biswas}}, \citenamefont {{Brown}}, \citenamefont {{Cinabro}}, \citenamefont
  {{Dawson}}, \citenamefont {{Dilday}}, \citenamefont {{Foley}}, \citenamefont
  {{Frieman}}, \citenamefont {{Garnavich}}, \citenamefont {{Hlozek}},
  \citenamefont {{Jha}}, \citenamefont {{Kuhlmann}}, \citenamefont {{Kunz}},
  \citenamefont {{Marriner}}, \citenamefont {{Miquel}}, \citenamefont
  {{Richmond}}, \citenamefont {{Riess}}, \citenamefont {{Schneider}},
  \citenamefont {{Sollerman}}, \citenamefont {{Taylor}},\ and\ \citenamefont
  {{Zhao}}}]{2013ApJ...763...88C}%
  \BibitemOpen
  \bibfield  {author} {\bibinfo {author} {\bibfnamefont {H.}~\bibnamefont
  {{Campbell}}}, \bibinfo {author} {\bibfnamefont {C.~B.}\ \bibnamefont
  {{D'Andrea}}}, \bibinfo {author} {\bibfnamefont {R.~C.}\ \bibnamefont
  {{Nichol}}}, \bibinfo {author} {\bibfnamefont {M.}~\bibnamefont {{Sako}}},
  \bibinfo {author} {\bibfnamefont {M.}~\bibnamefont {{Smith}}}, \bibinfo
  {author} {\bibfnamefont {H.}~\bibnamefont {{Lampeitl}}}, \bibinfo {author}
  {\bibfnamefont {M.~D.}\ \bibnamefont {{Olmstead}}}, \bibinfo {author}
  {\bibfnamefont {B.}~\bibnamefont {{Bassett}}}, \bibinfo {author}
  {\bibfnamefont {R.}~\bibnamefont {{Biswas}}}, \bibinfo {author}
  {\bibfnamefont {P.}~\bibnamefont {{Brown}}}, \bibinfo {author} {\bibfnamefont
  {D.}~\bibnamefont {{Cinabro}}}, \bibinfo {author} {\bibfnamefont {K.~S.}\
  \bibnamefont {{Dawson}}}, \bibinfo {author} {\bibfnamefont {B.}~\bibnamefont
  {{Dilday}}}, \bibinfo {author} {\bibfnamefont {R.~J.}\ \bibnamefont
  {{Foley}}}, \bibinfo {author} {\bibfnamefont {J.~A.}\ \bibnamefont
  {{Frieman}}}, \bibinfo {author} {\bibfnamefont {P.}~\bibnamefont
  {{Garnavich}}}, \bibinfo {author} {\bibfnamefont {R.}~\bibnamefont
  {{Hlozek}}}, \bibinfo {author} {\bibfnamefont {S.~W.}\ \bibnamefont {{Jha}}},
  \bibinfo {author} {\bibfnamefont {S.}~\bibnamefont {{Kuhlmann}}}, \bibinfo
  {author} {\bibfnamefont {M.}~\bibnamefont {{Kunz}}}, \bibinfo {author}
  {\bibfnamefont {J.}~\bibnamefont {{Marriner}}}, \bibinfo {author}
  {\bibfnamefont {R.}~\bibnamefont {{Miquel}}}, \bibinfo {author}
  {\bibfnamefont {M.}~\bibnamefont {{Richmond}}}, \bibinfo {author}
  {\bibfnamefont {A.}~\bibnamefont {{Riess}}}, \bibinfo {author} {\bibfnamefont
  {D.~P.}\ \bibnamefont {{Schneider}}}, \bibinfo {author} {\bibfnamefont
  {J.}~\bibnamefont {{Sollerman}}}, \bibinfo {author} {\bibfnamefont
  {M.}~\bibnamefont {{Taylor}}}, \ and\ \bibinfo {author} {\bibfnamefont
  {G.-B.}\ \bibnamefont {{Zhao}}},\ }\href {\doibase
  10.1088/0004-637X/763/2/88} {\bibfield  {journal} {\bibinfo  {journal}
  {\apj}\ }\textbf {\bibinfo {volume} {763}},\ \bibinfo {eid} {88} (\bibinfo
  {year} {2013})},\ \Eprint {http://arxiv.org/abs/1211.4480} {arXiv:1211.4480
  [astro-ph.CO]} \BibitemShut {NoStop}%
\bibitem [{\citenamefont {{Jones}}\ \emph {et~al.}(2018)\citenamefont
  {{Jones}}, \citenamefont {{Scolnic}}, \citenamefont {{Riess}}, \citenamefont
  {{Rest}}, \citenamefont {{Kirshner}}, \citenamefont {{Berger}}, \citenamefont
  {{Kessler}}, \citenamefont {{Pan}}, \citenamefont {{Foley}}, \citenamefont
  {{Chornock}}, \citenamefont {{Ortega}}, \citenamefont {{Challis}},
  \citenamefont {{Burgett}}, \citenamefont {{Chambers}}, \citenamefont
  {{Draper}}, \citenamefont {{Flewelling}}, \citenamefont {{Huber}},
  \citenamefont {{Kaiser}}, \citenamefont {{Kudritzki}}, \citenamefont
  {{Metcalfe}}, \citenamefont {{Tonry}}, \citenamefont {{Wainscoat}},
  \citenamefont {{Waters}}, \citenamefont {{Gall}}, \citenamefont {{Kotak}},
  \citenamefont {{McCrum}}, \citenamefont {{Smartt}},\ and\ \citenamefont
  {{Smith}}}]{2018ApJ...857...51J}%
  \BibitemOpen
  \bibfield  {author} {\bibinfo {author} {\bibfnamefont {D.~O.}\ \bibnamefont
  {{Jones}}}, \bibinfo {author} {\bibfnamefont {D.~M.}\ \bibnamefont
  {{Scolnic}}}, \bibinfo {author} {\bibfnamefont {A.~G.}\ \bibnamefont
  {{Riess}}}, \bibinfo {author} {\bibfnamefont {A.}~\bibnamefont {{Rest}}},
  \bibinfo {author} {\bibfnamefont {R.~P.}\ \bibnamefont {{Kirshner}}},
  \bibinfo {author} {\bibfnamefont {E.}~\bibnamefont {{Berger}}}, \bibinfo
  {author} {\bibfnamefont {R.}~\bibnamefont {{Kessler}}}, \bibinfo {author}
  {\bibfnamefont {Y.-C.}\ \bibnamefont {{Pan}}}, \bibinfo {author}
  {\bibfnamefont {R.~J.}\ \bibnamefont {{Foley}}}, \bibinfo {author}
  {\bibfnamefont {R.}~\bibnamefont {{Chornock}}}, \bibinfo {author}
  {\bibfnamefont {C.~A.}\ \bibnamefont {{Ortega}}}, \bibinfo {author}
  {\bibfnamefont {P.~J.}\ \bibnamefont {{Challis}}}, \bibinfo {author}
  {\bibfnamefont {W.~S.}\ \bibnamefont {{Burgett}}}, \bibinfo {author}
  {\bibfnamefont {K.~C.}\ \bibnamefont {{Chambers}}}, \bibinfo {author}
  {\bibfnamefont {P.~W.}\ \bibnamefont {{Draper}}}, \bibinfo {author}
  {\bibfnamefont {H.}~\bibnamefont {{Flewelling}}}, \bibinfo {author}
  {\bibfnamefont {M.~E.}\ \bibnamefont {{Huber}}}, \bibinfo {author}
  {\bibfnamefont {N.}~\bibnamefont {{Kaiser}}}, \bibinfo {author}
  {\bibfnamefont {R.-P.}\ \bibnamefont {{Kudritzki}}}, \bibinfo {author}
  {\bibfnamefont {N.}~\bibnamefont {{Metcalfe}}}, \bibinfo {author}
  {\bibfnamefont {J.}~\bibnamefont {{Tonry}}}, \bibinfo {author} {\bibfnamefont
  {R.~J.}\ \bibnamefont {{Wainscoat}}}, \bibinfo {author} {\bibfnamefont
  {C.}~\bibnamefont {{Waters}}}, \bibinfo {author} {\bibfnamefont {E.~E.~E.}\
  \bibnamefont {{Gall}}}, \bibinfo {author} {\bibfnamefont {R.}~\bibnamefont
  {{Kotak}}}, \bibinfo {author} {\bibfnamefont {M.}~\bibnamefont {{McCrum}}},
  \bibinfo {author} {\bibfnamefont {S.~J.}\ \bibnamefont {{Smartt}}}, \ and\
  \bibinfo {author} {\bibfnamefont {K.~W.}\ \bibnamefont {{Smith}}},\ }\href
  {\doibase 10.3847/1538-4357/aab6b1} {\bibfield  {journal} {\bibinfo
  {journal} {\apj}\ }\textbf {\bibinfo {volume} {857}},\ \bibinfo {eid} {51}
  (\bibinfo {year} {2018})},\ \Eprint {http://arxiv.org/abs/1710.00846}
  {arXiv:1710.00846} \BibitemShut {NoStop}%
\bibitem [{\citenamefont {{Rosswog}}\ \emph {et~al.}(2017)\citenamefont
  {{Rosswog}}, \citenamefont {{Feindt}}, \citenamefont {{Korobkin}},
  \citenamefont {{Wu}}, \citenamefont {{Sollerman}}, \citenamefont {{Goobar}},\
  and\ \citenamefont {{Martinez-Pinedo}}}]{2017CQGra..34j4001R}%
  \BibitemOpen
  \bibfield  {author} {\bibinfo {author} {\bibfnamefont {S.}~\bibnamefont
  {{Rosswog}}}, \bibinfo {author} {\bibfnamefont {U.}~\bibnamefont {{Feindt}}},
  \bibinfo {author} {\bibfnamefont {O.}~\bibnamefont {{Korobkin}}}, \bibinfo
  {author} {\bibfnamefont {M.-R.}\ \bibnamefont {{Wu}}}, \bibinfo {author}
  {\bibfnamefont {J.}~\bibnamefont {{Sollerman}}}, \bibinfo {author}
  {\bibfnamefont {A.}~\bibnamefont {{Goobar}}}, \ and\ \bibinfo {author}
  {\bibfnamefont {G.}~\bibnamefont {{Martinez-Pinedo}}},\ }\href {\doibase
  10.1088/1361-6382/aa68a9} {\bibfield  {journal} {\bibinfo  {journal}
  {Classical and Quantum Gravity}\ }\textbf {\bibinfo {volume} {34}},\ \bibinfo
  {eid} {104001} (\bibinfo {year} {2017})},\ \Eprint
  {http://arxiv.org/abs/1611.09822} {arXiv:1611.09822 [astro-ph.HE]}
  \BibitemShut {NoStop}%
\bibitem [{\citenamefont {{Scolnic}}\ \emph {et~al.}(2018)\citenamefont
  {{Scolnic}}, \citenamefont {{Kessler}}, \citenamefont {{Brout}},
  \citenamefont {{Cowperthwaite}}, \citenamefont {{Soares-Santos}},
  \citenamefont {{Annis}}, \citenamefont {{Herner}}, \citenamefont {{Chen}},
  \citenamefont {{Sako}}, \citenamefont {{Doctor}}, \citenamefont {{Butler}},
  \citenamefont {{Palmese}}, \citenamefont {{Diehl}}, \citenamefont
  {{Frieman}}, \citenamefont {{Holz}}, \citenamefont {{Berger}}, \citenamefont
  {{Chornock}}, \citenamefont {{Villar}}, \citenamefont {{Nicholl}},
  \citenamefont {{Biswas}}, \citenamefont {{Hounsell}}, \citenamefont
  {{Foley}}, \citenamefont {{Metzger}}, \citenamefont {{Rest}}, \citenamefont
  {{Garc{\'{\i}}a-Bellido}}, \citenamefont {{M{\"o}ller}}, \citenamefont
  {{Nugent}}, \citenamefont {{Abbott}}, \citenamefont {{Abdalla}},
  \citenamefont {{Allam}}, \citenamefont {{Bechtol}}, \citenamefont
  {{Benoit-L{\'e}vy}}, \citenamefont {{Bertin}}, \citenamefont {{Brooks}},
  \citenamefont {{Buckley-Geer}}, \citenamefont {{Carnero Rosell}},
  \citenamefont {{Carrasco Kind}}, \citenamefont {{Carretero}}, \citenamefont
  {{Castander}}, \citenamefont {{Cunha}}, \citenamefont {{D'Andrea}},
  \citenamefont {{da Costa}}, \citenamefont {{Davis}}, \citenamefont {{Doel}},
  \citenamefont {{Drlica-Wagner}}, \citenamefont {{Eifler}}, \citenamefont
  {{Flaugher}}, \citenamefont {{Fosalba}}, \citenamefont {{Gaztanaga}},
  \citenamefont {{Gerdes}}, \citenamefont {{Gruen}}, \citenamefont {{Gruendl}},
  \citenamefont {{Gschwend}}, \citenamefont {{Gutierrez}}, \citenamefont
  {{Hartley}}, \citenamefont {{Honscheid}}, \citenamefont {{James}},
  \citenamefont {{Johnson}}, \citenamefont {{Johnson}}, \citenamefont
  {{Krause}}, \citenamefont {{Kuehn}}, \citenamefont {{Kuhlmann}},
  \citenamefont {{Lahav}}, \citenamefont {{Li}}, \citenamefont {{Lima}},
  \citenamefont {{Maia}}, \citenamefont {{March}}, \citenamefont {{Marshall}},
  \citenamefont {{Menanteau}}, \citenamefont {{Miquel}}, \citenamefont
  {{Neilsen}}, \citenamefont {{Plazas}}, \citenamefont {{Sanchez}},
  \citenamefont {{Scarpine}}, \citenamefont {{Schubnell}}, \citenamefont
  {{Sevilla-Noarbe}}, \citenamefont {{Smith}}, \citenamefont {{Smith}},
  \citenamefont {{Sobreira}}, \citenamefont {{Suchyta}}, \citenamefont
  {{Swanson}}, \citenamefont {{Tarle}}, \citenamefont {{Thomas}}, \citenamefont
  {{Tucker}}, \citenamefont {{Walker}},\ and\ \citenamefont {{DES
  Collaboration}}}]{2018ApJ...852L...3S}%
  \BibitemOpen
  \bibfield  {author} {\bibinfo {author} {\bibfnamefont {D.}~\bibnamefont
  {{Scolnic}}}, \bibinfo {author} {\bibfnamefont {R.}~\bibnamefont
  {{Kessler}}}, \bibinfo {author} {\bibfnamefont {D.}~\bibnamefont {{Brout}}},
  \bibinfo {author} {\bibfnamefont {P.~S.}\ \bibnamefont {{Cowperthwaite}}},
  \bibinfo {author} {\bibfnamefont {M.}~\bibnamefont {{Soares-Santos}}},
  \bibinfo {author} {\bibfnamefont {J.}~\bibnamefont {{Annis}}}, \bibinfo
  {author} {\bibfnamefont {K.}~\bibnamefont {{Herner}}}, \bibinfo {author}
  {\bibfnamefont {H.-Y.}\ \bibnamefont {{Chen}}}, \bibinfo {author}
  {\bibfnamefont {M.}~\bibnamefont {{Sako}}}, \bibinfo {author} {\bibfnamefont
  {Z.}~\bibnamefont {{Doctor}}}, \bibinfo {author} {\bibfnamefont {R.~E.}\
  \bibnamefont {{Butler}}}, \bibinfo {author} {\bibfnamefont {A.}~\bibnamefont
  {{Palmese}}}, \bibinfo {author} {\bibfnamefont {H.~T.}\ \bibnamefont
  {{Diehl}}}, \bibinfo {author} {\bibfnamefont {J.}~\bibnamefont {{Frieman}}},
  \bibinfo {author} {\bibfnamefont {D.~E.}\ \bibnamefont {{Holz}}}, \bibinfo
  {author} {\bibfnamefont {E.}~\bibnamefont {{Berger}}}, \bibinfo {author}
  {\bibfnamefont {R.}~\bibnamefont {{Chornock}}}, \bibinfo {author}
  {\bibfnamefont {V.~A.}\ \bibnamefont {{Villar}}}, \bibinfo {author}
  {\bibfnamefont {M.}~\bibnamefont {{Nicholl}}}, \bibinfo {author}
  {\bibfnamefont {R.}~\bibnamefont {{Biswas}}}, \bibinfo {author}
  {\bibfnamefont {R.}~\bibnamefont {{Hounsell}}}, \bibinfo {author}
  {\bibfnamefont {R.~J.}\ \bibnamefont {{Foley}}}, \bibinfo {author}
  {\bibfnamefont {J.}~\bibnamefont {{Metzger}}}, \bibinfo {author}
  {\bibfnamefont {A.}~\bibnamefont {{Rest}}}, \bibinfo {author} {\bibfnamefont
  {J.}~\bibnamefont {{Garc{\'{\i}}a-Bellido}}}, \bibinfo {author}
  {\bibfnamefont {A.}~\bibnamefont {{M{\"o}ller}}}, \bibinfo {author}
  {\bibfnamefont {P.}~\bibnamefont {{Nugent}}}, \bibinfo {author}
  {\bibfnamefont {T.~M.~C.}\ \bibnamefont {{Abbott}}}, \bibinfo {author}
  {\bibfnamefont {F.~B.}\ \bibnamefont {{Abdalla}}}, \bibinfo {author}
  {\bibfnamefont {S.}~\bibnamefont {{Allam}}}, \bibinfo {author} {\bibfnamefont
  {K.}~\bibnamefont {{Bechtol}}}, \bibinfo {author} {\bibfnamefont
  {A.}~\bibnamefont {{Benoit-L{\'e}vy}}}, \bibinfo {author} {\bibfnamefont
  {E.}~\bibnamefont {{Bertin}}}, \bibinfo {author} {\bibfnamefont
  {D.}~\bibnamefont {{Brooks}}}, \bibinfo {author} {\bibfnamefont
  {E.}~\bibnamefont {{Buckley-Geer}}}, \bibinfo {author} {\bibfnamefont
  {A.}~\bibnamefont {{Carnero Rosell}}}, \bibinfo {author} {\bibfnamefont
  {M.}~\bibnamefont {{Carrasco Kind}}}, \bibinfo {author} {\bibfnamefont
  {J.}~\bibnamefont {{Carretero}}}, \bibinfo {author} {\bibfnamefont {F.~J.}\
  \bibnamefont {{Castander}}}, \bibinfo {author} {\bibfnamefont {C.~E.}\
  \bibnamefont {{Cunha}}}, \bibinfo {author} {\bibfnamefont {C.~B.}\
  \bibnamefont {{D'Andrea}}}, \bibinfo {author} {\bibfnamefont {L.~N.}\
  \bibnamefont {{da Costa}}}, \bibinfo {author} {\bibfnamefont
  {C.}~\bibnamefont {{Davis}}}, \bibinfo {author} {\bibfnamefont
  {P.}~\bibnamefont {{Doel}}}, \bibinfo {author} {\bibfnamefont
  {A.}~\bibnamefont {{Drlica-Wagner}}}, \bibinfo {author} {\bibfnamefont
  {T.~F.}\ \bibnamefont {{Eifler}}}, \bibinfo {author} {\bibfnamefont
  {B.}~\bibnamefont {{Flaugher}}}, \bibinfo {author} {\bibfnamefont
  {P.}~\bibnamefont {{Fosalba}}}, \bibinfo {author} {\bibfnamefont
  {E.}~\bibnamefont {{Gaztanaga}}}, \bibinfo {author} {\bibfnamefont {D.~W.}\
  \bibnamefont {{Gerdes}}}, \bibinfo {author} {\bibfnamefont {D.}~\bibnamefont
  {{Gruen}}}, \bibinfo {author} {\bibfnamefont {R.~A.}\ \bibnamefont
  {{Gruendl}}}, \bibinfo {author} {\bibfnamefont {J.}~\bibnamefont
  {{Gschwend}}}, \bibinfo {author} {\bibfnamefont {G.}~\bibnamefont
  {{Gutierrez}}}, \bibinfo {author} {\bibfnamefont {W.~G.}\ \bibnamefont
  {{Hartley}}}, \bibinfo {author} {\bibfnamefont {K.}~\bibnamefont
  {{Honscheid}}}, \bibinfo {author} {\bibfnamefont {D.~J.}\ \bibnamefont
  {{James}}}, \bibinfo {author} {\bibfnamefont {M.~W.~G.}\ \bibnamefont
  {{Johnson}}}, \bibinfo {author} {\bibfnamefont {M.~D.}\ \bibnamefont
  {{Johnson}}}, \bibinfo {author} {\bibfnamefont {E.}~\bibnamefont {{Krause}}},
  \bibinfo {author} {\bibfnamefont {K.}~\bibnamefont {{Kuehn}}}, \bibinfo
  {author} {\bibfnamefont {S.}~\bibnamefont {{Kuhlmann}}}, \bibinfo {author}
  {\bibfnamefont {O.}~\bibnamefont {{Lahav}}}, \bibinfo {author} {\bibfnamefont
  {T.~S.}\ \bibnamefont {{Li}}}, \bibinfo {author} {\bibfnamefont
  {M.}~\bibnamefont {{Lima}}}, \bibinfo {author} {\bibfnamefont {M.~A.~G.}\
  \bibnamefont {{Maia}}}, \bibinfo {author} {\bibfnamefont {M.}~\bibnamefont
  {{March}}}, \bibinfo {author} {\bibfnamefont {J.~L.}\ \bibnamefont
  {{Marshall}}}, \bibinfo {author} {\bibfnamefont {F.}~\bibnamefont
  {{Menanteau}}}, \bibinfo {author} {\bibfnamefont {R.}~\bibnamefont
  {{Miquel}}}, \bibinfo {author} {\bibfnamefont {E.}~\bibnamefont {{Neilsen}}},
  \bibinfo {author} {\bibfnamefont {A.~A.}\ \bibnamefont {{Plazas}}}, \bibinfo
  {author} {\bibfnamefont {E.}~\bibnamefont {{Sanchez}}}, \bibinfo {author}
  {\bibfnamefont {V.}~\bibnamefont {{Scarpine}}}, \bibinfo {author}
  {\bibfnamefont {M.}~\bibnamefont {{Schubnell}}}, \bibinfo {author}
  {\bibfnamefont {I.}~\bibnamefont {{Sevilla-Noarbe}}}, \bibinfo {author}
  {\bibfnamefont {M.}~\bibnamefont {{Smith}}}, \bibinfo {author} {\bibfnamefont
  {R.~C.}\ \bibnamefont {{Smith}}}, \bibinfo {author} {\bibfnamefont
  {F.}~\bibnamefont {{Sobreira}}}, \bibinfo {author} {\bibfnamefont
  {E.}~\bibnamefont {{Suchyta}}}, \bibinfo {author} {\bibfnamefont {M.~E.~C.}\
  \bibnamefont {{Swanson}}}, \bibinfo {author} {\bibfnamefont {G.}~\bibnamefont
  {{Tarle}}}, \bibinfo {author} {\bibfnamefont {R.~C.}\ \bibnamefont
  {{Thomas}}}, \bibinfo {author} {\bibfnamefont {D.~L.}\ \bibnamefont
  {{Tucker}}}, \bibinfo {author} {\bibfnamefont {A.~R.}\ \bibnamefont
  {{Walker}}}, \ and\ \bibinfo {author} {\bibnamefont {{DES Collaboration}}},\
  }\href {\doibase 10.3847/2041-8213/aa9d82} {\bibfield  {journal} {\bibinfo
  {journal} {\apjl}\ }\textbf {\bibinfo {volume} {852}},\ \bibinfo {eid} {L3}
  (\bibinfo {year} {2018})},\ \Eprint {http://arxiv.org/abs/1710.05845}
  {arXiv:1710.05845 [astro-ph.IM]} \BibitemShut {NoStop}%
\bibitem [{\citenamefont {{Cowperthwaite}}\ \emph {et~al.}(2018)\citenamefont
  {{Cowperthwaite}}, \citenamefont {{Villar}}, \citenamefont {{Scolnic}},\ and\
  \citenamefont {{Berger}}}]{2018arXiv181103098C}%
  \BibitemOpen
  \bibfield  {author} {\bibinfo {author} {\bibfnamefont {P.~S.}\ \bibnamefont
  {{Cowperthwaite}}}, \bibinfo {author} {\bibfnamefont {V.~A.}\ \bibnamefont
  {{Villar}}}, \bibinfo {author} {\bibfnamefont {D.~M.}\ \bibnamefont
  {{Scolnic}}}, \ and\ \bibinfo {author} {\bibfnamefont {E.}~\bibnamefont
  {{Berger}}},\ }\href@noop {} {\bibfield  {journal} {\bibinfo  {journal}
  {arXiv e-prints}\ } (\bibinfo {year} {2018})},\ \Eprint
  {http://arxiv.org/abs/1811.03098} {arXiv:1811.03098 [astro-ph.HE]}
  \BibitemShut {NoStop}%
\bibitem [{\citenamefont {{Setzer}}\ \emph {et~al.}(2018)\citenamefont
  {{Setzer}}, \citenamefont {{Biswas}}, \citenamefont {{Peiris}}, \citenamefont
  {{Rosswog}}, \citenamefont {{Korobkin}},\ and\ \citenamefont
  {{Wollaeger}}}]{2018arXiv181210492S}%
  \BibitemOpen
  \bibfield  {author} {\bibinfo {author} {\bibfnamefont {C.~N.}\ \bibnamefont
  {{Setzer}}}, \bibinfo {author} {\bibfnamefont {R.}~\bibnamefont {{Biswas}}},
  \bibinfo {author} {\bibfnamefont {H.~V.}\ \bibnamefont {{Peiris}}}, \bibinfo
  {author} {\bibfnamefont {S.}~\bibnamefont {{Rosswog}}}, \bibinfo {author}
  {\bibfnamefont {O.}~\bibnamefont {{Korobkin}}}, \ and\ \bibinfo {author}
  {\bibfnamefont {R.~T.}\ \bibnamefont {{Wollaeger}}},\ }\href@noop {}
  {\bibfield  {journal} {\bibinfo  {journal} {arXiv e-prints}\ ,\ \bibinfo
  {eid} {arXiv:1812.10492}} (\bibinfo {year} {2018})},\ \Eprint
  {http://arxiv.org/abs/1812.10492} {arXiv:1812.10492 [astro-ph.IM]}
  \BibitemShut {NoStop}%
\bibitem [{\citenamefont {Choma}\ \emph {et~al.}(2018)\citenamefont {Choma},
  \citenamefont {Monti}, \citenamefont {Gerhardt}, \citenamefont {Palczewski},
  \citenamefont {Ronaghi}, \citenamefont {Bhimji}, \citenamefont {Bronstein},
  \citenamefont {Klein}, \citenamefont {Bruna} \emph
  {et~al.}}]{choma2018graph}%
  \BibitemOpen
  \bibfield  {author} {\bibinfo {author} {\bibfnamefont {N.}~\bibnamefont
  {Choma}}, \bibinfo {author} {\bibfnamefont {F.}~\bibnamefont {Monti}},
  \bibinfo {author} {\bibfnamefont {L.}~\bibnamefont {Gerhardt}}, \bibinfo
  {author} {\bibfnamefont {T.}~\bibnamefont {Palczewski}}, \bibinfo {author}
  {\bibfnamefont {Z.}~\bibnamefont {Ronaghi}}, \bibinfo {author} {\bibfnamefont
  {W.}~\bibnamefont {Bhimji}}, \bibinfo {author} {\bibfnamefont {M.~M.}\
  \bibnamefont {Bronstein}}, \bibinfo {author} {\bibfnamefont {S.~R.}\
  \bibnamefont {Klein}}, \bibinfo {author} {\bibfnamefont {J.}~\bibnamefont
  {Bruna}},  \emph {et~al.},\ }\href@noop {} {\bibfield  {journal} {\bibinfo
  {journal} {arXiv preprint arXiv:1809.06166}\ } (\bibinfo {year}
  {2018})}\BibitemShut {NoStop}%
\bibitem [{\citenamefont {{Ismail Fawaz}}\ \emph {et~al.}(2018)\citenamefont
  {{Ismail Fawaz}}, \citenamefont {{Forestier}}, \citenamefont {{Weber}},
  \citenamefont {{Idoumghar}},\ and\ \citenamefont {{Muller}}}]{ismail:2018I}%
  \BibitemOpen
  \bibfield  {author} {\bibinfo {author} {\bibfnamefont {H.}~\bibnamefont
  {{Ismail Fawaz}}}, \bibinfo {author} {\bibfnamefont {G.}~\bibnamefont
  {{Forestier}}}, \bibinfo {author} {\bibfnamefont {J.}~\bibnamefont
  {{Weber}}}, \bibinfo {author} {\bibfnamefont {L.}~\bibnamefont
  {{Idoumghar}}}, \ and\ \bibinfo {author} {\bibfnamefont {P.-A.}\ \bibnamefont
  {{Muller}}},\ }\href@noop {} {\bibfield  {journal} {\bibinfo  {journal}
  {arXiv e-prints}\ ,\ \bibinfo {eid} {arXiv:1809.04356}} (\bibinfo {year}
  {2018})},\ \Eprint {http://arxiv.org/abs/1809.04356} {arXiv:1809.04356
  [cs.LG]} \BibitemShut {NoStop}%
\bibitem [{\citenamefont {George}\ and\ \citenamefont
  {Huerta}(2018)}]{geodf:2017a}%
  \BibitemOpen
  \bibfield  {author} {\bibinfo {author} {\bibfnamefont {D.}~\bibnamefont
  {George}}\ and\ \bibinfo {author} {\bibfnamefont {E.~A.}\ \bibnamefont
  {Huerta}},\ }\href {\doibase 10.1103/PhysRevD.97.044039} {\bibfield
  {journal} {\bibinfo  {journal} {Phys. Rev. D}\ }\textbf {\bibinfo {volume}
  {97}},\ \bibinfo {pages} {044039} (\bibinfo {year} {2018})},\ \Eprint
  {http://arxiv.org/abs/1701.00008} {arXiv:1701.00008 [astro-ph.IM]}
  \BibitemShut {NoStop}%
\bibitem [{\citenamefont {{George}}\ and\ \citenamefont
  {{Huerta}}(2018)}]{geodf:2017b}%
  \BibitemOpen
  \bibfield  {author} {\bibinfo {author} {\bibfnamefont {D.}~\bibnamefont
  {{George}}}\ and\ \bibinfo {author} {\bibfnamefont {E.~A.}\ \bibnamefont
  {{Huerta}}},\ }\href {\doibase 10.1016/j.physletb.2017.12.053} {\bibfield
  {journal} {\bibinfo  {journal} {Physics Letters B}\ }\textbf {\bibinfo
  {volume} {778}},\ \bibinfo {pages} {64} (\bibinfo {year} {2018})},\ \Eprint
  {http://arxiv.org/abs/1711.03121} {arXiv:1711.03121 [gr-qc]} \BibitemShut
  {NoStop}%
\bibitem [{\citenamefont {{Shen}}\ \emph {et~al.}(2017)\citenamefont {{Shen}},
  \citenamefont {{George}}, \citenamefont {{Huerta}},\ and\ \citenamefont
  {{Zhao}}}]{hshen:2017}%
  \BibitemOpen
  \bibfield  {author} {\bibinfo {author} {\bibfnamefont {H.}~\bibnamefont
  {{Shen}}}, \bibinfo {author} {\bibfnamefont {D.}~\bibnamefont {{George}}},
  \bibinfo {author} {\bibfnamefont {E.~A.}\ \bibnamefont {{Huerta}}}, \ and\
  \bibinfo {author} {\bibfnamefont {Z.}~\bibnamefont {{Zhao}}},\ }\href@noop {}
  {\bibfield  {journal} {\bibinfo  {journal} {ArXiv e-prints}\ } (\bibinfo
  {year} {2017})},\ \Eprint {http://arxiv.org/abs/1711.09919} {arXiv:1711.09919
  [gr-qc]} \BibitemShut {NoStop}%
\bibitem [{\citenamefont {{Wei}}\ and\ \citenamefont
  {{Huerta}}(2019)}]{wei:2019W}%
  \BibitemOpen
  \bibfield  {author} {\bibinfo {author} {\bibfnamefont {W.}~\bibnamefont
  {{Wei}}}\ and\ \bibinfo {author} {\bibfnamefont {E.~A.}\ \bibnamefont
  {{Huerta}}},\ }\href@noop {} {\bibfield  {journal} {\bibinfo  {journal}
  {arXiv e-prints}\ ,\ \bibinfo {eid} {arXiv:1901.00869}} (\bibinfo {year}
  {2019})},\ \Eprint {http://arxiv.org/abs/1901.00869} {arXiv:1901.00869
  [gr-qc]} \BibitemShut {NoStop}%
\bibitem [{\citenamefont {{Rebei}}\ \emph {et~al.}(2018)\citenamefont
  {{Rebei}}, \citenamefont {{Huerta}}, \citenamefont {{Wang}}, \citenamefont
  {{Habib}}, \citenamefont {{Haas}}, \citenamefont {{Johnson}},\ and\
  \citenamefont {{George}}}]{Rebei:2018R}%
  \BibitemOpen
  \bibfield  {author} {\bibinfo {author} {\bibfnamefont {A.}~\bibnamefont
  {{Rebei}}}, \bibinfo {author} {\bibfnamefont {E.~A.}\ \bibnamefont
  {{Huerta}}}, \bibinfo {author} {\bibfnamefont {S.}~\bibnamefont {{Wang}}},
  \bibinfo {author} {\bibfnamefont {S.}~\bibnamefont {{Habib}}}, \bibinfo
  {author} {\bibfnamefont {R.}~\bibnamefont {{Haas}}}, \bibinfo {author}
  {\bibfnamefont {D.}~\bibnamefont {{Johnson}}}, \ and\ \bibinfo {author}
  {\bibfnamefont {D.}~\bibnamefont {{George}}},\ }\href@noop {} {\bibfield
  {journal} {\bibinfo  {journal} {arXiv e-prints}\ ,\ \bibinfo {eid}
  {arXiv:1807.09787}} (\bibinfo {year} {2018})},\ \Eprint
  {http://arxiv.org/abs/1807.09787} {arXiv:1807.09787 [gr-qc]} \BibitemShut
  {NoStop}%
\bibitem [{\citenamefont {George}\ \emph {et~al.}(2018)\citenamefont {George},
  \citenamefont {Shen},\ and\ \citenamefont {Huerta}}]{shengeorge:PhysRevD97}%
  \BibitemOpen
  \bibfield  {author} {\bibinfo {author} {\bibfnamefont {D.}~\bibnamefont
  {George}}, \bibinfo {author} {\bibfnamefont {H.}~\bibnamefont {Shen}}, \ and\
  \bibinfo {author} {\bibfnamefont {E.~A.}\ \bibnamefont {Huerta}},\ }\href
  {\doibase 10.1103/PhysRevD.97.101501} {\bibfield  {journal} {\bibinfo
  {journal} {Phys. Rev. D}\ }\textbf {\bibinfo {volume} {97}},\ \bibinfo
  {pages} {101501} (\bibinfo {year} {2018})}\BibitemShut {NoStop}%
\bibitem [{\citenamefont {{Chua}}\ \emph {et~al.}(2018)\citenamefont {{Chua}},
  \citenamefont {{Galley}},\ and\ \citenamefont {{Vallisneri}}}]{AlvinC:2018}%
  \BibitemOpen
  \bibfield  {author} {\bibinfo {author} {\bibfnamefont {A.~J.~K.}\
  \bibnamefont {{Chua}}}, \bibinfo {author} {\bibfnamefont {C.~R.}\
  \bibnamefont {{Galley}}}, \ and\ \bibinfo {author} {\bibfnamefont
  {M.}~\bibnamefont {{Vallisneri}}},\ }\href@noop {} {\bibfield  {journal}
  {\bibinfo  {journal} {ArXiv e-prints}\ } (\bibinfo {year} {2018})},\ \Eprint
  {http://arxiv.org/abs/1811.05491} {arXiv:1811.05491 [astro-ph.IM]}
  \BibitemShut {NoStop}%
\bibitem [{\citenamefont {{Gabbard}}\ \emph {et~al.}(2018)\citenamefont
  {{Gabbard}}, \citenamefont {{Williams}}, \citenamefont {{Hayes}},\ and\
  \citenamefont {{Messenger}}}]{2018GN}%
  \BibitemOpen
  \bibfield  {author} {\bibinfo {author} {\bibfnamefont {H.}~\bibnamefont
  {{Gabbard}}}, \bibinfo {author} {\bibfnamefont {M.}~\bibnamefont
  {{Williams}}}, \bibinfo {author} {\bibfnamefont {F.}~\bibnamefont {{Hayes}}},
  \ and\ \bibinfo {author} {\bibfnamefont {C.}~\bibnamefont {{Messenger}}},\
  }\href {\doibase 10.1103/PhysRevLett.120.141103} {\bibfield  {journal}
  {\bibinfo  {journal} {Physical Review Letters}\ }\textbf {\bibinfo {volume}
  {120}},\ \bibinfo {eid} {141103} (\bibinfo {year} {2018})},\ \Eprint
  {http://arxiv.org/abs/1712.06041} {arXiv:1712.06041 [astro-ph.IM]}
  \BibitemShut {NoStop}%
\bibitem [{\citenamefont {{Fan}}\ \emph {et~al.}(2018)\citenamefont {{Fan}},
  \citenamefont {{Li}}, \citenamefont {{Li}}, \citenamefont {{Zhong}},\ and\
  \citenamefont {{Cao}}}]{Fan:2018}%
  \BibitemOpen
  \bibfield  {author} {\bibinfo {author} {\bibfnamefont {X.}~\bibnamefont
  {{Fan}}}, \bibinfo {author} {\bibfnamefont {J.}~\bibnamefont {{Li}}},
  \bibinfo {author} {\bibfnamefont {X.}~\bibnamefont {{Li}}}, \bibinfo {author}
  {\bibfnamefont {Y.}~\bibnamefont {{Zhong}}}, \ and\ \bibinfo {author}
  {\bibfnamefont {J.}~\bibnamefont {{Cao}}},\ }\href@noop {} {\bibfield
  {journal} {\bibinfo  {journal} {ArXiv e-prints}\ } (\bibinfo {year}
  {2018})},\ \Eprint {http://arxiv.org/abs/1811.01380} {arXiv:1811.01380
  [astro-ph.IM]} \BibitemShut {NoStop}%
\bibitem [{\citenamefont {{Gonz{\'a}lez}}\ and\ \citenamefont
  {{Guzm{\'a}n}}(2018)}]{Gonza:2018}%
  \BibitemOpen
  \bibfield  {author} {\bibinfo {author} {\bibfnamefont {J.~A.}\ \bibnamefont
  {{Gonz{\'a}lez}}}\ and\ \bibinfo {author} {\bibfnamefont {F.~S.}\
  \bibnamefont {{Guzm{\'a}n}}},\ }\href {\doibase 10.1103/PhysRevD.97.063001}
  {\bibfield  {journal} {\bibinfo  {journal} {\prd}\ }\textbf {\bibinfo
  {volume} {97}},\ \bibinfo {eid} {063001} (\bibinfo {year} {2018})},\ \Eprint
  {http://arxiv.org/abs/1803.06060} {arXiv:1803.06060 [astro-ph.HE]}
  \BibitemShut {NoStop}%
\bibitem [{\citenamefont {{Fujimoto}}\ \emph {et~al.}(2018)\citenamefont
  {{Fujimoto}}, \citenamefont {{Fukushima}},\ and\ \citenamefont
  {{Murase}}}]{Fuji:2018}%
  \BibitemOpen
  \bibfield  {author} {\bibinfo {author} {\bibfnamefont {Y.}~\bibnamefont
  {{Fujimoto}}}, \bibinfo {author} {\bibfnamefont {K.}~\bibnamefont
  {{Fukushima}}}, \ and\ \bibinfo {author} {\bibfnamefont {K.}~\bibnamefont
  {{Murase}}},\ }\href {\doibase 10.1103/PhysRevD.98.023019} {\bibfield
  {journal} {\bibinfo  {journal} {\prd}\ }\textbf {\bibinfo {volume} {98}},\
  \bibinfo {eid} {023019} (\bibinfo {year} {2018})},\ \Eprint
  {http://arxiv.org/abs/1711.06748} {arXiv:1711.06748 [nucl-th]} \BibitemShut
  {NoStop}%
\bibitem [{\citenamefont {{Li}}\ \emph {et~al.}(2017)\citenamefont {{Li}},
  \citenamefont {{Yu}},\ and\ \citenamefont {{Fan}}}]{LiYu:2017}%
  \BibitemOpen
  \bibfield  {author} {\bibinfo {author} {\bibfnamefont {X.}~\bibnamefont
  {{Li}}}, \bibinfo {author} {\bibfnamefont {W.}~\bibnamefont {{Yu}}}, \ and\
  \bibinfo {author} {\bibfnamefont {X.}~\bibnamefont {{Fan}}},\ }\href@noop {}
  {\bibfield  {journal} {\bibinfo  {journal} {ArXiv e-prints}\ } (\bibinfo
  {year} {2017})},\ \Eprint {http://arxiv.org/abs/1712.00356} {arXiv:1712.00356
  [astro-ph.IM]} \BibitemShut {NoStop}%
\bibitem [{\citenamefont {{Nakano}}\ \emph {et~al.}(2018)\citenamefont
  {{Nakano}} \emph {et~al.}}]{Nakano:2018}%
  \BibitemOpen
  \bibfield  {author} {\bibinfo {author} {\bibfnamefont {H.}~\bibnamefont
  {{Nakano}}} \emph {et~al.},\ }\href@noop {} {\bibfield  {journal} {\bibinfo
  {journal} {ArXiv e-prints}\ } (\bibinfo {year} {2018})},\ \Eprint
  {http://arxiv.org/abs/1811.06443} {arXiv:1811.06443 [gr-qc]} \BibitemShut
  {NoStop}%
\bibitem [{\citenamefont {Springenberg}\ \emph {et~al.}(2016)\citenamefont
  {Springenberg}, \citenamefont {Klein}, \citenamefont {Falkner},\ and\
  \citenamefont {Hutter}}]{NIPS2016_6117}%
  \BibitemOpen
  \bibfield  {author} {\bibinfo {author} {\bibfnamefont {J.~T.}\ \bibnamefont
  {Springenberg}}, \bibinfo {author} {\bibfnamefont {A.}~\bibnamefont {Klein}},
  \bibinfo {author} {\bibfnamefont {S.}~\bibnamefont {Falkner}}, \ and\
  \bibinfo {author} {\bibfnamefont {F.}~\bibnamefont {Hutter}},\ }in\ \href
  {http://papers.nips.cc/paper/6117-bayesian-optimization-with-robust-bayesian-neural-networks.pdf}
  {\emph {\bibinfo {booktitle} {Advances in Neural Information Processing
  Systems 29}}},\ \bibinfo {editor} {edited by\ \bibinfo {editor}
  {\bibfnamefont {D.~D.}\ \bibnamefont {Lee}}, \bibinfo {editor} {\bibfnamefont
  {M.}~\bibnamefont {Sugiyama}}, \bibinfo {editor} {\bibfnamefont {U.~V.}\
  \bibnamefont {Luxburg}}, \bibinfo {editor} {\bibfnamefont {I.}~\bibnamefont
  {Guyon}}, \ and\ \bibinfo {editor} {\bibfnamefont {R.}~\bibnamefont
  {Garnett}}}\ (\bibinfo  {publisher} {Curran Associates, Inc.},\ \bibinfo
  {year} {2016})\ pp.\ \bibinfo {pages} {4134--4142}\BibitemShut {NoStop}%
\bibitem [{\citenamefont {Gal}\ and\ \citenamefont
  {Ghahramani}(2016)}]{gal2016dropout}%
  \BibitemOpen
  \bibfield  {author} {\bibinfo {author} {\bibfnamefont {Y.}~\bibnamefont
  {Gal}}\ and\ \bibinfo {author} {\bibfnamefont {Z.}~\bibnamefont
  {Ghahramani}},\ }in\ \href@noop {} {\emph {\bibinfo {booktitle}
  {international conference on machine learning}}}\ (\bibinfo {year} {2016})\
  pp.\ \bibinfo {pages} {1050--1059}\BibitemShut {NoStop}%
\bibitem [{\citenamefont {Gal}\ and\ \citenamefont
  {Ghahramani}(2015)}]{gal2015bayesian}%
  \BibitemOpen
  \bibfield  {author} {\bibinfo {author} {\bibfnamefont {Y.}~\bibnamefont
  {Gal}}\ and\ \bibinfo {author} {\bibfnamefont {Z.}~\bibnamefont
  {Ghahramani}},\ }in\ \href@noop {} {\emph {\bibinfo {booktitle} {Proc. ICLR
  workshop track}}}\ (\bibinfo {year} {2015})\BibitemShut {NoStop}%
\bibitem [{\citenamefont {{George}}\ \emph {et~al.}(2017)\citenamefont
  {{George}}, \citenamefont {{Shen}},\ and\ \citenamefont {{Huerta}}}]{dgNIPS}%
  \BibitemOpen
  \bibfield  {author} {\bibinfo {author} {\bibfnamefont {D.}~\bibnamefont
  {{George}}}, \bibinfo {author} {\bibfnamefont {H.}~\bibnamefont {{Shen}}}, \
  and\ \bibinfo {author} {\bibfnamefont {E.~A.}\ \bibnamefont {{Huerta}}},\
  }\href@noop {} {\bibfield  {journal} {\bibinfo  {journal} {ArXiv e-prints}\ }
  (\bibinfo {year} {2017})},\ \Eprint {http://arxiv.org/abs/1711.07468}
  {arXiv:1711.07468 [astro-ph.IM]} \BibitemShut {NoStop}%
\bibitem [{\citenamefont {{Abadi}}\ \emph {et~al.}(2016)\citenamefont
  {{Abadi}}, \citenamefont {{Agarwal}} \emph {et~al.}}]{abadi2016tensorflow}%
  \BibitemOpen
  \bibfield  {author} {\bibinfo {author} {\bibfnamefont {M.}~\bibnamefont
  {{Abadi}}}, \bibinfo {author} {\bibfnamefont {A.}~\bibnamefont {{Agarwal}}},
  \emph {et~al.},\ }\href@noop {} {\bibfield  {journal} {\bibinfo  {journal}
  {ArXiv e-prints}\ } (\bibinfo {year} {2016})},\ \Eprint
  {http://arxiv.org/abs/1603.04467} {arXiv:1603.04467 [cs.DC]} \BibitemShut
  {NoStop}%
\bibitem [{\citenamefont {Paszke}\ \emph {et~al.}(2017)\citenamefont {Paszke},
  \citenamefont {Gross}, \citenamefont {Chintala}, \citenamefont {Chanan},
  \citenamefont {Yang}, \citenamefont {DeVito}, \citenamefont {Lin},
  \citenamefont {Desmaison}, \citenamefont {Antiga},\ and\ \citenamefont
  {Lerer}}]{paszke2017automatic}%
  \BibitemOpen
  \bibfield  {author} {\bibinfo {author} {\bibfnamefont {A.}~\bibnamefont
  {Paszke}}, \bibinfo {author} {\bibfnamefont {S.}~\bibnamefont {Gross}},
  \bibinfo {author} {\bibfnamefont {S.}~\bibnamefont {Chintala}}, \bibinfo
  {author} {\bibfnamefont {G.}~\bibnamefont {Chanan}}, \bibinfo {author}
  {\bibfnamefont {E.}~\bibnamefont {Yang}}, \bibinfo {author} {\bibfnamefont
  {Z.}~\bibnamefont {DeVito}}, \bibinfo {author} {\bibfnamefont
  {Z.}~\bibnamefont {Lin}}, \bibinfo {author} {\bibfnamefont {A.}~\bibnamefont
  {Desmaison}}, \bibinfo {author} {\bibfnamefont {L.}~\bibnamefont {Antiga}}, \
  and\ \bibinfo {author} {\bibfnamefont {A.}~\bibnamefont {Lerer}},\ }in\
  \href@noop {} {\emph {\bibinfo {booktitle} {NIPS-W}}}\ (\bibinfo {year}
  {2017})\BibitemShut {NoStop}%
\bibitem [{\citenamefont {{Comet User Guide}}()}]{comet}%
  \BibitemOpen
  \bibfield  {author} {\bibinfo {author} {\bibnamefont {{Comet User Guide}}},\
  }\href@noop {} {}\bibinfo {note}
  {\url{https://portal.xsede.org/sdsc-comet}}\BibitemShut {NoStop}%
\bibitem [{\citenamefont {{Bridges User Guide}}()}]{bridges}%
  \BibitemOpen
  \bibfield  {author} {\bibinfo {author} {\bibnamefont {{Bridges User
  Guide}}},\ }\href@noop {} {}\bibinfo {note}
  {\url{https://www.psc.edu/bridges/user-guide/gpu-use}}\BibitemShut {NoStop}%
\bibitem [{\citenamefont {{Frontera}}()}]{frontera}%
  \BibitemOpen
  \bibfield  {author} {\bibinfo {author} {\bibnamefont {{Frontera}}},\
  }\href@noop {} {}\bibinfo {note}
  {\url{https://www.tacc.utexas.edu/systems/frontera}}\BibitemShut {NoStop}%
\bibitem [{\citenamefont {{Huerta}}\ \emph
  {et~al.}(2018{\natexlab{b}})\citenamefont {{Huerta}}, \citenamefont {{Haas}},
  \citenamefont {{Jha}}, \citenamefont {{Neubauer}},\ and\ \citenamefont
  {{Katz}}}]{HuertaES}%
  \BibitemOpen
  \bibfield  {author} {\bibinfo {author} {\bibfnamefont {E.~A.}\ \bibnamefont
  {{Huerta}}}, \bibinfo {author} {\bibfnamefont {R.}~\bibnamefont {{Haas}}},
  \bibinfo {author} {\bibfnamefont {S.}~\bibnamefont {{Jha}}}, \bibinfo
  {author} {\bibfnamefont {M.}~\bibnamefont {{Neubauer}}}, \ and\ \bibinfo
  {author} {\bibfnamefont {D.~S.}\ \bibnamefont {{Katz}}},\ }\href@noop {}
  {\bibfield  {journal} {\bibinfo  {journal} {ArXiv e-prints}\ } (\bibinfo
  {year} {2018}{\natexlab{b}})},\ \Eprint {http://arxiv.org/abs/1810.03056}
  {arXiv:1810.03056 [cs.DC]} \BibitemShut {NoStop}%
\bibitem [{\citenamefont {Huerta}\ \emph {et~al.}(2017)\citenamefont {Huerta},
  \citenamefont {Haas}, \citenamefont {Fajardo}, \citenamefont {Katz},
  \citenamefont {Anderson}, \citenamefont {Couvares}, \citenamefont {Willis},
  \citenamefont {Bouvet}, \citenamefont {Enos}, \citenamefont {Kramer},
  \citenamefont {Leong},\ and\ \citenamefont {Wheeler}}]{8109152}%
  \BibitemOpen
  \bibfield  {author} {\bibinfo {author} {\bibfnamefont {E.~A.}\ \bibnamefont
  {Huerta}}, \bibinfo {author} {\bibfnamefont {R.}~\bibnamefont {Haas}},
  \bibinfo {author} {\bibfnamefont {E.}~\bibnamefont {Fajardo}}, \bibinfo
  {author} {\bibfnamefont {D.~S.}\ \bibnamefont {Katz}}, \bibinfo {author}
  {\bibfnamefont {S.}~\bibnamefont {Anderson}}, \bibinfo {author}
  {\bibfnamefont {P.}~\bibnamefont {Couvares}}, \bibinfo {author}
  {\bibfnamefont {J.}~\bibnamefont {Willis}}, \bibinfo {author} {\bibfnamefont
  {T.}~\bibnamefont {Bouvet}}, \bibinfo {author} {\bibfnamefont
  {J.}~\bibnamefont {Enos}}, \bibinfo {author} {\bibfnamefont {W.~T.~C.}\
  \bibnamefont {Kramer}}, \bibinfo {author} {\bibfnamefont {H.~W.}\
  \bibnamefont {Leong}}, \ and\ \bibinfo {author} {\bibfnamefont
  {D.}~\bibnamefont {Wheeler}},\ }in\ \href {\doibase 10.1109/eScience.2017.47}
  {\emph {\bibinfo {booktitle} {2017 IEEE 13th International Conference on
  e-Science (e-Science)}}}\ (\bibinfo {year} {2017})\ pp.\ \bibinfo {pages}
  {335--344}\BibitemShut {NoStop}%
\bibitem [{\citenamefont {{Weitzel}}\ \emph {et~al.}(2017)\citenamefont
  {{Weitzel}}, \citenamefont {{Bockelman}}, \citenamefont {{Brown}},
  \citenamefont {{Couvares}}, \citenamefont {{W{\"u}rthwein}},\ and\
  \citenamefont {{Fajardo Hernandez}}}]{2017Weitzel}%
  \BibitemOpen
  \bibfield  {author} {\bibinfo {author} {\bibfnamefont {D.}~\bibnamefont
  {{Weitzel}}}, \bibinfo {author} {\bibfnamefont {B.}~\bibnamefont
  {{Bockelman}}}, \bibinfo {author} {\bibfnamefont {D.~A.}\ \bibnamefont
  {{Brown}}}, \bibinfo {author} {\bibfnamefont {P.}~\bibnamefont {{Couvares}}},
  \bibinfo {author} {\bibfnamefont {F.}~\bibnamefont {{W{\"u}rthwein}}}, \ and\
  \bibinfo {author} {\bibfnamefont {E.}~\bibnamefont {{Fajardo Hernandez}}},\
  }\href@noop {} {\bibfield  {journal} {\bibinfo  {journal} {arXiv e-prints}\
  ,\ \bibinfo {eid} {arXiv:1705.06202}} (\bibinfo {year} {2017})},\ \Eprint
  {http://arxiv.org/abs/1705.06202} {arXiv:1705.06202 [cs.DC]} \BibitemShut
  {NoStop}%
\bibitem [{\citenamefont {Belkin}\ \emph {et~al.}(2018)\citenamefont {Belkin},
  \citenamefont {Haas}, \citenamefont {Arnold}, \citenamefont {Leong},
  \citenamefont {Huerta}, \citenamefont {Lesny},\ and\ \citenamefont
  {Neubauer}}]{shifter}%
  \BibitemOpen
  \bibfield  {author} {\bibinfo {author} {\bibfnamefont {M.}~\bibnamefont
  {Belkin}}, \bibinfo {author} {\bibfnamefont {R.}~\bibnamefont {Haas}},
  \bibinfo {author} {\bibfnamefont {G.~W.}\ \bibnamefont {Arnold}}, \bibinfo
  {author} {\bibfnamefont {H.~W.}\ \bibnamefont {Leong}}, \bibinfo {author}
  {\bibfnamefont {E.~A.}\ \bibnamefont {Huerta}}, \bibinfo {author}
  {\bibfnamefont {D.}~\bibnamefont {Lesny}}, \ and\ \bibinfo {author}
  {\bibfnamefont {M.}~\bibnamefont {Neubauer}},\ }in\ \href {\doibase
  10.1145/3219104.3219145} {\emph {\bibinfo {booktitle} {Proceedings of the
  Practice and Experience on Advanced Research Computing}}},\ \bibinfo {series
  and number} {PEARC '18}\ (\bibinfo  {publisher} {ACM},\ \bibinfo {address}
  {New York, NY, USA},\ \bibinfo {year} {2018})\ pp.\ \bibinfo {pages}
  {43:1--43:8}\BibitemShut {NoStop}%
\bibitem [{\citenamefont {{Street}}\ \emph {et~al.}(2018)\citenamefont
  {{Street}}, \citenamefont {{Bowman}}, \citenamefont {{Saunders}},\ and\
  \citenamefont {{Boroson}}}]{street:2018SPIE}%
  \BibitemOpen
  \bibfield  {author} {\bibinfo {author} {\bibfnamefont {R.~A.}\ \bibnamefont
  {{Street}}}, \bibinfo {author} {\bibfnamefont {M.}~\bibnamefont {{Bowman}}},
  \bibinfo {author} {\bibfnamefont {E.~S.}\ \bibnamefont {{Saunders}}}, \ and\
  \bibinfo {author} {\bibfnamefont {T.}~\bibnamefont {{Boroson}}},\ }in\ \href
  {\doibase 10.1117/12.2312293} {\emph {\bibinfo {booktitle} {Software and
  Cyberinfrastructure for Astronomy V}}},\ \bibinfo {series} {Society of
  Photo-Optical Instrumentation Engineers (SPIE) Conference Series}, Vol.\
  \bibinfo {volume} {10707}\ (\bibinfo {year} {2018})\ p.\ \bibinfo {pages}
  {1070711},\ \Eprint {http://arxiv.org/abs/1806.09557} {arXiv:1806.09557
  [astro-ph.IM]} \BibitemShut {NoStop}%
\bibitem [{\citenamefont {{Astropy}}(2018)}]{astropy}%
  \BibitemOpen
  \bibfield  {author} {\bibinfo {author} {\bibnamefont {{Astropy}}},\
  }\href@noop {} {\enquote {\bibinfo {title} {The astropy project},}\ }
  (\bibinfo {year} {2018}),\ \bibinfo {note}
  {\url{http://www.astropy.org/}}\BibitemShut {NoStop}%
\bibitem [{\citenamefont {{yt Community}}(2018)}]{ytproject}%
  \BibitemOpen
  \bibfield  {author} {\bibinfo {author} {\bibnamefont {{yt Community}}},\
  }\href@noop {} {\enquote {\bibinfo {title} {yt project},}\ } (\bibinfo {year}
  {2018}),\ \bibinfo {note} {\url{http://www.astropy.org/}}\BibitemShut
  {NoStop}%
\bibitem [{\citenamefont {{IceCube}}(2018)}]{icerepo}%
  \BibitemOpen
  \bibfield  {author} {\bibinfo {author} {\bibnamefont {{IceCube}}},\
  }\href@noop {} {\enquote {\bibinfo {title} {Icecube open source software},}\
  } (\bibinfo {year} {2018}),\ \bibinfo {note}
  {\url{https://github.com/IceCubeOpenSource}}\BibitemShut {NoStop}%
\bibitem [{\citenamefont {{Porter}}(2008)}]{Porter:2008ar}%
  \BibitemOpen
  \bibfield  {author} {\bibinfo {author} {\bibfnamefont {F.~C.}\ \bibnamefont
  {{Porter}}},\ }\href@noop {} {\bibfield  {journal} {\bibinfo  {journal}
  {arXiv e-prints}\ ,\ \bibinfo {eid} {arXiv:0804.0380}} (\bibinfo {year}
  {2008})},\ \Eprint {http://arxiv.org/abs/0804.0380} {arXiv:0804.0380
  [physics.data-an]} \BibitemShut {NoStop}%
\bibitem [{\citenamefont {{Software Carpentry}}(2018)}]{scar}%
  \BibitemOpen
  \bibfield  {author} {\bibinfo {author} {\bibnamefont {{Software
  Carpentry}}},\ }\href@noop {} {\enquote {\bibinfo {title} {Teaching basic lab
  skills for research computing},}\ } (\bibinfo {year} {2018}),\ \bibinfo
  {note} {\url{https://software-carpentry.org}}\BibitemShut {NoStop}%
\bibitem [{\citenamefont {{Elmer}}\ \emph {et~al.}(2017)\citenamefont
  {{Elmer}}, \citenamefont {{Neubauer}},\ and\ \citenamefont
  {{Sokoloff}}}]{S2I2HEP-SPj}%
  \BibitemOpen
  \bibfield  {author} {\bibinfo {author} {\bibfnamefont {P.}~\bibnamefont
  {{Elmer}}}, \bibinfo {author} {\bibfnamefont {M.}~\bibnamefont {{Neubauer}}},
  \ and\ \bibinfo {author} {\bibfnamefont {M.~D.}\ \bibnamefont {{Sokoloff}}},\
  }\href@noop {} {\bibfield  {journal} {\bibinfo  {journal} {arXiv e-prints}\
  ,\ \bibinfo {eid} {arXiv:1712.06592}} (\bibinfo {year} {2017})},\ \Eprint
  {http://arxiv.org/abs/1712.06592} {arXiv:1712.06592 [physics.comp-ph]}
  \BibitemShut {NoStop}%
\bibitem [{\citenamefont {{Albrecht}}\ \emph {et~al.}(2017)\citenamefont
  {{Albrecht}} \emph {et~al.}}]{CWP}%
  \BibitemOpen
  \bibfield  {author} {\bibinfo {author} {\bibfnamefont {J.}~\bibnamefont
  {{Albrecht}}} \emph {et~al.},\ }\href@noop {} {\bibfield  {journal} {\bibinfo
   {journal} {arXiv e-prints}\ ,\ \bibinfo {eid} {arXiv:1712.06982}} (\bibinfo
  {year} {2017})},\ \Eprint {http://arxiv.org/abs/1712.06982} {arXiv:1712.06982
  [physics.comp-ph]} \BibitemShut {NoStop}%
\bibitem [{\citenamefont {{Allen}}\ \emph {et~al.}(2018)\citenamefont
  {{Allen}}, \citenamefont {{Anderson}}, \citenamefont {{Blaufuss}},
  \citenamefont {{Bloom}}, \citenamefont {{Brady}}, \citenamefont
  {{Burke-Spolaor}}, \citenamefont {{Cenko}}, \citenamefont {{Connolly}},
  \citenamefont {{Couvares}}, \citenamefont {{Fox}}, \citenamefont {{Gal-Yam}},
  \citenamefont {{Gezari}}, \citenamefont {{Goodman}}, \citenamefont {{Grant}},
  \citenamefont {{Groot}}, \citenamefont {{Guillochon}}, \citenamefont
  {{Hanna}}, \citenamefont {{Hogg}}, \citenamefont {{Holley-Bockelmann}},
  \citenamefont {{Howell}}, \citenamefont {{Kaplan}}, \citenamefont
  {{Katsavounidis}}, \citenamefont {{Kowalski}}, \citenamefont {{Lehner}},
  \citenamefont {{Muthukrishna}}, \citenamefont {{Narayan}}, \citenamefont
  {{Peek}}, \citenamefont {{Saha}}, \citenamefont {{Shawhan}},\ and\
  \citenamefont {{Taboada}}}]{whitepaper:SCIMMA}%
  \BibitemOpen
  \bibfield  {author} {\bibinfo {author} {\bibfnamefont {G.}~\bibnamefont
  {{Allen}}}, \bibinfo {author} {\bibfnamefont {W.}~\bibnamefont {{Anderson}}},
  \bibinfo {author} {\bibfnamefont {E.}~\bibnamefont {{Blaufuss}}}, \bibinfo
  {author} {\bibfnamefont {J.~S.}\ \bibnamefont {{Bloom}}}, \bibinfo {author}
  {\bibfnamefont {P.}~\bibnamefont {{Brady}}}, \bibinfo {author} {\bibfnamefont
  {S.}~\bibnamefont {{Burke-Spolaor}}}, \bibinfo {author} {\bibfnamefont
  {S.~B.}\ \bibnamefont {{Cenko}}}, \bibinfo {author} {\bibfnamefont
  {A.}~\bibnamefont {{Connolly}}}, \bibinfo {author} {\bibfnamefont
  {P.}~\bibnamefont {{Couvares}}}, \bibinfo {author} {\bibfnamefont
  {D.}~\bibnamefont {{Fox}}}, \bibinfo {author} {\bibfnamefont
  {A.}~\bibnamefont {{Gal-Yam}}}, \bibinfo {author} {\bibfnamefont
  {S.}~\bibnamefont {{Gezari}}}, \bibinfo {author} {\bibfnamefont
  {A.}~\bibnamefont {{Goodman}}}, \bibinfo {author} {\bibfnamefont
  {D.}~\bibnamefont {{Grant}}}, \bibinfo {author} {\bibfnamefont
  {P.}~\bibnamefont {{Groot}}}, \bibinfo {author} {\bibfnamefont
  {J.}~\bibnamefont {{Guillochon}}}, \bibinfo {author} {\bibfnamefont
  {C.}~\bibnamefont {{Hanna}}}, \bibinfo {author} {\bibfnamefont {D.~W.}\
  \bibnamefont {{Hogg}}}, \bibinfo {author} {\bibfnamefont {K.}~\bibnamefont
  {{Holley-Bockelmann}}}, \bibinfo {author} {\bibfnamefont {D.~A.}\
  \bibnamefont {{Howell}}}, \bibinfo {author} {\bibfnamefont {D.}~\bibnamefont
  {{Kaplan}}}, \bibinfo {author} {\bibfnamefont {E.}~\bibnamefont
  {{Katsavounidis}}}, \bibinfo {author} {\bibfnamefont {M.}~\bibnamefont
  {{Kowalski}}}, \bibinfo {author} {\bibfnamefont {L.}~\bibnamefont
  {{Lehner}}}, \bibinfo {author} {\bibfnamefont {D.}~\bibnamefont
  {{Muthukrishna}}}, \bibinfo {author} {\bibfnamefont {G.}~\bibnamefont
  {{Narayan}}}, \bibinfo {author} {\bibfnamefont {J.~E.~G.}\ \bibnamefont
  {{Peek}}}, \bibinfo {author} {\bibfnamefont {A.}~\bibnamefont {{Saha}}},
  \bibinfo {author} {\bibfnamefont {P.}~\bibnamefont {{Shawhan}}}, \ and\
  \bibinfo {author} {\bibfnamefont {I.}~\bibnamefont {{Taboada}}},\ }\href@noop
  {} {\bibfield  {journal} {\bibinfo  {journal} {arXiv e-prints}\ ,\ \bibinfo
  {eid} {arXiv:1807.04780}} (\bibinfo {year} {2018})},\ \Eprint
  {http://arxiv.org/abs/1807.04780} {arXiv:1807.04780 [astro-ph.IM]}
  \BibitemShut {NoStop}%
\bibitem [{\citenamefont {{DESC}}(2018)}]{desc}%
  \BibitemOpen
  \bibfield  {author} {\bibinfo {author} {\bibnamefont {{DESC}}},\ }\href@noop
  {} {\enquote {\bibinfo {title} {Dark energy science collaboration},}\ }
  (\bibinfo {year} {2018}),\ \bibinfo {note}
  {\url{http://lsst-desc.org/}}\BibitemShut {NoStop}%
\end{thebibliography}%
\bibliographystyle{apsrev4-1}
\end{document}